\documentclass[11pt]{article}
\usepackage[T1]{fontenc}
\usepackage{lmodern}
\usepackage{axodraw2}
\usepackage{epsfig}
\usepackage{amsfonts}
\usepackage{amsmath,amssymb}
\usepackage{bbm,bm}
\usepackage{mathrsfs} %
\usepackage{listings}
\usepackage{cite}
\usepackage{multirow}
\allowdisplaybreaks[1]
\usepackage[
  colorlinks=true,
  linkcolor={red!50!black},
  citecolor={blue!50!black},
  filecolor=black,
  urlcolor=black,
  breaklinks=true
]{hyperref}
\usepackage{orcidlink}

\usepackage{algorithm}
\usepackage{algpseudocode}

\usepackage{color}
\definecolor{codeblue}{rgb}{0,0,0.5}
\definecolor{codered}{rgb}{0.6,0,0}
\definecolor{codegreen}{rgb}{0,0.5,0}

\usepackage{caption}
\DeclareCaptionFormat{listingformat}{\colorbox{black}{\color{white}\thelstlisting}}
\graphicspath{{.}{./figures/}}

\usepackage{tikz}
\usetikzlibrary{shapes,arrows,shadows,automata,positioning}
\newdimen\nodeDist
\def\WallGo{{\tt WallGo}}
\def\WallGoMatrix{\WallGo{\tt Matrix}}
\def\WallGoCollision{\WallGo{\tt Collision}}
\def\WallGoUrl{https://github.com/Wall-Go/WallGo}

\def\WallGoVersion{{\tt v1.1.1}}
\def\WallGoOldVersion{{\tt v1.0.0}}
\def\WallGoMatrixVersion{{\tt v1.1.0}}
\def\WallGoMatrixOldVersion{{\tt v1.0.0}}
\def\WallGoCollisionVersion{{\tt v1.1.0}}

\usepackage[normalem]{ulem} %

\usepackage{breakurl}
\usepackage{pifont}
\newcommand*\colourcheck[1]{%
  \expandafter\newcommand\csname #1check\endcsname{\textcolor{#1}{\ding{52}}}%
}
\colourcheck{blue}
\colourcheck{green}

\renewcommand{\vec}[1]{{\bf #1}}

\newcommand{\dd}{{\rm d}}

\newcommand{\alphan}{\alpha_{\rm n}}

\newcommand{\LamD}{\bmu}

\newcommand{\vw}{v_w}
\newcommand{\vJ}{v_\rmii{$J$}}
\newcommand{\cs}{c_s}

\usepackage{amsmath}
\usepackage{xparse}

\newcommand{\abs}[1]{\left\vert #1 \right\vert}

\newcommand{\Tc}{T_{\rm c}}
\newcommand{\Tn}{T_{\rm n}}
\newcommand{\vn}{v_{\rm n}}
\newcommand{\vcoll}{v_{\rm coll}}

\newcommand{\mW}{m_\rmii{$W$}}

\newcommand{\mA}{m_\rmii{$A$}}
\newcommand{\mH}{m_\rmii{$H$}}
\newcommand{\meff}{m_\rmi{eff}}

\newcommand{\gY}{y_t}

\newcommand{\bmu}{\bar\mu}

\newcommand{\supsetsim}{\mathrel{\ooalign{\raise.4ex\hbox{$\supset$}\cr$\raise-.9ex\hbox{$\sim$}$}}}

\def\lsi{\raise0.3ex\hbox{$<$\kern-0.75em\raise-1.1ex\hbox{$\sim$}}}
\def\gsi{\raise0.3ex\hbox{$>$\kern-0.75em\raise-1.1ex\hbox{$\sim$}}}

\newcommand{\nn}{\nonumber \\}
\newcommand{\rmi}[1]{{\mbox{\scriptsize #1}}}
\newcommand{\rmii}[1]{{\mbox{\tiny\rm{#1}}}}

\newcommand{\re}{\mathop{\mbox{Re}}}

\newcommand{\Tint}[1]{{\hbox{$\sum$}\!\!\!\!\!\!\!\int\,}_{\!\!\!\!\raise-0.9ex\hbox{$\scriptstyle{#1}$}}}
\newcommand{\Tinti}[1]{{{\Sigma}\!\!\!\!\raise0.3ex\hbox{$\int$}_\rmii{${#1}$}}}
\newcommand{\Tintip}[1]{{{\Sigma'}\!\!\!\!\!\raise0.3ex\hbox{$\int$}_\rmii{${#1}$}}}

\newcommand{\Veff}{{V_{\rmi{eff}}}} %

\newcommand{\deltabar}{\raise-0.02em\hbox{$\bar{}$}\hspace*{-0.8mm}{\delta}}

\NewDocumentCommand{\pdv}{O{} m m}{%
  \frac{\partial^{#1} #2}{\partial #3^{#1}}%
}

\NewDocumentCommand{\dv}{O{} m m}{%
  \frac{{\rm d}^{#1} #2}{{\rm d} #3^{#1}}%
}

\newcommand{\vev}{vacuum expectation value}
\renewcommand{\vev}{VEV}

\def\backtick{\char18}
\lstdefinestyle{backtickavailable}{literate={`}{\backtick}1, escapechar=@}

\def\Lwidth{1}

\newcommand{\picc}[1]{\;\parbox[c]{60pt}{\begin{picture}(60,30)(0,0)
\SetWidth{1.0}\SetScale{1.0} #1 \end{picture}}\;}
\newcommand{\picb}[1]{\;\parbox[c]{45pt}{\begin{picture}(45,30)(0,0)
\SetWidth{1.0}\SetScale{1.0} #1 \end{picture}}\;}

\def\Lglii(#1,#2)(#3,#4){
  \ZigZag(#1,#2)(#3,#4){\Lwidth}
  {#1 #3 sub #1 #3 sub mul #2 #4 sub #2 #4 sub mul add sqrt Ldensity mul}}
\def\Lgliii(#1,#2)(#3,#4){
  \Gluon(#1,#2)(#3,#4){\Lwidth}
  {#1 #3 sub #1 #3 sub mul #2 #4 sub #2 #4 sub mul add sqrt Ldensity mul}}
\def\Lglx(#1,#2)(#3,#4){\Photon(#1,#2)(#3,#4){\Lwidth}
  {#1 #3 sub #1 #3 sub mul #2 #4 sub #2 #4 sub mul add sqrt Ldensity mul}}
\def\Lxx(#1,#2)(#3,#4){\DashLine(#1,#2)(#3,#4){3}}
\def\Lqu(#1,#2)(#3,#4){\ArrowLine(#1,#2)(#3,#4)}
\def\Luq(#1,#2)(#3,#4){\ArrowLine(#3,#4)(#1,#2)}

\def\Textb(#1,#2,#3){\Text(#1,#2)[b]{{$\scriptstyle #3$}}}
\def\Texttl(#1,#2,#3){\Text(#1,#2)[tl]{{$\scriptstyle #3$}}}
\def\Textbl(#1,#2,#3){\Text(#1,#2)[bl]{{$\scriptstyle #3$}}}
\def\Texttr(#1,#2,#3){\Text(#1,#2)[tr]{{$\scriptstyle #3$}}}
\def\Textbr(#1,#2,#3){\Text(#1,#2)[br]{{$\scriptstyle #3$}}}

\def\Vtxvn(#1,#2,#3,#4,#5,#6,#7,#8){\picb{
  #1(22.5 45 cos 22.5 mul add,15 45 sin 22.5 mul add)(22.5,15)%
  #2(22.5 135 cos 22.5 mul add,15 135 sin 22.5 mul add)(22.5,15)%
  #3(22.5 225 cos 22.5 mul add,15 225 sin 22.5 mul add)(22.5,15)%
  #4(22.5 315 cos 22.5 mul add,15 315 sin 22.5 mul add)(22.5,15)%
  \Textb(45,22.5,#5)%
  \Textb(0,22.5,#6)%
  \Textb(0,0,#7)%
  \Textb(45,0,#8)%
  \Vertex(22.5,15){3}}}

\def\VtxvSn(#1,#2,#3,#4,#5,#6,#7,#8,#9){\picc{
  #1(40 45 cos 20 mul add,15 45 sin 20 mul add)(40,15)%
  #2(20 135 cos 20 mul add,15 135 sin 20 mul add)(20,15)%
  #3(20 225 cos 20 mul add,15 225 sin 20 mul add)(20,15)%
  #4(40 315 cos 20 mul add,15 315 sin 20 mul add)(40,15)%
  #5(20,15)(40,15)
  \Textbl(45,30,#6)%
  \Textbl(0,30,#7)%
  \Texttl(0,0,#8)%
  \Texttl(45,0,#9)%
}}
\def\VtxvTn(#1,#2,#3,#4,#5,#6,#7,#8,#9){\picb{
  #1(20 45 cos 20 mul add,15 45 sin 20 mul add)(20,25)%
  #2(20 135 cos 20 mul add,15 135 sin 20 mul add)(20,25)%
  #3(20 225 cos 20 mul add,15 225 sin 20 mul add)(20,5)%
  #4(20 315 cos 20 mul add,15 315 sin 20 mul add)(20,5)%
  #5(20,25)(20,5)%
  \Textbr(40,30,#6)%
  \Textbl(0,30,#7)%
  \Texttl(0,0,#8)%
  \Texttr(40,0,#9)%
}}
\def\VtxvUn(#1,#2,#3,#4,#5,#6,#7,#8,#9){\picc{%
  #1(%
    40 45 cos 20 mul add,15 45 sin 20 mul add)(%
    40 45 cos 20 mul add 20 sub 0.45 mul 20 add,15 45 sin 20 mul add 5 sub 0.45 mul 5 add)%
  #1(
    40 45 cos 20 mul add 20 sub 0.35 mul 20 add,15 45 sin 20 mul add 5 sub 0.35 mul 5 add)(%
    20,5)%
  #2(20 135 cos 20 mul add,15 135 sin 20 mul add)(20,25)%
  #3(20 225 cos 20 mul add,15 225 sin 20 mul add)(20,5)%
  #4(40 315 cos 20 mul add,15 315 sin 20 mul add)(20,25)%
  #5(20,25)(20,5)%
  \Textbl(50,30,#6)%
  \Textbl(0,30,#7)%
  \Texttl(0,0,#8)%
  \Texttl(50,0,#9)%
}}

\makeatletter \@addtoreset{equation}{section} \makeatother
\renewcommand{\theequation}{\arabic{section}.\arabic{equation}}
\makeatletter
\renewcommand\section{\@startsection{section}{1}{\z@}%
  {-5.5ex \@plus -1ex \@minus -.2ex}%
  {2.3ex \@plus.2ex}%
  {\normalfont\large\bfseries}}
\renewcommand\subsection{\@startsection{subsection}{2}{\z@}%
  {-3.25ex\@plus -1ex \@minus -.2ex}%
  {1.5ex \@plus .2ex}%
  {\normalfont\normalsize\bfseries}}
\renewcommand\thesection{\@arabic\c@section}
\renewcommand\thesubsection{\thesection.\@arabic\c@subsection}
\renewcommand{\@seccntformat}[1]{%
  \csname the#1\endcsname.\hspace{1.0em}}
\makeatother

\begin{document}

\flushbottom

\begin{titlepage}

  {
\flushright
CERN-TH-2025-221\\
  }

\begin{centering}
\includegraphics[height=3cm]{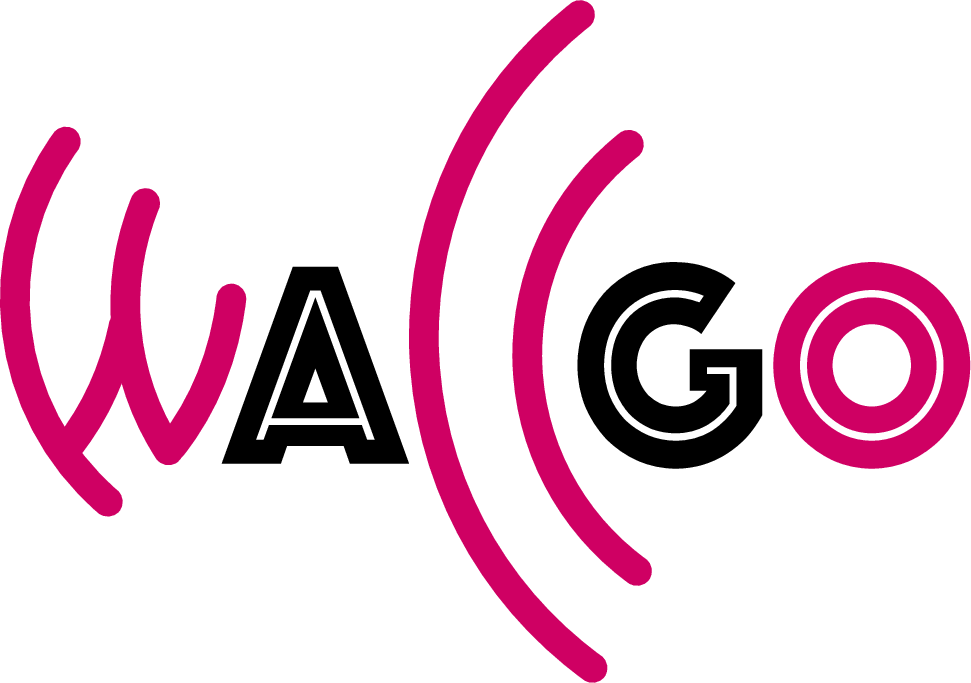}
\vfill

{\Large\bf
    \WallGo{} investigates: \\
  \large\bf
  Theoretical uncertainties in the bubble wall velocity
}

\vspace{0.6cm}

\renewcommand{\thefootnote}{\fnsymbol{footnote}}
Jorinde van de Vis%
\orcidlink{0000-0002-8110-1983},%
$^{\rm a,}$%
\footnotemark[1]
Philipp Schicho%
\orcidlink{0000-0001-5869-7611},%
$^{\rm b,}$%
\footnotemark[2]
Lauri Niemi%
\orcidlink{0000-0001-8068-4366},%
$^{\rm c,}$%
\footnotemark[3]
\\
Benoit Laurent%
\orcidlink{0000-0002-1306-3620},%
$^{\rm d,e,}$%
\footnotemark[4]
Joonas Hirvonen%
\orcidlink{0000-0002-5350-7556},%
$^{\rm f,}$%
\footnotemark[5]
Oliver Gould%
\orcidlink{0000-0002-7815-3379} %
$^{\rm f,}$%
\footnotemark[6]

\vspace{0.5cm}

$^\rmi{a}$%
{\em
  Theoretical Physics Department, CERN,\\
  1 Esplanade des Particules, CH-1211 Geneva 23, Switzerland
}

\vspace{0.3cm}

$^\rmi{b}$%
{\em
  D\'epartement de Physique Th\'eorique, Universit\'e de Gen\`eve,\\
  24 quai Ernest Ansermet, CH-1211 Gen\`eve 4, Switzerland
}

\vspace{0.3cm}

$^\rmi{c}$%
{\em
  CSC -- IT Center for Science Ltd, P.O. Box 405, Espoo, FI-02101, Finland
}

\vspace{0.3cm}

$^\rmi{d}$%
{\em
  McGill University, Department of Physics, 3600 University St.,
  Montr\'eal, QC H3A2T8 Canada
}

\vspace{0.3cm}

$^\rmi{e}$%
{\em
  Perimeter Institute for Theoretical Physics, Waterloo, Ontario N2L 2Y5, Canada
}

\vspace{0.3cm}

$^\rmi{f}$%
{\em
  School of Physics and Astronomy, University of Nottingham,\\
  Nottingham NG7 2RD, United Kingdom
}

\vspace{0.3cm}

\vfill

\mbox{\bf Abstract}

\end{centering}

\vspace*{0.3cm}

\noindent
We examine theoretical uncertainties in
state-of-the-art calculations of the bubble wall velocity during
first-order cosmological phase transitions.
By utilising the software \WallGo{} 
for two extensions of the Standard Model,
we find several $\mathcal{O}(1)$ uncertainties arising from
the number of particles taken out of equilibrium,
the logarithmically and power enhanced collision integrals,
the treatment of thermal masses,
the nucleation temperature,
the $\tanh$ ansatz, and
the perturbative order of the effective potential.
However, we show that the linearisation of the Boltzmann equations is generally a good approximation with much smaller associated errors.
We further clarify the limitations of the quasiparticle approximation in regions with negative mass squared.
This study provides a detailed uncertainty budget and
highlights where future efforts should be directed to improve
the reliability of wall velocity and hence gravitational wave predictions.

\end{titlepage}
\clearpage

{\hypersetup{hidelinks}
\tableofcontents
}
\renewcommand{\thefootnote}{\arabic{footnote}}
\setcounter{footnote}{0}

{\hypersetup{hidelinks}
\renewcommand{\thefootnote}{\fnsymbol{footnote}}
\footnotetext[1]{jorinde.van.de.vis@cern.ch}
\footnotetext[2]{philipp.schicho@unige.ch}
\footnotetext[3]{lauri.niemi@csc.fi}
\footnotetext[4]{blaurent@perimeterinstitute.ca}
\footnotetext[5]{joonas.hirvonen@nottingham.ac.uk}
\footnotetext[6]{oliver.gould@nottingham.ac.uk}
}
\vfill

\clearpage

\section{Introduction}
\label{sec:intro}

First-order phase transitions in the early Universe are fascinating possible
phenomena that offer a window into high-energy physics beyond the reach of the cosmic microwave background,
big bang nucleosynthesis and terrestrial particle physics experiments. 
Such transitions, potentially tied to the electroweak scale or entirely new physics sectors, 
proceed through the nucleation and expansion of true-vacuum bubbles within a metastable false vacuum. 
The expansion velocity of these bubbles strongly affects the dynamics of the transition and its observable signatures. 
In particular, it governs the efficiency of baryogenesis~\cite{Cline:2020jre, Cline:2021dkf, Dorsch:2021ubz, vandeVis:2025efm},  certain dark matter generation scenarios~\cite{Baker:2019ndr, Azatov:2021ifm, Giudice:2024tcp} and it 
affects the spectrum of stochastic gravitational waves~\cite{Espinosa:2010hh, Hindmarsh:2016lnk, Hindmarsh:2017gnf, Hindmarsh:2019phv, Caprini:2019egz}. 
An accurate prediction of the bubble wall velocity is thus essential to connect new physics models to phenomenology.
Since the computation of the wall velocity is computationally rather involved, 
many predictions treat the wall velocity as a free parameter, or resort to simple estimates.
This introduces a significant error in the phenomenological predictions.

In~\cite{Ekstedt:2024fyq}, we presented \WallGo{}: a software package for computing the bubble wall
velocity $\vw$, meant to significantly reduce the uncertainty discussed above. 
\WallGo{} implements the full computation of $\vw$:
from matrix element generation and evaluation of collision integrals, to solving the scalar
field equation of motion (EOM) coupled with the Boltzmann equations, using the method developed in~\cite{Laurent:2022jrs}. 
The model describing the
phase transition and the surrounding primordial plasma is fully user-defined, making
\WallGo{} applicable to a huge range of Beyond the Standard Model scenarios that feature
a first-order phase transition.

In addition to providing a user-friendly tool for computing the wall velocity, 
our motivation for developing \WallGo{} was to improve the theoretical
framework for computing $\vw$.
State-of-the-art calculations~\cite{Moore:1995si,Lewicki:2021pgr, Laurent:2022jrs, Dorsch:2023tss, Ekstedt:2024fyq, Dorsch:2024jjl, Branchina:2025adj} involve several simplifying assumptions, and obtaining a
reliable result may require going beyond some of them.
To efficiently allocate computational and intellectual resources,
it is crucial to identify the dominant sources of uncertainty in the calculation.
\WallGo{} provides a systematic framework for assessing and reducing these uncertainties.

This article represents a first step toward the latter objective.
We investigate the uncertainties in the wall velocity arising from various sources,
focusing on two specific models:
\begin{itemize}
  \item
    the Standard Model (SM) extended by a real scalar singlet,
    in the region of parameter space where the phase transition proceeds in two steps,
  \item
    the Inert Doublet Model (IDM), in the regime where the transition is radiatively induced.
\end{itemize}
In this context, we examine several approximations commonly used in the literature to compute
the wall velocity.
We assess their impact on the calculation by estimating the resulting
uncertainties in both
the total pressure on the wall $\Delta P$ and
the wall velocity $\Delta \vw$.
Both errors can be related to each other through
\begin{equation}
\label{eq:vwError}
  \Delta \vw \approx \Delta P\left(\frac{{\rm d}P}{{\rm d}\vw}\right)^{-1}
  \,.
\end{equation}
According to this relation,
the sensitivity of $\vw$ to various approximations is expected to
be small when ${\rm d}P/{\rm d}\vw$ is large.
This is typically the case in the range
$\vw \in [\cs, \vJ]$, where hydrodynamic effects generate a strong shock wave that causes the
pressure to rise rapidly.
An example of such a pressure curve is shown in fig.~2 of~\cite{Laurent:2022jrs}.

Since the initial release of \WallGo{}~\WallGoOldVersion{}~\cite{Ekstedt:2024fyq},
its codebase has undergone several important updates.
This manuscript is accompanied by the release of
\begin{align}
  \text{\tt \WallGo{}~[\WallGoVersion{}]}
  \,,&&&
  \href{https://github.com/Wall-Go/WallGo/releases/tag/v1.1.1}{%
    \text{\tt github.com/Wall-Go/WallGo/releases/tag/v1.1.1}}
    \,,
  \nn
  \text{\tt \WallGoMatrix{}~[\WallGoMatrixVersion{}]}
  \,,&&&
  \href{https://github.com/Wall-Go/WallGoMatrix/releases/tag/v1.1.0}{%
    \text{\tt github.com/Wall-Go/WallGoMatrix/releases/tag/v1.1.0}}
    \,,
  \nn
  \text{\tt\WallGoCollision{}~[\WallGoCollisionVersion{}]}
  \,,&&&
  \href{https://github.com/Wall-Go/WallGoCollision/releases/tag/v1.1.0}{%
    \text{\tt github.com/Wall-Go/WallGoCollision/releases/tag/v1.1.0}}
  \,.
  \nonumber
\end{align}
The changes to the code are summarised in
appendices~\ref{sec:wallGoUpdate},
\ref{sec:wallGoMatrixUpdate}, and
\ref{sec:wallGoCollisionUpdate}, respectively.

After introducing the models in
sec.~\ref{sec:models}, we investigate theoretical uncertainties
arising from the Boltzmann equations in
sec.~\ref{sec:Boltzmann},
followed by an analysis
of uncertainties related to the scalar field equations of motion in
sec.~\ref{sec:scalar_fields}.
In this work, we do not repeat the overview of the theoretical framework for the computation of $\vw$, 
but we refer the reader to~\cite{Laurent:2022jrs, Ekstedt:2024fyq} for details.

\section{Models}
\label{sec:models}
Although the electroweak phase transition (EWPT) is a crossover in the SM, it is of first order in many
well-motivated extensions, in particular in models with new fields in the scalar sector.
We will discuss two examples of such extensions that we will use as benchmarks for the computation of the wall velocity.

\subsection{Standard Model coupled to a singlet scalar}
One of the simplest extensions of the SM with a first-order phase transition is the scalar singlet extension (xSM), where the Higgs field
is coupled to a new real scalar field $s$, which is a singlet under the SM gauge group.
To reduce the number of new parameters and simplify the analysis, we impose a $\mathbb{Z}_2$
symmetry under which $s \to -s$.
The tree-level scalar potential then takes the form
\begin{equation}
V_0(\Phi,s) = 
    \mu_h^2 \Phi^\dagger \Phi
  + \lambda_h (\Phi^\dagger \Phi)^2
  + \frac{1}{2} \mu_s^2 s^2
  + \frac{1}{4} \lambda_s s^4
  + \frac{1}{2} \lambda_{hs} s^2 \Phi^\dagger \Phi
  \,.
\end{equation}
This potential introduces three new parameters beyond the SM:
the singlet mass
$m_s^2 = -\lambda_{hs}^{ }\mu_h^2/(2\lambda_h) + \mu_s^2$ (at tree level),
its self-coupling
$\lambda_s$, and
the Higgs-singlet coupling $\lambda_{hs}$.

Further details on the full effective potential, including thermal and Coleman-Weinberg
corrections, can be found in~\cite{Ekstedt:2024fyq}.
In this work, we focus on the region of parameter space where
the EWPT is of first order and proceeds via a two-step process.
In the first step, the singlet scalar acquires a vacuum expectation value (\vev),
spontaneously breaking its $\mathbb{Z}_2$ symmetry.
In the second step, the Higgs field obtains a \vev,
breaking the electroweak symmetry, while the singlet \vev{} returns to zero.
This trajectory in field space enables a first-order EWPT facilitated by bubble nucleation.
Here, we are interested in the region of the parameter space where the second step is of first order.

In secs.~\ref{sec:Boltzmann} and \ref{sec:scalar_fields},
we present several plots for the full two-step region of the xSM with $\lambda_s = 1$.
The dataset corresponds to the one used in~\cite{Laurent:2022jrs} and uses
the nucleation temperatures obtained there.
In addition, we will present some results for two sets of benchmark points
{\tt BM:xSM1} and
{\tt BM:xSM2}
presented in table~\ref{tab:xSM}.
\begin{table}[]
\centering
\begin{tabular}{|l|cccc|}
\hline
 & $m_s$~[GeV] & $\lambda_s$ & $\lambda_{hs}$  & $\alphan$ \\
 \hline
 \hline
  {\tt BM:xSM1} & 100 & 1 & $0.7$, \dots, $0.9\hphantom{1} $
           & $3\times10^{-3}$, \dots, $3\times10^{-2}$
           \\
  {\tt BM:xSM2} & 85  & 1 & $0.7$, \dots, $0.81$
           & $ 6\times 10^{-3}$, \dots, $3\times 10^{-2}$
           \\ \hline
\end{tabular}
\caption{%
  Values of the singlet mass, couplings and phase transition strength
  at nucleation $\alphan$ for
  the xSM benchmark points.
  The singlet mass is in GeV.
}
\label{tab:xSM}
\end{table}

\paragraph*{Collision terms.}
In figs.~\ref{fig:linCriterion},
\ref{fig:collisionMultiplier},
\ref{fig:supercooling:daisy},
\ref{fig:tanhError}, and
\ref{fig:tanhErrorScan},
we consider just the top quark out of equilibrium, and QCD matrix elements and collision terms. 
In fig.~\ref{fig:vwOutOfEq} we consider larger sets of out-of-equilibrium particles and interactions,
as detailed in sec.~\ref{sec:Boltzmann:particles:inout}.
In all cases, the collisions are computed with symmetric phase asymptotic masses
(arising from thermal self-energies evaluated near the light cone, 
and neglecting the field-dependent mass) in the propagators:
\begin{align}
\label{eq:asympqg}
  m_{g,\infty}^{2} &= g_s^2 T^2
  &&
  (\text{gluons})
  \,,
  &
  m_{q,\infty}^2 = \frac{g_s^2T^2}{3}
  &&
  (\text{quarks})
  \,.
\end{align}
In all graphs presented in this paper,
we used a basis size of $\tt N = 11$ Chebyshev basis polynomials 
in the momentum directions (see also~\cite{Laurent:2022jrs, Ekstedt:2024fyq}). 
We used a spatial grid size of $\tt M = 30$ in fig.~\ref{fig:vwOutOfEq}, and $\tt M=40$ in the other xSM figures.

\subsection{Inert doublet model}
\label{sec:IDM}

The inert doublet model (IDM) is a specific realization of the two-Higgs-doublet model in
which the new scalar doublet $\chi$ interacts only with the SM Higgs field $\Phi$ and
the gauge sector.
Crucially, none of the components of $\chi$ acquires a \vev{} at zero temperature.
In this work, we adopt the implementation of~\cite{Blinov:2015vma},
which we briefly summarise below.%
\footnote{
  In contrast, our earlier work~\cite{Ekstedt:2024fyq} followed~\cite{Jiang:2022btc},
  which employs a different renormalization and tree-level matching scheme.
}

The tree-level potential is given by
\begin{align}
  V_0 =
      \mu_1^2 \Phi^\dagger \Phi
    + \mu_2^2 \chi^\dagger \chi
    &
    + \lambda_1 (\Phi^\dagger \Phi)^2
    + \lambda_2 (\chi^\dagger \chi)^2
    \nn &
    + \lambda_3 \Phi^\dagger \Phi \chi^\dagger \chi
    + \lambda_4 \Phi^\dagger \chi \chi^\dagger \Phi
    + \biggl[ \frac{\lambda_5}{2} (\Phi^\dagger \chi)^2
        + \mbox{H.c.}
    \biggr]
  \,,
\end{align}
with $\Phi$ and $\chi$ parameterised by
\begin{align}
  \Phi &= \begin{pmatrix}G^+ \\ \frac{1}{\sqrt 2} (v + h + iG^0) \end{pmatrix}
  \,,&
  \chi &= \begin{pmatrix}H^+ \\ \frac{1}{\sqrt 2} (\phi + H + iA) \end{pmatrix}
  \,,
\end{align}
corresponding to four new degrees of freedom beyond the SM:
a CP-even scalar $H$,
a CP-odd scalar $A$, and
a pair of charged scalars $H^\pm$.
We denote the
background value of the SM Higgs field by $v$, and
that of the inert doublet by $\phi$.
In the IDM, we choose parameters such that $\phi$ remains zero throughout the transition.
The corresponding tree-level masses are given by
\begin{align}
  \overline m_{h}^2 &= \mu_1^2 + 3 \lambda_1 v^2
  \,, &
  \overline m_{\rmii{$G^\pm$},\rmii{$G^0$}}^2 &= \mu_1^2 +  \lambda_1 v^2
  \,, &
  \overline m_{\rmii{$H$}}^2 &= \mu_2^2 + \frac 1 2 (\lambda_3 + \lambda_4 + \lambda_5) v^2
  \,,\nn
  \overline m_{\rmii{$A$}}^2 &= \mu_2^2 + \frac 1 2 (\lambda_3 + \lambda_4 - \lambda_5) v^2
  \,, &
  \overline m_{\rmii{$H^\pm$}}^2&= \mu_2^2 + \frac 1 2 \lambda_3 v^2
  \,,
\end{align}
where the bar indicates tree-level quantities.
Since a first-order phase transition in the IDM
typically requires sizeable values of the couplings $\lambda_{3,4,5}$,
the scalar masses receive significant one-loop corrections.
To account for this, we perform a one-loop
matching of the Lagrangian parameters to the physical input parameters.%
\footnote{%
  Large values of the couplings also challenge the validity of the high-temperature
  perturbative expansion;
  see~\cite{Laine:2017hdk,Kainulainen:2019kyp,Navarrete:2025yxy}.
  A computation of the wall velocity including
  higher-order corrections to the effective potential is left for future work.
}
Following~\cite{Blinov:2015vma}, we take as input the zero-temperature physical masses
$\mH$, $\mA$, and $m_{\rmii{$H^\pm$}}$, along with the couplings
$\lambda_2$ and
$\lambda_\rmii{$L$} = (\lambda_3 + \lambda_4 + \lambda_5)/2$, as well as the
Higgs mass and \vev{} evaluated at the scale of the $Z$ boson mass.
We then extract the couplings $\lambda_{3,4,5}$ and the mass parameter $\mu_2^2$ using the
one-loop mass relations given in appendix~A of~\cite{Goudelis:2013uca}.
The relevant Passarino-Veltman functions~\cite{tHooft:1978jhc,Denner:1991kt}
are computed from:%
\begin{align}
  A(m^2) &= \int_\mathcal{P} \frac{1}{\mathcal{P}^2-m^2}
  \, ,&
  B(\mathcal{K}^2,m_1^{ },m_2^{ }) &=
    \int_\mathcal{P}
    \frac{1}{
      [\mathcal{P}^2-m_1^2]
      [(\mathcal{K}-\mathcal{P})^2-m_2^2]}
    \,,
\end{align}
where
$\int_\mathcal{P} \equiv \int \frac{{\rm d}p^0}{2\pi} \int_\vec{p}$
denotes the $D = (d+1)$-dimensional integration measure in Minkowski space-time,
with four-momenta $\mathcal{K}$ and $\mathcal{P}$, and
$\int_\vec{p} \equiv \int \frac{{\rm d}^d \vec{p}}{(2\pi)^d}$
its $d$-dimensional spatial component.
For $D=4-2\epsilon$
one can expand
$A = \sum_{n=-1}^{\infty} A^{(n)}\epsilon^n$ and
$B = \sum_{n=-1}^{\infty} B^{(n)}\epsilon^n$.
The finite parts of the Passarino-Veltman functions are%
\footnote{%
  By following the conventions of~\cite{Goudelis:2013uca,Denner:1991kt},
  the loop integrals $A^{(0)}$ and $B^{(0)}$ are defined with a modified
  integration measure,
  $\int'_\mathcal{P} \equiv (4\pi)^2 \int_\mathcal{P}$.
  In~\cite{Goudelis:2013uca}, the arguments of $A^{(0)}$ and $B^{(0)}$
  were inconsistently given as squared masses,
  while, as in our convention, they should correspond to the masses themselves.
}
\begin{align}
  A^{(0)}(m;\LamD^2) &=
    m^2
    \left(\ln{\frac{m^2}{\LamD^2}} + 1\right)
 \,,
 \nn[2mm]
  \qquad B^{(0)}(\LamD^2;m_1,m_2) &=
  \begin{cases}
      - \frac{m_1^2}{m_1^2 - m_2^2 } \ln{\frac{m_1^2}{\LamD^2}}
      + \frac{m_2^2}{m_1^2 - m_2^2 } \ln{\frac{m_2^2}{\LamD^2}}
      & m_1 \neq m_2\\[1mm]
      - \ln{\frac{m_1^2}{\LamD^2}}
      & m_1 = m_2
  \end{cases}
 \,.
\end{align}
Here, $\LamD$ denotes the energy scale at which the matching is performed,
and the expression for $B^{(0)}$ assumes zero loop momentum.
For a consistent one-loop treatment,
the momentum squared should be set to the mass squared,
as done in~\cite{Laine:2017hdk, Kainulainen:2019kyp}.
To reproduce the benchmark points of~\cite{Blinov:2015vma},
we follow their approach and defer a more consistent implementation of
the matching relations.

To make the matching procedure numerically simpler,
we substitute the physical masses, rather than the tree-level masses,
in the one-loop functions appearing in the matching relations for the masses of the new scalars. 
Strictly speaking, these should be the tree-level masses, but the two approaches are equivalent up to two-loop level.
Because of the sizeable couplings, this leads to some differences with the values obtained in~\cite{Blinov:2015vma}.

After obtaining the values of $\lambda_{3,4,5}$ and $\mu_2^2$ at the matching scale,
we determine the Higgs parameters $\lambda_1$ and $\mu_1$ by simultaneously setting
the measured Higgs mass
to the one-loop expression given in~\cite{Goudelis:2013uca} and by
demanding that the symmetry-breaking minimum of
the one-loop effective potential lies at $v_0 = 246.22$~GeV.
The gauge couplings and top-quark Yukawa are determined from the tree-level masses.
The couplings are evolved to the scale $\mu = v_0$, where the finite-temperature potential is evaluated, 
using the RG-equations reported in appendix~A of~\cite{Blinov:2015vma}.

The effective potential is implemented following~\cite{Blinov:2015vma}, comprising
the tree-level term,
the one-loop Coleman-Weinberg correction, and
the finite-temperature contribution.
We use the full thermal functions $J_{B/F}$ without expansion and
apply Arnold-Espinosa (daisy) resummation.
Contributions from electroweak gauge bosons, the top quark, and the scalar sector are included,
while those from the bottom quark and tau lepton are omitted,
unlike in~\cite{Blinov:2015vma}.
The nucleation temperature is determined using
{\tt FindBounce}~\cite{Guada:2020xnz}
to compute the bounce action $S_b$ from the one-loop potential,
applying the nucleation criterion $S_b = 127$~\cite{Caprini:2019egz}.

We generate two sets of benchmark points
from BM~1 and BM~3 of~\cite{Blinov:2015vma} by varying
the masses of the heavy scalars $\mA$ and $m_{\rmii{$H^\pm$}}$.
The resulting benchmarks are summarised in table~\ref{tab:IDM}.
The values of the couplings and the nucleation temperatures used in
the computation can be found in~\cite{ZenodoData}.
\begin{table}[]
\centering
\begin{tabular}{|l|ccccc|}
\hline
 & $\mH$~[GeV] & $\mA, m_{\rmii{$H^\pm$}}$~[GeV] & $\lambda_{\rmii{$L$}}$ & $\lambda_2$ & $\alphan$
 \\
 \hline
 \hline
  {\tt BM:IDM1} & 66
                & 265, \dots, 360
                & $1.07 \times 10^{-2}$
                & $10^{-2}$
                & $2.2 \times 10^{-3}$, \dots, $5.5 \times 10^{-3}$
                \\
  {\tt BM:IDM3} & 5
                & 230, \dots, 270
                & $-6 \times 10^{-3}$
                & $10^{-2}$
                & $1.0 \times 10^{-3}$, \dots, $4.6 \times 10^{-3}$
                \\ \hline
\end{tabular}
\caption{%
  Values of the masses, couplings and phase transition strength
  at nucleation $\alphan$ for the IDM benchmark points.
  All masses are in GeV.
  }
\label{tab:IDM} 
\end{table}

\paragraph*{Collision terms.}
In the following sections, we allow the top quark,
the weak gauge bosons, and
the heavy scalars $A$, $H^\pm$ to go out of equilibrium.
We consider strong interactions and weak interactions (we treat the $W$ and $Z$ as identical). 
Yukawa interactions are neglected in the scattering, but scalar-gauge and all scalar-scalar interactions are included.
Note that this is different to~\cite{Jiang:2022btc,Ekstedt:2024fyq}, where only $\lambda_3$ was considered in the collisions.

For the top and $W$, the dominant interactions depend on the strong and weak gauge couplings,
which remain relatively constant between the benchmarks.
The scalar interactions, however, depend on the scalar couplings, which vary by up to $50\%$
between the smallest and largest values of $\mA$.
In addition, scalar interactions depend on the Higgs \vev, which varies across the bubble wall.
Since \WallGoCollision{} 
currently requires this scale to be a constant,
we approximate it by evaluating the collision integrals at two constant \vev{} values $\vcoll$:
the broken phase minimum at the nucleation temperature,
$\vn = v(\Tn)$, and half that value,
{\em viz.}
\begin{equation}
\label{eq:VEV:col:factor}
  \vcoll \to [0.5,1.0] \times \vn
  \,.
\end{equation}
Since the computation of the collision integrals is numerically expensive,
we restrict to two coupling sets,
both corresponding to the first benchmark point
(with $\mA = 265$~GeV for {\tt IDM1} and $\mA = 230$~GeV for {\tt IDM3})
and with the two choices of $\vcoll$ described above.%

The symmetric phase asymptotic masses appearing in
the propagators are given by~\cite{Moore:1995si}
\begin{align}
\label{eq:asympIDM}
  m_{\rmii{$W$},\infty}^{2} &= g_w^2 T^2
  &&
  (\text{weak bosons})
  \,,
  &
  m_{l,\infty}^2 &= \frac{3}{16} g_w^2 T^2
  &&
  (\text{leptons})
  \,,
\end{align}
including the ones
for the quark and gluon from eq.~\eqref{eq:asympqg}.
For the corresponding scalar degrees of freedom
\begin{align}
\label{eq:asympIDM:scalars}
  m_{\rmii{$A$,$H^{\pm}$},\infty}^2 &= \frac{T^2}{12} \Bigl(
        6\lambda_2
      + 2\lambda_3
      + \lambda_4 
      + \frac{9}{4} g_w^2
      \Bigr)
    \,,
  &&&
  (A, H^{\pm})
  \nn
  m_{h,\infty}^2 &= \frac{T^2}{12}\Bigl(
          6\lambda_1
        + 2\lambda_3
        + \lambda_4
        + \frac{9}{4} g_w^2
        + 3 \gY^2
        \Bigr)
  \,,
  &&&
  (\text{Higgs})
\end{align}
we neglect the contributions from the U(1) coupling in the asymptotic masses,
since we also ignore U(1) interactions in the collisions.
Note that for the effective potential,
we then include again the effect of U(1) contributions in the thermal masses,
following the approach of~\cite{Blinov:2015vma},
as this effect contributes to the equation of motion at lower order.

Just like for the xSM, we used a basis size of {\tt N = 11} Chebyshev basis polynomials 
in the momentum directions.
In the position direction, we used {\tt M = 25} basis polynomials.

\section{Boltzmann equation}
\label{sec:Boltzmann}

The Boltzmann equation describes the evolution of distribution functions
out of thermal equilibrium,
and plays a central role in determining the friction acting on the bubble wall.
It is a partial integro-differential equation living in a three-dimensional space
(one spatial and two momentum dimensions) and contains high-dimensional integrals
over the distribution functions. Consequently,
solving the Boltzmann equation poses a significant challenge
and typically requires advanced numerical methods.

In this section,
we study key approximations often made to simplify this problem
focusing on
the linearisation of the Boltzmann equation,
the content of out-of-equilibrium particles, 
higher orders in collision integrals, as well as
approximations of particle masses.

\subsection{Linearisation of the Boltzmann equation}
\label{sec:Boltzmann:linearisation}

A common strategy to simplify the Boltzmann equation
is to linearise it in the deviation from equilibrium.
This approximation assumes that the distribution function
remains close to its equilibrium form,
so that higher-order nonlinear corrections are small and can be neglected.
Here,
we investigate the accuracy of this approximation
and determine the regime in which it can be trusted.

In general, the Boltzmann equation can be expressed as
\begin{align}
  \mathcal{L}[f]+\mathcal{C}[f]=0
  \,.
\end{align}
While $\mathcal{L}[f]$ is a linear operator describing the free streaming
of particles,
the collision operator $\mathcal{C}[f]$, which accounts for interactions among plasma species,
is generally nonlinear.
For processes involving $N$ particles,
$\mathcal{C}[f]$ is an integral over a polynomial of degree $N$
in the distribution functions.
Expanding the distribution function around an equilibrium profile as
$f = f_{\rm eq} + \delta f$,
the collision operator can be expanded as%
\footnote{%
  The zeroth-order term, $\mathcal{C}[f_{\rm eq}]$,
  vanishes due to a balance of gain and loss terms.
}
\begin{align}
\label{eq:CExpansion}
  \mathcal{C}[f]=\sum_{n=1}^N \mathcal{C}_{n}[\delta f^n]
  \,,
\end{align}
where each operator $\mathcal{C}_{n}[X]$ is linear in $X$.
Here and in the following $\delta f^n$ depends on $n$ three-momenta, e.g. $\delta f^2(\vec p, \vec q,z) = \delta f(\vec p,z) \delta f(\vec q, z)$.
Linearising the Boltzmann equation then amounts to retaining only the first term,
{\em viz.}
\begin{align}
\label{eq:CLinear}
  \mathcal{C}[f]\approx \mathcal{C}_{1}[\delta f]
  \,.
\end{align}

To assess the validity of the linearised approximation~\eqref{eq:CLinear},
we aim to derive a criterion that quantifies its accuracy.
There are two limiting cases where the linearisation becomes exact.
First, in the regime of large $\mathcal{C}$,
local thermal equilibrium (LTE) is maintained near the wall,
implying $\delta f \to 0$ and rendering the nonlinear terms negligible.
Second, in the regime of small $\mathcal{C}$,
even though $\delta f$ may be large,
the only nonlinear term in the Boltzmann equation, $\mathcal{C}[f]$, vanishes,
so the equation becomes effectively linear.
This limit corresponds to the so-called
ballistic regime~\cite{Liu:1992tn,BarrosoMancha:2020fay,Wang:2024wcs,Ai:2024btx}.%
\footnote{%
  The term \emph{ballistic regime} is used with two distinct meanings in the wall-velocity literature.
  In~\cite{Bodeker:2009qy}, it denotes the regime of ultrarelativistic bubble walls,
  whereas here it refers to the regime of infrequent (or weak) collisions.
  See also~\cite{Ai:2024btx} for a comparison.
}
We note that, although the linearisation becomes exact in this limit, \WallGo{} is not expected to work optimally in this regime as the ballistic distribution function has sharp discontinuities that cannot be resolved by \WallGo{}.

\begin{figure}[t]
    \centering
    \includegraphics[width=0.5\textwidth]{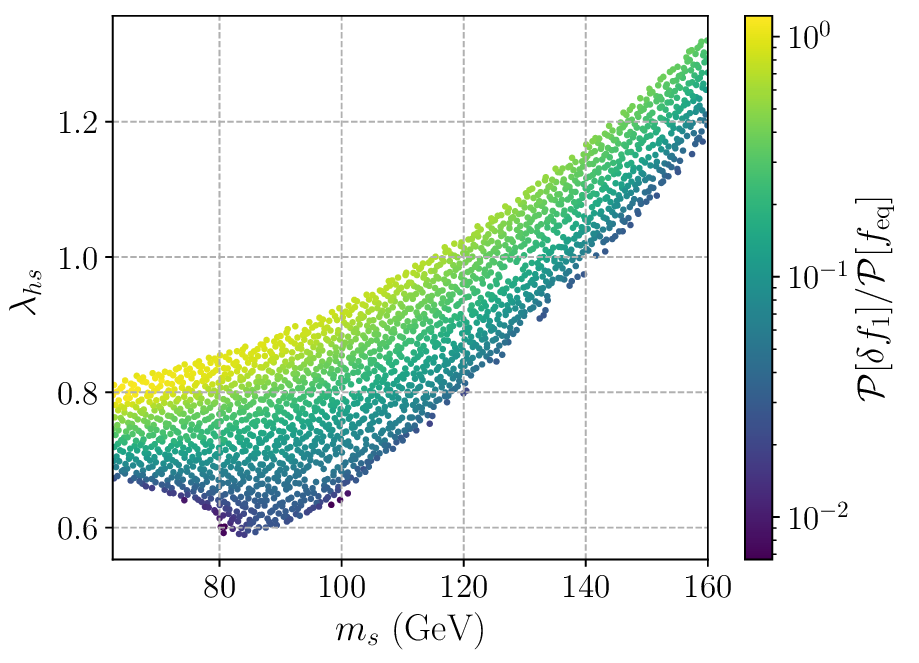}%
    \includegraphics[width=0.5\textwidth]{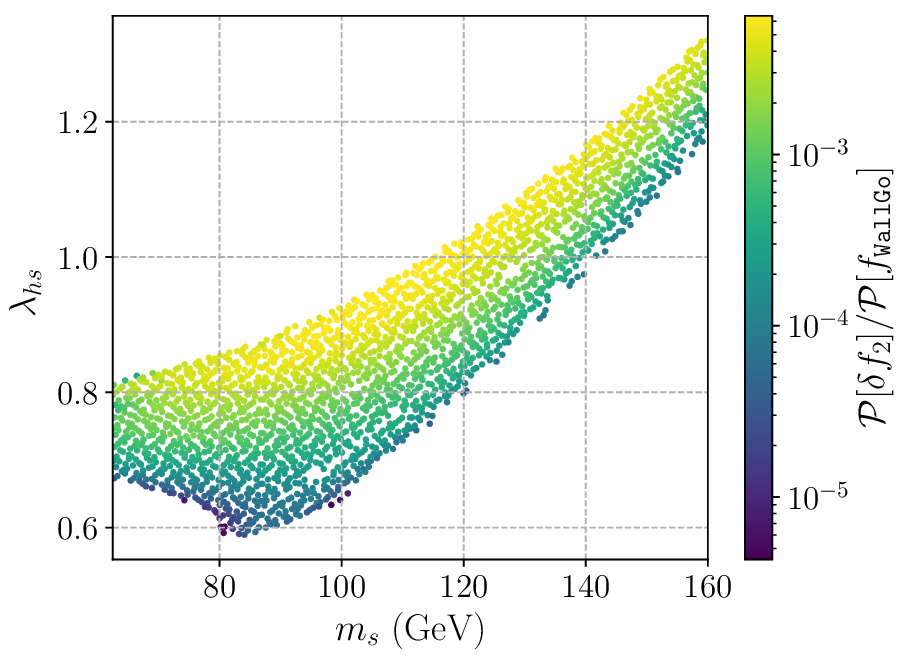}
    \caption{%
      Linearisation criteria across the xSM scan,
      computed by \WallGo{} and stored in the {\tt WallGoResults} object.
      Left: {\tt linearizationCriterion1}, based on the ratio $\mathcal P[\delta f_1]/\mathcal P[f_{\rm eq}]$.
      Right: {\tt linearizationCriterion2}, based on $\mathcal P[\delta f_2]/\mathcal P[f_\WallGo]$,
      vanishing in both the LTE and ballistic limits.
    }
    \label{fig:linCriterion}
\end{figure}

A simple criterion is to compare the magnitude of $\delta f$ to $f_{\rm eq}$,
for instance by evaluating the ratio of the corresponding pressures,
$\mathcal P[\delta f]/\mathcal P[f_{\rm eq}]$, where
\begin{align}
  \mathcal P[X] = \frac{1}{2}\int\! {\rm d}z\, \partial_z(m^2)
    \int_{\vec{p}}\,\frac{X}{E}
  \,.
\end{align}
In \WallGo{}, this pressure ratio is stored in
a {\tt WallGoResults} object as {\tt linearizationCriterion1}.
However, note that this quantity should not be considered as a criterion 
for the linearisation to be valid, as we discuss below.%
\footnote{%
  This criterion was introduced in \WallGo{}~\WallGoOldVersion{},
  and we keep the name for backwards compatibility.
}
Instead, it carries information about the importance of the out-of-equilibrium contributions.
It has been evaluated across the xSM scan and is shown in fig.~\ref{fig:linCriterion} (left).
For most of the scan, the criterion remains below 0.1, but
it can reach values of order unity.

Naively, a large value of {\tt linearizationCriterion1} suggests a breakdown of the linearisation of the Boltzmann equation
in that region of the parameter space.
However, as previously discussed,
a robust criterion should vanish both in the LTE and ballistic limits,
whereas $\mathcal P[\delta f]/\mathcal P[f_{\rm eq}] \sim \mathcal{O}(1)$ in the latter.
Thus, while this ratio provides a useful measure of
the magnitude of out-of-equilibrium contributions,
it does not reliably indicate whether the linearisation remains valid.

A more robust criterion can be derived by examining how
a solution to the linearised Boltzmann equation is affected by
the second-order correction $\mathcal{C}_2[\delta f^2]$.
Assuming $\delta f = \delta f_1 + \delta f_2$, where
$\delta f_1$ solves the linearised equation and
$\delta f_2 = \mathcal{O}(\delta f_1^2)$ represents the leading nonlinear correction,
the Boltzmann equation for the $\mathcal{O}(\delta f_1^2)$ terms becomes
\begin{align}
    \mathcal{L}[\delta f_2^{ }]
  + \mathcal{C}_1[\delta f_2^{ }] =
  - \mathcal{C}_2[\delta f_1^{2}]
  \,.
\end{align}
This equation is formally analogous to
the linear Boltzmann equation with a source term $-\mathcal{C}_2[\delta f_1^2]$.
Treating the Liouville operator $\mathcal{L}$ and collision operators $\mathcal{C}_n$ as tensors,
and the distribution functions as vectors (as implemented in \WallGo{}),
the solution reads
\begin{align}
\label{eq:deltaF1}
  \delta f_1 &= -(\mathcal{L}+\mathcal{C}_1)^{-1}\cdot\mathcal{L}\cdot f_{\rm eq}
  \,,\\
\label{eq:deltaF2}
  \delta f_2 &= -(\mathcal{L}+\mathcal{C}_1)^{-1}\cdot\mathcal{C}_2\cdot \delta f_1^2
  \,.
\end{align}
\WallGo{} cannot compute $\delta f_2$ exactly,
as computing the operator $\mathcal{C}_2$ is not implemented in version \WallGoVersion{}.
However, since our goal is to estimate the magnitude of $\delta f_2$
rather than compute it precisely,
we assume that all operators $\mathcal{C}_n$ are of comparable size and approximate
\begin{align}
  \delta f_2^{ } \sim
    -(\mathcal{L}+\mathcal{C}_1)^{-1}\cdot\mathcal{C}_1^{ }\cdot \delta f_1^2
  \,,
\end{align}
which is calculable by \WallGo{}, taking the square of $\delta f_1$ element wise.

More precisely, \WallGo{} computes the pressure ratio $\mathcal P[\delta f_2]/\mathcal P[f_\WallGo{}]$
(stored in the {\tt WallGoResults} object as {\tt linearizationCriterion2}),
where $f_\WallGo{} = f_{\rm eq} + \delta f_1$ is the total distribution function used
to evaluate the wall velocity.
This improved criterion, shown for the xSM scan in fig.~\ref{fig:linCriterion} (right),
yields significantly smaller values than the previous one,
reaching at most $\mathcal P[\delta f_2]/\mathcal P[f_\WallGo{}] \simeq 0.005$ across the scan.
This indicates that the error from linearising the Boltzmann equation remains
below $1\%$ in the xSM,
confirming that the approximation is well controlled.

It is possible to get an analytic estimate of how the linearisation error varies with the relative size of $\mathcal{L}$ and $\mathcal{C}$ by taking $\mathcal{L}\sim F$ and $\mathcal{C}\sim\Gamma$, with $F$ the typical force on the particles and $\Gamma$ the typical collision rate.
From eqs.~\eqref{eq:deltaF1} and~\eqref{eq:deltaF2},
we obtain the approximate solutions
\begin{align}
  \delta f_1 &\sim a\,\frac{F}{F+\Gamma}\,f_{\rm eq}
  \,,&
  \delta f_2 &\sim b\,\frac{\Gamma F^2}{(F+\Gamma)^3}\,f_{\rm eq}^2
  \,,
\end{align}
where $a$ and $b$ are $\mathcal{O}(1)$ coefficients.
Neglecting the contribution from $\delta f_2$ therefore induces an error
\begin{align}
  \frac{\delta f_2}{f_\WallGo{}}\sim b\frac{\Gamma F^2}{((1+a)F+\Gamma)(F+\Gamma)^2}f_{\rm eq}
  \,.
\end{align}
As expected, this predicts the error to vanish in both $F / \Gamma \to0$ and $F / \Gamma\to\infty$ limits.
Moreover, taking $a=b=1$, the maximum error is obtained for
$F/\Gamma = (1+\sqrt{5})/2 \approx 1.62$
(the golden ratio),
in which case
\begin{align}
  \max\left\vert\frac{\delta f_2}{f_\WallGo{}}\right\vert\sim 0.09
  \,,
\end{align}
where we have taken $f_{\rm eq}\sim 1$.
Interestingly, this indicates that the error caused by neglecting the nonlinear terms of the Boltzmann equation should never be larger than the \WallGo{} solution $f_\WallGo{}$. Therefore, linearising the Boltzmann equation should always be a reasonable approximation, even for stronger phase transitions.

This conclusion aligns with the findings of~\cite{DeCurtis:2024hvh},
where a full calculation including all orders of $\delta f$
in the Boltzmann equation was performed.
Despite $\delta f/f$ reaching values of $\mathcal{O}(0.1)$ and as much as 0.4 for their parameter points,
the differences between the linearised and nonlinear calculations were found to be
orders of magnitude smaller.

\subsection{Particles in and out of equilibrium}
\label{sec:Boltzmann:particles:inout}
In principle, all particles in the plasma are pushed out of equilibrium by their change in mass due to the passing bubble wall.
Even massless particles, like the gluon, are slightly pushed out of equilibrium by scattering with other out-of-equilibrium particles.
Typically, it is computationally not feasible to solve the Boltzmann equation for the entire plasma particle content. 
A more economical approach is to focus on the particles that give the strongest contribution to the friction
and assume that the rest of the plasma remains in equilibrium.

In this subsection, we investigate the importance of solving the Boltzmann equations for several particles in the plasma
and compare the impact of different interaction types between the plasma particles.
Since including additional matrix elements in the collision integrals increases the computational cost, one 
usually focuses on the interactions with the largest coupling strength. 
This could imply that only QCD interactions are considered in the collision terms,
or only QCD and weak interactions. 
In this subsection, we vary these assumptions and examine how they affect $\vw$.

\begin{figure}[t]
  \centering
    \includegraphics[width=0.5\textwidth]{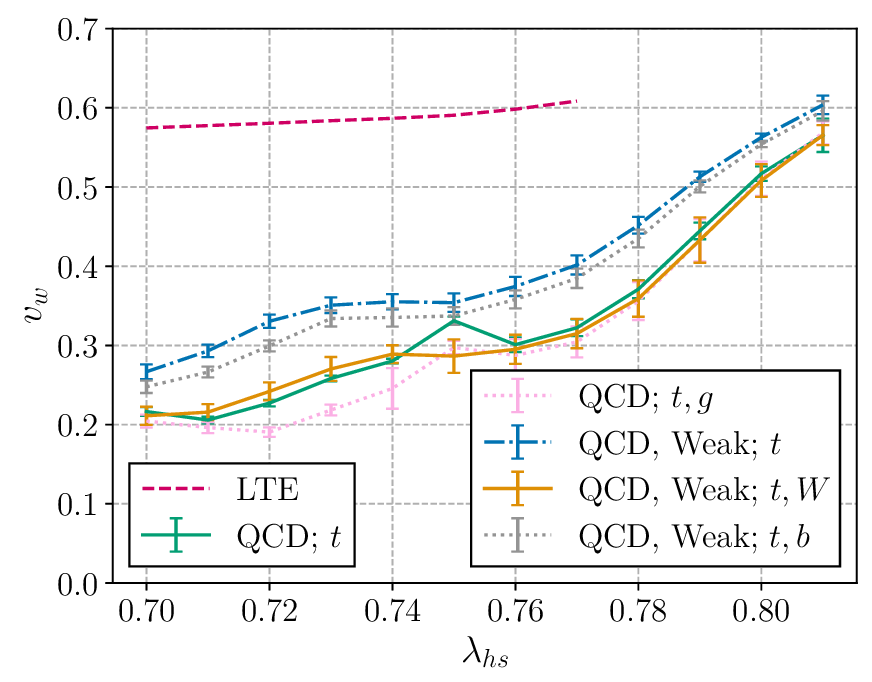}%
    \includegraphics[width=0.5\textwidth]{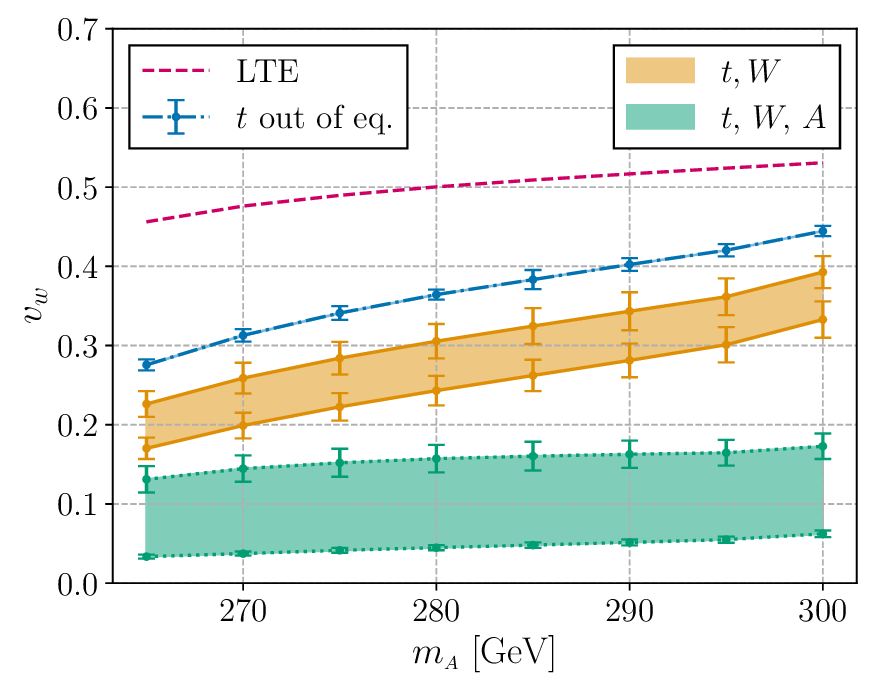}
  \caption{%
    The wall velocity for
    {\tt BM:xSM2} (left) and
    {\tt BM:IDM1} (right) for different choices of out-of-equilibrium particles,
    as indicated in the legend.
    Labels denote:
    $t$ for top quark,
    $g$ for gluon,
    $W$ for $W$ boson
    $b$ for bottom quark, and
    $A$ for the $A$ and $H^\pm$ scalars in the IDM.
    In the right panel,
    ``QCD'' and
    ``QCD, Weak'' indicate whether only QCD or both QCD and
    weak interactions
    were included in the collision terms.
    For the IDM, the bands reflect the uncertainty in $\vw$ arising from the choice of
    \vev{} used in
    the collision terms $\vcoll \to [0.5,1.0] \times \vn$;
    see eq.~\eqref{eq:VEV:col:factor}.
  }
  \label{fig:vwOutOfEq}
\end{figure}
\paragraph{Out-of-equilibrium particles in the xSM.}
The left panel of fig.~\ref{fig:vwOutOfEq} shows $\vw$ as a function of $\lambda_{hs}$ for
{\tt BM:xSM2}, comparing different sets of out-of-equilibrium particles and interactions.
For the solid green and dotted pink line, only QCD interactions are included, with
asymptotic masses as in eq.~\eqref{eq:asympqg}.
The blue dot-dashed, yellow solid and grey dotted lines contain also weak interactions.
Scattering with leptons is included, but interactions with Higgs bosons are neglected.
The symmetric phase asymptotic mass for the leptons is given in eq.~\eqref{eq:asympIDM},
and for the weak gauge bosons it is given by
\begin{equation}
  \mW^2 = \frac{11}{12}g_w^2 T^2
  \,.
\end{equation}

Since the phase transition strength increases with $\lambda_{hs}$, $\vw$ also grows with
$\lambda_{hs}$, largely independent of the out-of-equilibrium particle content.  
The magenta dashed line indicates the LTE result, which provides an upper bound on $\vw$ and
depends only mildly on $\lambda_{hs}$.
For $\lambda_{hs} > 0.76$, the LTE friction becomes insufficient
to provide a sufficient backreaction force,
and the solution becomes a runaway.

The runaway behaviour disappears when friction from out-of-equilibrium particles is considered. 
The solid green line shows $\vw$ when only the top quark is out of equilibrium, with collision terms computed using only QCD interactions. 
The dotted pink line shows $\vw$ when the gluon is also allowed to be out of equilibrium. 
We see that the wall velocity becomes slightly smaller than when only the top is out of equilibrium. 
The difference is small, as the massless gluon does not directly provide friction on the bubble wall,
but only indirectly via its interactions with the top quark distribution.

In the blue dot-dashed line the top quark is still the only out-of-equilibrium particle, but we have also included weak interactions in the collisions.
Since the top quark can now relax back to equilibrium more efficiently than with QCD interactions only,
the friction force decreases, and $\vw$ becomes closer to the LTE result.
The gray dashed line shows the effect of allowing bottom quarks out of equilibrium as well.
As the bottom quark Yukawa coupling is almost two orders of magnitude smaller than the top Yukawa coupling,
the additional friction from including the bottom quarks is relatively small, and
the result is very close to the result with only the top quark out of equilibrium, as expected.

Lastly, the solid orange line shows $\vw$ for out-of-equilibrium top quarks and $W$ bosons,
with QCD and weak interactions. 
Since the $W$ boson has a significant mass and contributes 9 degrees of freedom 
(the $W$ and $Z$ are treated as identical),
its contribution to the friction is expected to be significant.
Indeed, $\vw$ is significantly smaller than when only the friction from the out-of-equilibrium top quark is considered.
Interestingly, the result is close to the solid green line, where only the top was out of equilibrium
and weak interactions were neglected.
We believe this to be a coincidence.
The value of $\vw$ varies most strongly between the different approaches for smaller values of $\vw$.
The reason is that for $\vw$ close to the Jouguet velocity $\vJ$,
the pressure varies very strongly as a function of $\vw$.
The same absolute change in pressure therefore results in a smaller change in $\vw$ close to $\vJ$ than for $\vw \ll \vJ$.

We also tested setting the top quark and all the light quarks out of equilibrium,
which resulted in a negligibly small change in the wall velocity.
This is due to the light quarks not producing friction on the bubble wall directly (not shown in this graph).
Setting
the top quark,
light quarks \emph{and} gluon
out of equilibrium seemed to make the code unstable,
most likely due to the large number of degrees of freedom in the light quarks and the strong coupling between them and the gluon.
However, from the above tests, we expect the effects on the wall velocity to be small.

\paragraph{Out-of-equilibrium particles in the IDM.}
The right panel of fig.~\ref{fig:vwOutOfEq} displays $\vw$ as a function of the mass of the new heavy scalar $\mA$ in {\tt BM:IDM1}. 
The phase transition strength increases with the mass of $\mA$,
yielding a (slight) increase of $\vw$ with the phase transition strength in all approximations.
However, the strength of the friction from the heavy scalars also increases with $\mA$,
which explains why $\vw$ increases more slowly with $\mA$ when the heavy scalars are out of equilibrium.
Compared to the large-$\lambda_{hs}$ points in {\tt BM:xSM2},
the phase transition is weaker, and even in LTE,
we always observe that $\vw < 1$.
For all lines, we have included QCD and weak interactions,
as well as portal interactions between the scalars and the Higgs.

The dash-dotted blue line in fig.~\ref{fig:vwOutOfEq} shows $\vw$ when only
the top quark is treated as out of equilibrium.
Including the friction from the $W$ boson (orange line) suppresses $\vw$ by roughly one third.
However, there is a significant uncertainty in $\vw$, due to
collisions involving the scalar particles.
The interactions of the $W$ with the scalar sector depend on the varying Higgs \vev;
however, in the current implementation of \WallGoCollision{}, this \vev,
denoted $\vcoll$, must be taken as constant.
In the upper orange line of fig.~\ref{fig:vwOutOfEq},
we set this scale to the Higgs \vev{} in the broken phase at
the nucleation temperature for $\mA = 265$~GeV, i.e.\ $\vcoll = \vn$;
see also eq.~\eqref{eq:VEV:col:factor}.
For the lower line, we instead decreased this value by $\vcoll = \vn/2$,
corresponding approximately to the Higgs \vev{} in the middle of the bubble wall.
For the top quark being out of equilibrium, no such \vev-dependent interactions contribute.

Including the friction from the heavy scalars $A$ and $H^{\pm}$ further suppresses $\vw$, 
and the result depends even more sensitively on the assumed $\vcoll$.
For $\vcoll = \vn/2$, the heavy scalars equilibrate less efficiently, leading to stronger friction.
The total friction from the three heavy scalar degrees of freedom scales with $\lambda_3$,
which varies from
$\lambda_3 = 2.4$ for $\mA = 265$~GeV to
$\lambda_3 = 3.1$ for $\mA = 300$~GeV.
Moreover, the force driving the particles
away from equilibrium is also proportional to $\lambda_3$.
This large coupling could therefore explain why the heavy scalars significantly
slow down the wall when their collision terms are small.
Reducing the uncertainty in $\vw$ will require a more accurate treatment of
scalar collision terms, 
which we plan to implement in a future version of \WallGo{}.

\subsection{Infrared sensitivity of collision integrals at finite temperature}
\label{sec:Boltzmann:enhancements}

Squared matrix elements for $2\to 2$ scattering processes are proportional to
four powers of three-point couplings, or
two powers of four-point couplings,
multiplied by a dimensionless function of the Mandelstam variables,
{\em viz.}\ $|M_{ab\to cd}|^2 \propto g^4 F(s,t,u)$.
The collision terms in the Boltzmann equation are momentum integrals of these squared matrix elements,
weighted by combinations of distribution functions.
They have mass dimension 2 in our convention and therefore naively scale as $\mathcal{C}\sim g^4 T^2 \delta f$,
given that all particles are close to equilibrium with $\delta f \lesssim f_\text{eq}$.
The factor of $\delta f$ arises because gain and loss terms precisely cancel at equilibrium.
By equating the kinetic and collision terms of the Boltzmann equation,
\begin{equation}
  p^\mu \partial_\mu \delta f \sim E \frac{\partial}{\partial t} \delta f \sim
  \frac{T}{\tau} \delta f \sim \mathcal{C}
  \,,
\end{equation}
defining $\tau$ as a typical timescale over which $\delta f$ varies,
and
noting that the typical Boltzmann particles have an energy $E\sim T$,
we see that the corresponding collision rate (or scattering rate) is
$\Gamma \equiv 1/\tau \sim g^4 T$.

At high temperatures,
this naive scaling of the collision integrals can be modified by the presence of scale hierarchies.
Such hierarchies arise when the scattering process involves
the exchange of a soft particle,
or equivalently, a small momentum transfer.
For massless particles, these enhancements manifest as infrared (IR) divergences.
However, the medium strongly affects the propagation of the soft exchanged particle through
{\em hard thermal loop} (HTL) effects~\cite{Blaizot:2001nr}.
These medium effects
dynamically screen the interaction and regulate apparent IR divergences
at transfer momenta of $\mathcal{O}(gT)$,
by modifying particle self-energies in the IR, inducing an effective mass
$\meff = \mathcal{O}(gT)$~\cite{Biondini:2025ihi,Ghiglieri:2016xye,Ekstedt:2023oqb}.
Scalar background fields may also modify the effective mass at this order.

We classify the collision integrals in the Boltzmann
equations into three categories according to their IR behaviour
at small momentum transfer:
\begin{itemize}
  \item[(P)]
    {\em power enhanced},
    naively quadratically divergent in the infrared and
    scaling as\\
    $\mathcal{C}_\rmii{P}/\delta f \sim g^4 T^4 / \meff^2 \sim g^2 T^2$.
  \item[(LL)]
    {\em leading-logarithmic enhanced}, naively logarithmically divergent in the infrared and
    scaling as\\
    $\mathcal{C}_\rmii{LL}/\delta f \sim g^4 T^2 \ln(T/\meff) \sim g^4 T^2 \ln (1/g)$~\cite{Arnold:2000dr,Arnold:2003zc}.
  \item[(NLL)]
    {\em next-to-leading-logarithmic}, infrared finite and
    scaling as\\
    $\mathcal{C}_\rmii{NLL}/\delta f \sim g^4 T^2$.
\end{itemize}
At leading order in an expansion in $1/\ln(1/g)$, all other contributions
but $\mathcal{C}_\rmii{P}$ and $\mathcal{C}_\rmii{LL}$ can be neglected.

To understand these kinematic enhancements, consider
the collision integral for the process $ab\to cd$,
\begin{align}
  \mathcal{C}^\text{lin}_{ab\to cd}[\delta f] &= 
  \frac{1}{4}
  \int_{\vec{p}_2,\vec{p}_3,\vec{p}_4} 
  \frac{1}{2E_22E_32E_4}
  (2\pi)^4 \delta^{4}(p_1 + p_2 - p_3 - p_4)
  \, |M_{ab\to cd}|^2
  \nn &
  \qquad\qquad
  \times
      f_\text{eq}^a f_\text{eq}^b (1\pm f_\text{eq}^c)(1\pm f_\text{eq}^d)
      \Bigl[
          \chi^a(\vec{p}_1)
        + \chi^b(\vec{p}_2)
        - \chi^c(\vec{p}_3)
        - \chi^d(\vec{p}_4)
      \Bigr]
  \,, 
  \label{eq:linearizedCollisionIntegral}
\end{align}
where the single-particle energies are defined as
$E_i = \sqrt{\vec{p}_i^{2} + m_i^2}$,
the $\pm$ signs correspond to bosons ($+$) and fermions ($-$),
and we have introduced
\begin{equation}
  \chi^a(\vec{p}_i) = \frac{\delta f^a(\vec{p}_i)}{f_\text{eq}^a(1\pm f_\text{eq}^a)}
  \, ,
\end{equation}
in terms of $\delta f^a(\vec{p}_i)\equiv f^a(\vec{p}_i)-f_\text{eq}^a$,
the deviation from the equilibrium distribution for
particle $a$.
We have left the energy dependence of the equilibrium distribution functions implicit.

First, let us neglect possible cancellations among
the four terms inside the square brackets in eq.~\eqref{eq:linearizedCollisionIntegral}.
In this case, matrix elements that scale as
$1/t^2$ (or $1/u^2$) for $t \ll s$ (or $u \ll s$) lead to power-enhanced contributions, through integrals with the following form in the IR
\begin{equation}
\label{eq:power_divergence}
  \int_{\Lambda_\rmii{IR}^2} \frac{\mathrm{d}t}{t^2}\sim \frac{1}{\Lambda_\rmii{IR}^{2}}
  \,.
\end{equation}
Here, $\Lambda_\rmii{IR}$ denotes the infrared cutoff scale
that regulates the otherwise divergent IR behaviour.
Similarly, matrix elements scaling as $1/t$ (or $1/u$) lead to logarithmic
enhancements through momentum integrals that,
in the IR, take the form
\begin{equation}
\label{eq:log_divergence}
  \int_{\Lambda_\rmii{IR}^2} \frac{\mathrm{d}t}{t}\sim \ln \frac{\Lambda_\rmii{IR}^2}{T^2}
  \,.
\end{equation}
Here the temperature appears through the equilibrium distribution functions, not shown in eq.~\eqref{eq:log_divergence}, which suppress higher energy contributions.
See~\cite{Arnold:2000dr, Arnold:2003zc, Ghiglieri:2015ala} for explicit computations showing how the forms of integral in eqs.~\eqref{eq:power_divergence} and \eqref{eq:log_divergence} arise from eq.~\eqref{eq:linearizedCollisionIntegral}.

Let's consider now the IR parts of $s$-channel processes.
When $s$ is small, the other Mandelstam variables are also small, satisfying
$t, u = \mathcal{O}(s)$.
For models with only marginal (or irrelevant) couplings,
because the matrix elements are dimensionless, in the regime of small $s$, it follows that they cannot be larger than $|M_{ab \to cd}|^2 = \mathcal{O}(s^0)$. 
In addition, the IR phase space measure for $s$-channel processes contains a factor of $s$ in the numerator, leading to the scaling
\begin{equation}
\label{eq:log_divergence}
  \int_{\Lambda_\rmii{IR}^2} s\, \mathrm{d}s \sim \Lambda_\rmii{IR}^4
  \,.
\end{equation}
Consequently, the IR contribution from an $s$-channel process with marginal couplings
scales parametrically as
$\mathcal{C}_{\rmii{IR},s}/\delta f \sim g^8 T^2$.
This is smaller than the UV contribution from $s$-channel processes,
i.e.\ where $s\sim T^2$, which follows the naive scaling $\mathcal{C}_{\rmii{UV},s}/\delta f \sim g^4 T^2$.

If there are relevant couplings, which in four dimensions means scalar 3-point couplings, then $s$-channel matrix elements can be enhanced in the infrared, with the largest contribution from $2\to2$ scattering growing as $|M_{ab \to cd}|^2 = \mathcal{O}(s^{-2})$.
Nevertheless, the reduced IR phase space of $s$-channel processes means that this can at most lead to a logarithmic enhancement,
\begin{equation}
\label{eq:log_divergence}
  \int_{\Lambda_\rmii{IR}^2} \frac{s\, \mathrm{d}s}{s^2} \sim \ln \frac{\Lambda_\rmii{IR}^2}{T^2}
  \,.
\end{equation}
Note however that whenever such a logarithmic $s$-channel enhancement arises, there will be power enhancements from the corresponding $t$- and $u$-channels, so that this enhancement is always a subleading effect.
Hence, we do not include these IR-enhanced $s$-channels in any of the numerical results presented in this work.

If some of the particles $a, b, c, d$ are identical, there can be cancellations between the gain and loss terms in square brackets in eq.~\eqref{eq:linearizedCollisionIntegral}, leading to a suppression of the collision integral.
Consider the case in which particles $a$ and $c$ are identical, and particles $b$ and $d$ are identical.
Then the loss and gain terms are determined by the same functions.
For the enhanced region of phase space, with small momentum transfer in the $t$-channel, $\vec{p}\equiv \vec{p}_3-\vec{p}_1=\vec{p}_2-\vec{p}_4$, these functions will be evaluated at near-identical arguments, and so can be Taylor expanded,
\begin{align}\label{eq:cancel}
    \chi^a(\vec{p}_1) + \chi^b(\vec{p}_2)
    -\chi^a(\vec{p}_1 + \mathbf{p})
    -\chi^b(\vec{p}_2 - \mathbf{p}) =&
      \vec{p} \cdot \nabla_{\vec{p}}
      \left[\chi^a(\vec{p}_1 + \mathbf{p}) + \chi^b(\vec{p}_2 - \mathbf{p})\right]
      _{\vec{p}=\vec{0}}
    + \mathcal{O}(\vec{p}^2)
    \,.
\end{align}
Since in the enhanced region
$t = \mathcal{O}(\vec{p}^2) = \mathcal{O}(\Lambda_\rmii{IR}^2)$,
the vanishing of the $\mathbf{p}^0$ term in the Taylor expansion suppresses the $t$-channel by one power of $\Lambda_\rmii{IR}/T$.
Further, the terms linear in $\vec{p}$ 
are expected to vanish; see appendix~D of~\cite{Ghiglieri:2015ala}. 
The end result is that power enhancements are reduced to logarithmic enhancements, and
logarithmic enhancements are removed altogether.
See also~\cite{Ghiglieri:2016xye} for a similar discussion for neutrino
interaction rates.

By contrast, the $u$-channel momentum transfer, $\vec{p}_1 - \vec{p}_4$,
is not softened when particles $a \equiv c$ and $b \equiv d$,
but instead when particles $a \equiv d$ and $b \equiv c$.
By inspecting which particle pairs are identical,
one can determine which channels exhibit softening of the divergence
and which remain unsuppressed.

To identify the different contributions P, LL, and NLL,
\WallGoMatrix{} provides a leading-logarithmic truncation filter
that keeps both power-enhanced and leading-logarithmic terms.
This can be enabled by setting the default option
\verb!TruncateAtLeadingLog->True!
when generating the matrix elements via {\tt ExportMatrixElements}.%
\footnote{%
  By default, \WallGoMatrix{} simplifies expressions under
  the assumption that all variables (couplings, masses) are real.
}
The precise implementation of
algorithm~\ref{alg:divergence-structure} used in \WallGoMatrix{}
is described in detail in
appendix~\ref{sec:leadingLogAlgorithm}.
In this framework, both LL and P enhanced
contributions can be tagged by setting
\verb!TagLeadingLog -> True!.
The version of \WallGoMatrix{}~\WallGoMatrixVersion{}
employs this LL approximation scheme from
algorithm~\ref{alg:divergence-structure} via particle identification,
automatically classifying both LL and P enhanced contributions.
This approximation
remains the standard in computations of the bubble wall velocity.
Going beyond this approximation poses significant challenges~\cite{Arnold:2000dr,Arnold:2001ms,Arnold:2002ja,Arnold:2002zm}.
Some of them are related to going to NLL,
\begin{itemize}
  \item
    including $s$-channels in $2\to 2$ processes;
  \item
    including $1 \to 2$ and $2 \to 1$ processes,
    including those that are kinematically forbidden in vacuum but enter at
    subleading logarithmic order in a thermal bath;%
    \footnote{%
      For multi-scalar theories with cubic interactions
      $\lambda_{ijk}S_i S_j S_k$,
      $1 \to 2$ and $2 \to 1$ vacuum processes can occur,
      contributing at order $\mathcal{C}/\delta f \sim \lambda_{ijk}^2$.
      In comparison, the corresponding $2 \to 2$ scattering processes scale as
      $\mathcal{C}/\delta f \sim \lambda_{ijk}^4 T^2 / m_\text{eff}^2$.%
    }
  \item
    resumming the full set of $1+N \to 2+N$ and $2+N \to 1+N$ processes
    to capture the Landau-Pomeranchuk-Migdal (LPM) effect;
\end{itemize}
while others require to evaluate LL and P enhanced terms more carefully,
\begin{itemize}
  \item
    accounting for the full momentum-dependent screening effects of the medium,
    rather than approximating them via an asymptotic mass $\meff \sim m_\infty \sim \mathcal{O}(gT)$, as well as including background field dependence in the mass.
\end{itemize}
In fact, both the momentum-dependent screening effects and the background field dependence of masses are needed to get the P enhanced terms correct at LL.
Strictly, the approximations made in \WallGo{}
(and in all the existing literature on wall velocities)
are consequently only  correct at LL in cases where there are no P enhanced terms.

Below we discuss the subtleties and relative importance of the different contributions
(power enhanced, leading-logarithmic enhanced, and next-to-leading-logarithmic) in more detail.

\subsubsection{Demonstrating scaling of infrared-enhanced contributions in quark scattering}
\label{sec:powerDiv}

\begin{figure}
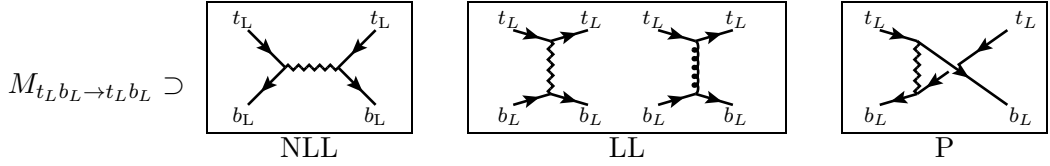

\begin{align*}
   M_{t_\rmii{$L$} b_\rmii{$L$} \to t_\rmii{$L$} b_\rmii{$L$}} &\supset
      \begin{array}{c}
        \fbox{%
          \begin{minipage}{2.5cm}
            \vspace{0.6em}
            \centering
            $\displaystyle
            \VtxvSn(\Lqu,\Lqu,\Luq,\Luq,\Lglii,
                \;\;\;t_\rmii{L},
                t_\rmii{L}\;\;,
                b_\rmii{L}\;\;,
                \;\;\;\\b_\rmii{L}
               )$
            \vspace{0.6em}
            \hspace{-2mm}
          \end{minipage}%
        }\\[-2pt]
        \text{NLL}
      \end{array}
      \quad
      \begin{array}{c}
        \fbox{%
          \begin{minipage}{4cm}
            \vspace{0.6em}
            \centering
            $\displaystyle
            \VtxvTn(\Luq,\Lqu,\Lqu,\Luq,\Lglii,
                  \;\;\;t_\rmii{$L$},
                  t_\rmii{$L$}\;\;,
                  b_\rmii{$L$}\;\;,
                  \;\;\;\\b_\rmii{$L$}
                 )
               $
            $\displaystyle
            \VtxvTn(\Luq,\Lqu,\Lqu,\Luq,\Lgliii,
                  \;\;\;t_\rmii{$L$},
                  t_\rmii{$L$}\;\;,
                  b_\rmii{$L$}\;\;,
                  \;\;\;\\b_\rmii{$L$}
                 )
               $
            \vspace{0.6em}
            \hspace{-2mm}
          \end{minipage}%
        }\\[-2pt]
        \text{LL}
      \end{array}
      \quad
      \begin{array}{c}
        \fbox{%
          \begin{minipage}{2.5cm}
            \vspace{0.6em}
            \centering
            $\displaystyle
            \VtxvUn(\Lqu,\Lqu,\Luq,\Luq,\Lglii,
                \;\;\;t_\rmii{$L$},
                t_\rmii{$L$}\;\;,
                b_\rmii{$L$}\;\;,
                \;\;\;\\b_\rmii{$L$}
                )$
            \vspace{0.6em}
            \hspace{-1mm}
          \end{minipage}%
        }\\[-2pt]
        \text{P}
      \end{array}
\end{align*}
\caption{%
  Contributing $2 \to 2$ scattering processes for
  $t_\rmii{L} b_\rmii{L} \to t_\rmii{L} b_\rmii{L}$ in the SM.
  Zig-zag lines denote $\mathrm{SU}(2)$ vector bosons ($Z$, $W^{\pm}$),
  curly lines represent gluons,
  and directed lines correspond to quarks ($t_\rmii{L}$, $b_\rmii{L}$).
  While there is only a single $t$-channel LL gluonic contribution~\cite{Arnold:2003zc},
  the weak sector receives
  power-enhanced (P) contributions from the $u$-channel and
  NLL contributions from the $s$-channel diagrams.%
}
\label{fig:tb->tb:powerEnhanced}
\end{figure}

A typical example of power-enhanced collision terms in the SM is scattering between left-handed quarks, such as
$t_\rmii{$L$} b_\rmii{$L$} \to t_\rmii{$L$} b_\rmii{$L$}$.
Since these distributions differ under Yukawa interactions,
it is then justified to treat $t_\rmii{$L$}$ and $b_\rmii{$L$}$ as separate
out-of-equilibrium distributions, $\delta f_{t_L}$ and $\delta f_{b_L}$.
Moreover, the top quark is driven out of equilibrium by the bubble wall more strongly than the
bottom quark.
While one could also allow quark distributions to differ in QCD due to their
Higgs-induced masses, because gluon emission vertices do not change quark flavour, any IR enhancements are softened to at most logarithmic~\cite{Arnold:2003zc}.

Following algorithm~\ref{alg:divergence-structure},
the $u$-channel of a processes in which
the incoming and outgoing particles satisfy
$a \neq d$ or
$b \neq c$
gives rise to power-enhanced IR contributions.%
\footnote{%
  In previous wall-velocity computations,
  such terms did not appear because the corresponding processes
  were neglected; see e.g.~\cite{Kozaczuk:2015owa,Jiang:2022btc}.
}
Hence, in contrast to pure QCD,
$W$ boson emission does change particle flavour, and thus
can lead to power enhancements
that dominate and significantly increase the collision rate;
see fig.~\ref{fig:tb->tb:powerEnhanced}.
Since these power enhanced weak interaction processes allow
the top and bottom distributions to 
equilibrate, 
they significantly reduce the friction from the top quark.
This is confirmed by  observing the significant difference between 
the green solid line (top quark only out of equilibrium)
and the blue dot-dashed line (top out of equilibrium, including weak interactions)
in the left panel of fig.~\ref{fig:vwOutOfEq}.
One may think that the power-enhanced collision terms make it
essential to solve for the out-of-equilibrium distribution of the bottom, 
and that the IR enhancement might get softened because 
the top and bottom distributions become more similar.
However, the grey dotted line in the left panel of
fig.~\ref{fig:vwOutOfEq} indicates that this effect is not very pronounced
in the xSM, since allowing the bottom quark to go out of equilibrium
does not lead to a significant change in $\vw$.

\begin{figure}[t]
  \centering
  \includegraphics[width=0.5\textwidth]{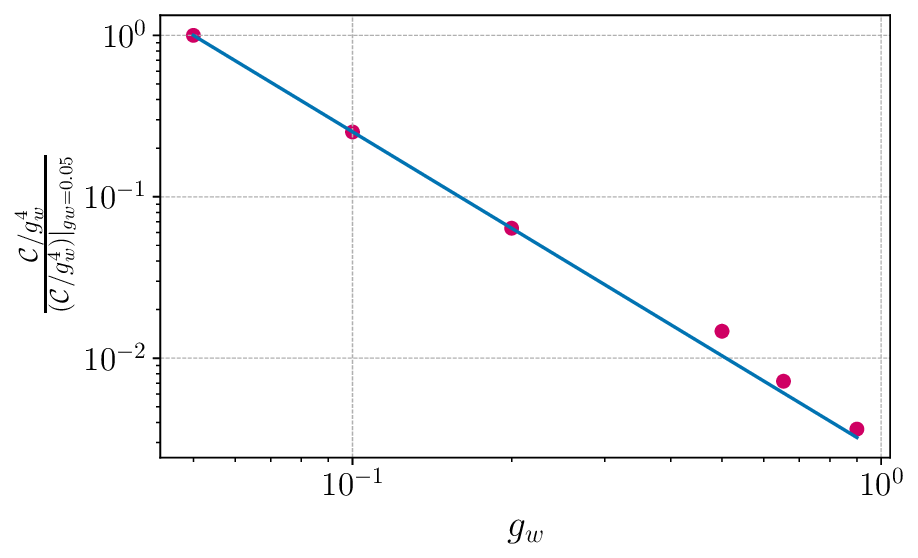}%
  \includegraphics[width=0.5\textwidth]{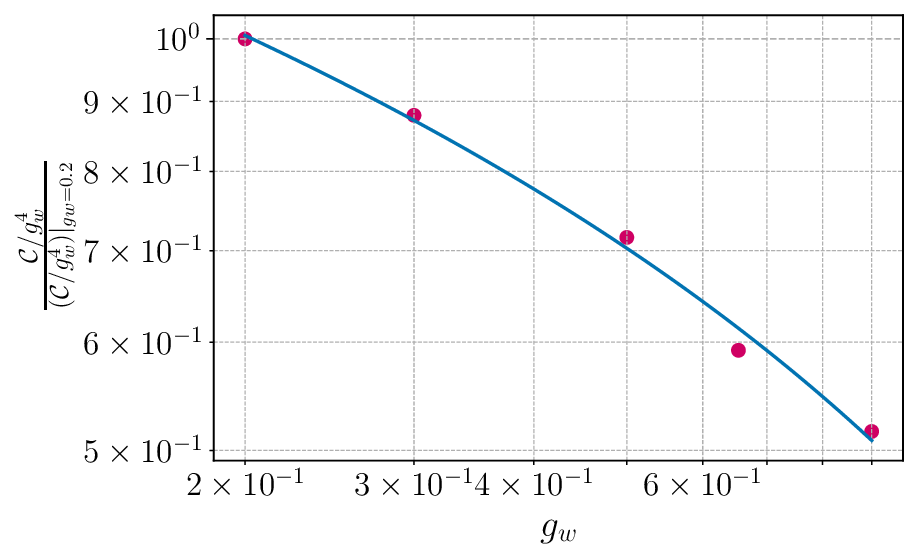}%
  \caption{%
  Estimate of the collision integral in the limit where $\mathcal C \propto 1/\delta f$,
  computed in the Standard Model with a Higgs mass of $\mH = 34$~GeV,
  and only out-of-equilibrium, weakly interacting, left-handed top quarks.
  Left: power-enhanced terms.
  Right: logarithmically enhanced terms.
  See the main text for details on the plotted quantities.
  }
  \label{fig:compare:P:LL}
\end{figure}
To demonstrate the scaling of the collision terms, we compute the 
out-of-equilibrium contribution
\begin{equation}
  \Delta_{00}^{t_L} = \int_\vec{p} \frac{1}{E} \delta f^{t_L}(p^\mu, z)
  \,,
\end{equation}
 of the left-handed top quark, in the Standard Model with a Higgs mass of $34 \, {\rm GeV}$, as implemented in~\cite{Ekstedt:2024fyq}, 
 while varying the $g_w$ appearing in the matrix elements,
and keeping $g_w$ constant everywhere else in the computation.
 We consider only weak interactions with other quarks and $W$ bosons (we ignore the leptons),
 and we consider two sets of matrix elements: power-enhanced and logarithmically enhanced.
 Since in the limit of large collisions, the out-of-equilibrium distributions 
 become inversely proportional to the collision term, $\Delta_{00}^{t_L} \propto 1/\mathcal C$,
 we use a large {\tt collisionMultiplier} (i.e.\ $100/g_w^2$ for power enhanced and 100/$g_w^4$ for log enhanced) to force the system into this regime.
We set the wall velocity $\vw = 0.1 $ by hand, and determine $\Delta_{00}^{t_L}$ by iteratively solving
the background equation of motion to determine the fluid profile and the wall parameters, and 
the Boltzmann equation.
Figure~\ref{fig:compare:P:LL} shows the estimated value of $\mathcal C/g_w^4$, obtained from the inverse
of $\Delta_{00}^{t_L}$, as a ratio to the value 
for the smallest value of $g_w$.
Since we use a $g_w$-dependent {\tt collisionMultiplier}, 
we multiply the result of the Boltzmann equation for $\Delta_{00}^{t_L}$ by {\tt collisionMultiplier} again at the end 
to undo this scaling.

The left panel of fig.~\ref{fig:compare:P:LL} shows an estimate of the collisions when we include only power-enhanced terms.
In our simplified set-up, these are caused by scattering of the top with other quarks,
and e.g. $t_{\rmii{$L$}} b_\rmii{$L$} WW$ interactions.
The dots show the estimated collision term, and the solid line shows a fit to a power law,
$\propto g_w^{-a}$ with power $a \approx 1.98 $ demonstrating the scaling 
predicted by eq.~(\ref{eq:power_divergence}). 

The right panel of fig.~\ref{fig:compare:P:LL} shows an estimate of the collisions for log-enhanced terms.
We used the same set-up as above, but kept only the log-enhanced terms. We set the terms
from diagrams with a quark propagator to zero by hand, as the asymptotic quark mass depends on $g_s$, but not $g_w$.
Moreover, we remove terms proportional to $s^2$ from the numerator, as these terms also contain terms that are beyond leading log.
The solid line shows a fit to $a \ln (1/g) + b$, and we find $a \approx 0.33$ and $b \approx 0.47 $, indeed confirming the logarithmic scaling.
For smaller values of $g_w$, we do not recover the logarithmic behaviour. The reason is that $\Delta^{t_L}_{00}$ crosses zero between $g_w = 0.05$
and $g_w = 0.1$ causing deviations from the predicted scaling.

Below, we will attempt to quantify the overall uncertainty arising from the
approximate treatment of such divergences.
A more rigorous treatment of power-divergent contributions is deferred to future work.

\subsubsection{Estimating the full leading order by using a collision multiplier}
\label{sec:BLL}

In models with power enhancements, without using full HTL self-energies there is an unavoidable $\mathcal{O}(1)$ error in the collision integrals, as argued above.
Even in models without power enhancements, this error only reduces to $\mathcal{O}(1/\ln \frac{\meff}{T})=\mathcal{O}(1/\ln g^{-1})$, which goes to zero very slowly as $g\to 0$.
So for models with fixed, mildly weak couplings, such as the electroweak sector,
LL expressions are in practice always subject to numerically
$\mathcal{O}(1)$ uncertainties in the collision integrals.
In the absence of a full leading-order computation,
that includes a careful treatment
of power enhanced contributions, as well as LL and NLL terms,
we would like to estimate the impact of these uncertainties for the bubble wall speed.

\begin{figure}[t]
  \centering
  \includegraphics[width=0.5\textwidth]{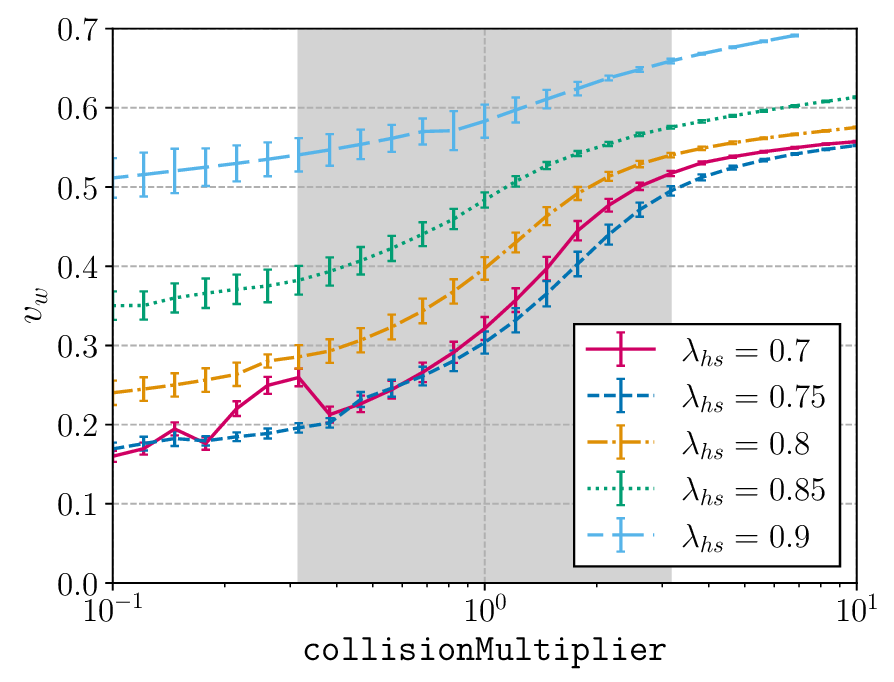}%
  \includegraphics[width=0.5\textwidth]{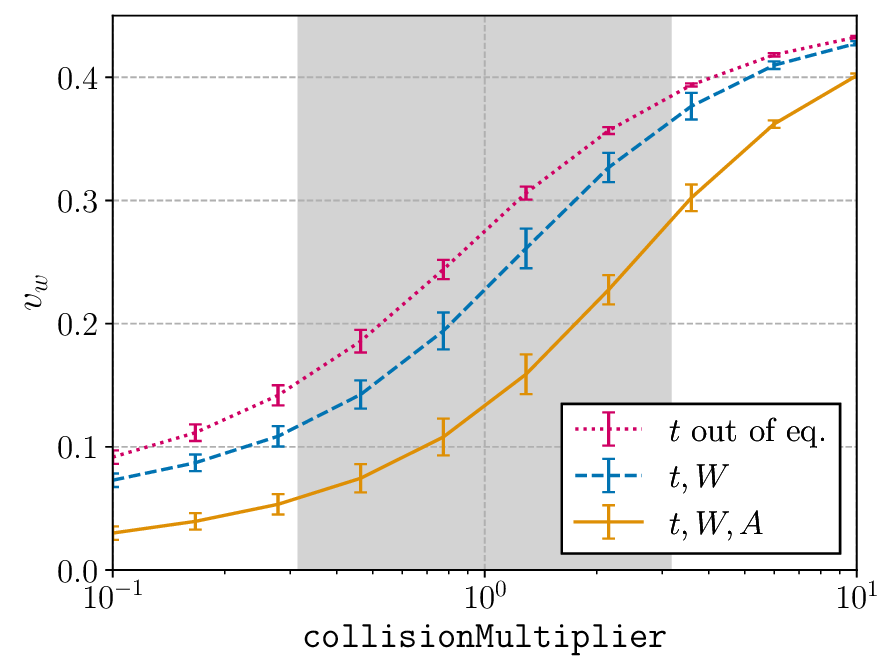}
  \caption{%
    Bubble wall velocity $\vw$ as a function of {\tt collisionMultiplier}.
    Left: {\tt BM:xSM1} with the top quark, $t$, out of equilibrium and varying $\lambda_{hs}$.
    Right: {\tt BM:IDM1} for $\mA = 265$~GeV and
    different out-of-equilibrium particle sets,
    as indicated in the legend.
    The gray band indicates the range ${\tt collisionMultiplier} \in [10^{-1/2}, 10^{1/2}]$.
  }
  \label{fig:collisionMultiplier}
\end{figure}

Previous computations have shown the effects of such NLL terms
to yield $\mathcal O(25\%)$ corrections to transport coefficients~\cite{Arnold:2003zc}, and
an $\mathcal O(200\%)$ enhancement to
the dilepton production rate~\cite{Anisimov:2010gy, Ghiglieri:2016xye, Ghiglieri:2014kma}.
To estimate the effect of this uncertainty on the wall speed,
we introduce a rescaling factor, {\tt collisionMultiplier},
that multiplies all included collision integrals.
Scanning over the range of values
${{\tt collisionMultiplier} \in [10^{-1}, 10]}$
yields a conservative estimate of the uncertainty.
For our final error estimates
presented in the concluding table~\ref{tab:uncertainties_summarised}, however,
we restrict to the narrower interval $[10^{-1/2}, 10^{1/2}]$,
while the full variation of
$\vw({\tt collisionMultiplier})$ is displayed in
fig.~\ref{fig:collisionMultiplier}.

In the xSM (left panel of fig.~\ref{fig:collisionMultiplier}), weaker transitions with smaller $\lambda_{hs}$
exhibit a stronger sensitivity to {\tt collisionMultiplier},
with $\vw$ varying by roughly a factor of  $\sim 2-8$ across the full range.
For stronger transitions, where the wall moves faster,
this dependence is significantly reduced and
$\vw$ remains large regardless of the collision strength.
In all cases, $\vw$ increases monotonically with {\tt collisionMultiplier},
since stronger collisions keep the distribution functions closer to LTE,
thereby reducing friction.

In the IDM, fig.~\ref{fig:collisionMultiplier} (right) shows the dependence on
{\tt collisionMultiplier} for
$\mA = 265$~GeV and
for different sets of out-of-equilibrium particles.
We have used the collision integrals with $\vcoll = \vn$, 
(cf.\ eq.~\eqref{eq:VEV:col:factor}).
Again, $\vw$ increases monotonously with {\tt collisionMultiplier}.
As expected, at large {\tt collisionMultiplier},
the wall speed becomes largely insensitive to the choice of out-of-equilibrium particles.
The variation $\Delta\vw$ due to the choice of particles is
\begin{align}
\label{eq:IDM3:collisionMultiplier:0.1}
 &\text{\tt collisionMultiplier = 0.1}
\,,&
\Delta\vw &= \mathcal{O}(100\%)
\,,\\
&\text{\tt collisionMultiplier = 10}
\,,&
\Delta\vw &= \mathcal{O}(10\%)
\,.
\end{align}
For eq.~\eqref{eq:IDM3:collisionMultiplier:0.1},
$\vw$ can vary by more than a factor of 2 depending on the out-of-equilibrium particle content.
This highlights again the importance of including all relevant species,
especially in scenarios where heavy particles are weakly interacting and
slowly equilibrate.

\subsubsection{Estimating full NLL contributions from scalar quartic interaction}
\label{sec:NLL:scalar:quartic}

To further investigate the effect of NLL terms,
we keep the LL matrix elements but now include
all possible scalar quartic interactions
in the $S_i S_j \to S_k S_l$ channel.
These interactions arise at tree level as
\begin{align}
\label{eq:general:quartic}
  M_{S_i S_j \to S_k S_l} \supset
  \Vtxvn(\Lxx,\Lxx,\Lxx,\Lxx,{k},{i},{j},{l})
  \;\sim\; \lambda_{ijkl}
  \,,
\end{align}
where the external dashed fields denote scalars $S_i$,
and $i,j,k,l$ label the corresponding scalar indices,
and $\lambda_{ijkl}$ is the quartic coupling tensor~\cite{Ekstedt:2024fyq}.
The amplitudes in eq.~\eqref{eq:general:quartic}
contribute to matrix elements $|M|^2 \supsetsim \lambda_{ijkl}^2$,
which goes beyond the LL approximation.
We therefore use them as a proxy to estimate
the numerical robustness of the wall velocity determination,
noting that a full NLL computation requires all contributions listed
in sec.~\ref{sec:Boltzmann:enhancements}.

To estimate the order of such additional contributions,
we consider
collision integrals corresponding to the matrix elements of the $SS\to SS$ channel
\begin{align}
\label{eq:quartic:matrix:element:scaling}
    \left|
        \scalebox{0.5}{$\VtxvTn(\Lxx,\Lxx,\Lxx,\Lxx,\Lglx,{},{},{},{})$}
      + \scalebox{0.5}{$\VtxvUn(\Lxx,\Lxx,\Lxx,\Lxx,\Lglx,{},{},{},{})$}
      + \scalebox{0.5}{$\VtxvTn(\Lxx,\Lxx,\Lxx,\Lxx,\Lxx,{},{},{},{})$}
      + \scalebox{0.5}{$\VtxvUn(\Lxx,\Lxx,\Lxx,\Lxx,\Lxx,{},{},{},{})$}
      + \dots
    \right|^2
    &\sim
    \mathcal{O} \biggl(
      g^4 \ln (1/g),
      \biggl(\frac{\lambda_{ijkl} \vcoll}{T} \biggr)^{\!4} \ln \frac{T}{\meff}\biggr)
    \,,\nn
    \left| \scalebox{0.5}{$\Vtxvn(\Lxx,\Lxx,\Lxx,\Lxx,{},{},{},{})$} \right|^2
    &\sim
    \mathcal{O} \Bigl( \lambda_{ijkl}^2 \Bigr)
  \,,
\end{align}
where the diagrams represent the integrated collision terms
rather than the bare matrix elements
barring a factor $T^2 \delta f$.
For simplicity, we assume $\vcoll/T \sim \mathcal{O}(1)$,
and neglect possible further
parametric suppressions or enhancements from cubic scalar interactions.
There are three scenarios to consider:
\begin{itemize}
  \item[(i)]
    $\lambda_{ijkl} \lesssim g^2 \lesssim 1$.
    The scalar quartic NLL and scalar exchange LL
    contributions are
    subleading compared to the vector-induced LL and power enhanced contributions;
  \item[(ii)]
    $g^2 \lesssim \lambda_{ijkl} \lesssim 1$.
    The scalar quartic NLL contributions can become comparable to
    the LL vector-induced contributions
    and should be included for a more accurate determination of $\vw$;
  \item[(iii)]
    $g^2 \lesssim 1 \lesssim \lambda_{ijkl}$.
    The scalar-induced LL contributions dominate over both quartic contributions and
    the LL vector-induced contributions.
\end{itemize}
In models with comparatively large scalar quartic couplings,
eq.~\eqref{eq:quartic:matrix:element:scaling} and
point~(ii) imply that quartic NLL contributions
can become comparable to, and potentially as significant as,
the leading-logarithmic or power enhanced terms.
This happens for example in the
IDM~\cite{Kainulainen:2019kyp} in the region with a strong first-order phase transition.
In such models (cf.\ point~(iii)), 
cubic interactions (yielding contributions $\propto \lambda_{ijkl}^4$)
can dominate.

\begin{figure}[t]
  \centering
    \includegraphics[width=0.5\textwidth]{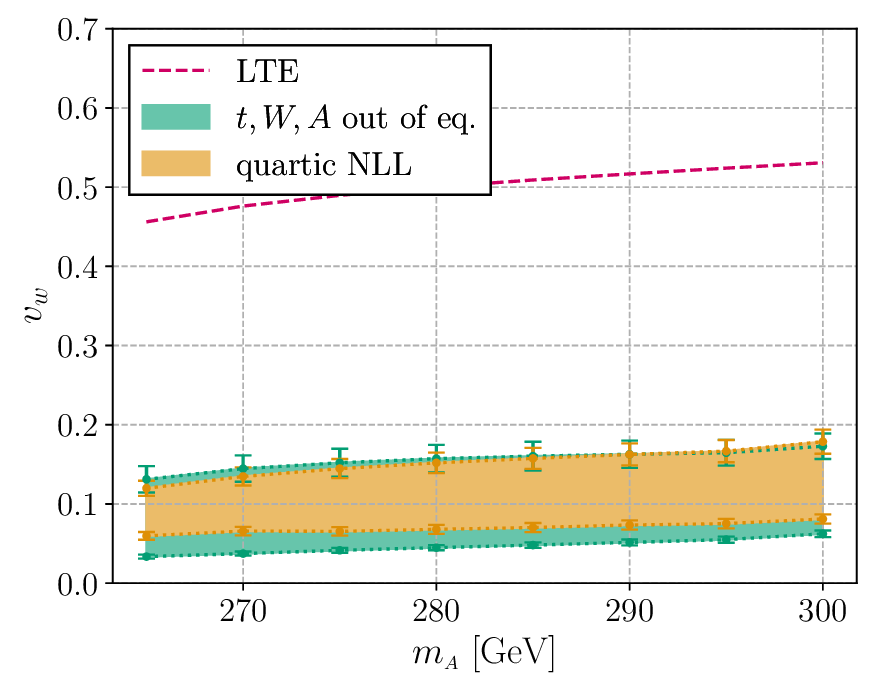}%
  \caption{%
    The wall velocity for
    {\tt BM:IDM1}
    by putting top, $W$ and $A$ particles out of equilibrium.
    Quartic NLL contributions of the form~\eqref{eq:general:quartic} are either
    included (orange) or
    excluded (green) in the matrix elements.
    The bands are defined as in fig.~\ref{fig:vwOutOfEq}.
  }
  \label{fig:vwOutOfEqNLL}
\end{figure}
In this study, we focus again on the IDM, considering the benchmark point
{\tt BM:IDM1} of table~\ref{tab:IDM}. 
The dominant quartic coupling in this setup is $\lambda_3 \gtrsim 2$, 
which enters for example in the scalar scattering process 
$A A \rightarrow h h$. 
Note that, while a large coupling such as $\lambda_3 \gtrsim 2$ 
leads to slow perturbative convergence,
thanks to $4\pi$ loop suppression factors, perturbative unitarity is intact.
We have numerically confirmed that no Landau pole appears  
for energy scales below 100 TeV.  

In fig.~\ref{fig:vwOutOfEqNLL}, we explore the impact of including pure scalar
quartic NLL contributions, while also taking the $t$, $W$, and $A$ particles
out of equilibrium,
since quartic interactions as in eq.~\eqref{eq:general:quartic} contribute only
when all external lines are scalars.
Including these NLL effects typically increases $\vw$ slightly, as the additional 
interactions enhance the equilibration of the heavy scalars and thereby reduce 
the corresponding friction. 
For $\vcoll = \vn \sim \Tn$ and $\lambda_{3}, \dots, \lambda_{5} \sim \mathcal{O}(1)$,
the LL term is enhanced by
nearly an order of magnitude compared to the NLL term.
The inclusion of the NLL terms therefore modifies $\vw$ only at the 
level of $\mathcal{O}(10^{-2})$,
which is expected given point~(iii) above.

While parametric suppression due to $\vcoll/T$ was not discussed in points~(i)--(iii) above,
we see how it can become relevant in practice in fig.~\ref{fig:vwOutOfEqNLL}.
For smaller values of $\vcoll$, the impact of the NLL quartic 
terms becomes more pronounced
as the decreased ratio $\vcoll/T$
effectively suppresses the $\lambda_{ijkl}$ terms in the LL contribution by a 
factor of $(1/2)^4$.
Consequently, the inclusion of the LL term produces a more pronounced 
relative change in $\vw$.

\subsection{Particle masses}
\label{sec:Boltzmann:negativeMsq}

In studies employing Boltzmann equations to determine the wall velocity
in cosmological phase transitions, comparatively little attention
has been devoted to the precise form of the particle masses
entering these equations.
Here, we discuss the relevant masses for these particles,
based on results from the high-temperature QCD literature.
These studies indicate that the effective masses entering the Boltzmann equations
differ from the conventional assumptions, implying that existing computations
may carry potentially significant systematic uncertainties.
A detailed theoretical investigation of the appropriate mass definition
for use in Boltzmann equations is deferred to future work.

In this section, we will also focus on the particular case of a negative mass squared parameter appearing in the Boltzmann equation,
which initially drew our attention to the more refined treatment of the particle masses,
as the thermal corrections can be crucial for the Boltzmann equation to be well-behaved.

\subsubsection{Asymptotic masses for particles in the Boltzmann equation}
\label{sec:asymptoticMasses}

It has long been appreciated that describing the behaviour of a scalar field undergoing a phase transition
requires taking into consideration thermal effects, most prominently on the thermal mass of the field, i.e. the static screening mass~\cite{Arnold:1992rz,Kajantie:1995dw}.
Similarly to the scalar field, the masses of the particles obtain corrections from the thermal medium.
The thermally corrected mass affects the dynamics of the particles and hence should appear, at least at some order,
in the Boltzmann equation describing these particles. Below, we will show that the thermal corrections
to particle masses can be quite important to the wall velocity.

The derivation of the Boltzmann equation from e.g.\ the Kadanoff-Baym equations requires
the so-called quasi particle approximation.
References~\cite{Jeon:1994if, Blaizot:2001nr} demonstrated for a real scalar and for hot QCD that this approximation fixes the particles on shell with their mass given by
the {\em asymptotic} masses, and not the vacuum masses,%
\footnote{%
  The name \emph{asymptotic mass} stems from the fact that it corresponds to
  the mass of an on-shell particle in the limit of infinite three-momentum.%
}
in contrast to the usual assumption in wall velocity computations.
At leading order, the thermal part of the asymptotic masses can be obtained from the usual static screening masses
as half of (double) the static thermal correction to the mass squared for gauge (fermion) fields.
For scalar fields, the two masses coincide,
which is important for stability of the system as we will discuss in
sec.~\ref{sec:negativeMassSquared}.
Beyond leading order, further perturbative and non-perturbative
corrections to the asymptotic masses
are generated, as in QCD, for example~\cite{Ghiglieri:2023ies,Ghiglieri:2021bom}.

In this section, we refer to the \emph{full} asymptotic mass, which includes both the
field-dependent mass and the leading thermal correction.
The linearised collision term is computed as in the other parts of the paper. Hence, this asymptotic mass
is different to the \emph{symmetric phase} asymptotic masses appearing in the scattering matrix elements.
Also, the external particles in the collisions have been assumed to be ultrarelativistic.
The comparison between the asymptotic mass and vacuum mass is made in the force term of the Liouville operator
and the source term of the off-equilibrium particles.
In the other parts of this work, we have followed the conventions in the literature, and used vacuum masses
 in the Liouville operator and the source term.
\begin{figure}[t]
  \centering
  \includegraphics[width=0.5\textwidth]{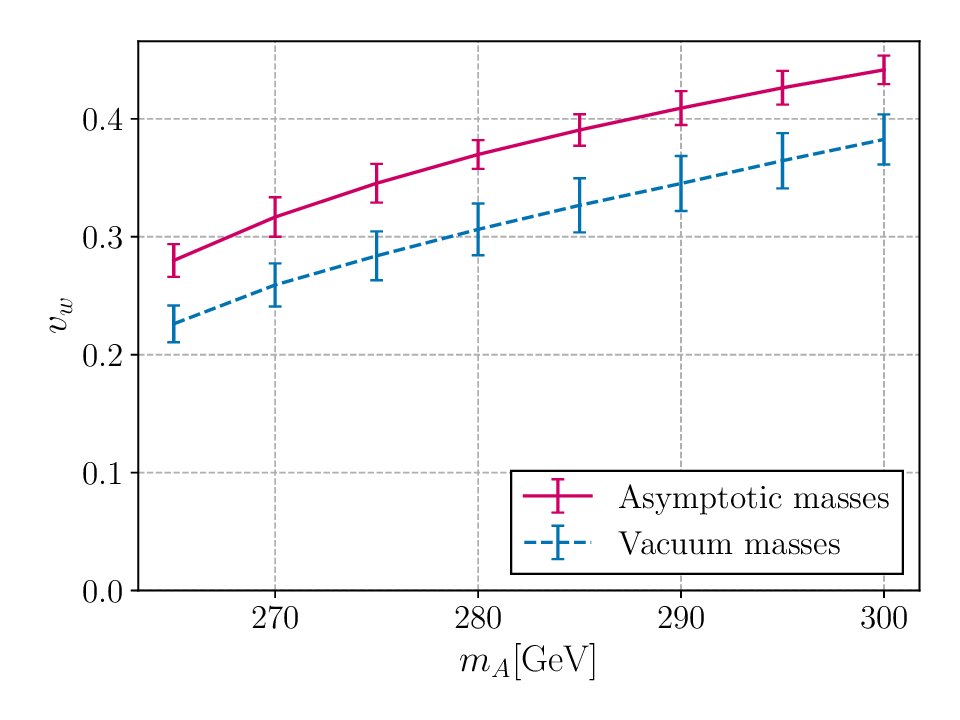}%
  \includegraphics[width=0.5\textwidth]{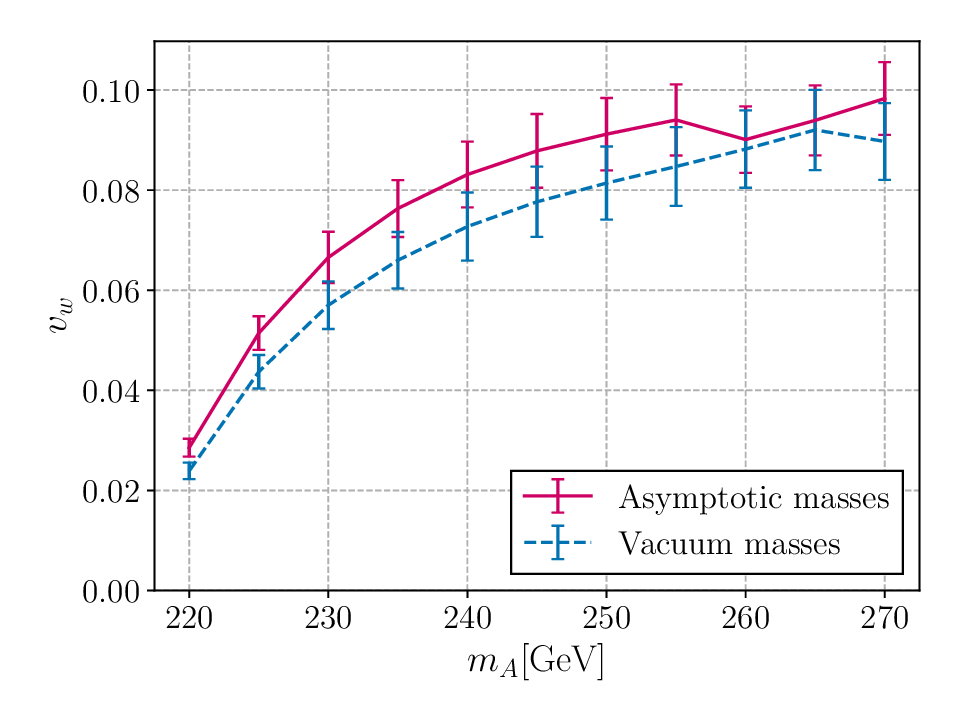}
  \caption{%
    Wall velocities for
    {\tt BM:IDM1} with top and $W$ out of equilibrium (left) and
    {\tt BM:IDM3} with top, $W$ and $A$ out of equilibrium (right) for different
    choices of particle masses, with $\vcoll=vn$.
    Using asymptotic thermal masses in the Liouville operator increases the wall velocity
    by approximately 10--25\%, primarily due to a reduced source term leading to smaller deviations
    from equilibrium in the distribution functions.}
  \label{fig:differentMassesIDM}
\end{figure}
Figure~\ref{fig:differentMassesIDM} shows that replacing the vacuum masses by the
asymptotic masses in the Boltzmann equation could have a substantial impact on the wall velocity.
When the masses of the out-of-equilibrium particles are set to their 
asymptotic values,
the resulting wall velocity can increase by 25\%.
This effect is primarily due to a reduced source term, $\mathcal S_a\propto f_\text{eq}'(E)$ in Eq.~\eqref{eq:Boltzmann:outofeq} below.
The derivative is smaller due to the increased, field-independent mass,
which leads to smaller deviations from equilibrium in the distribution functions.
These findings highlight a potential source of error in earlier wall velocity computations
that neglected thermal mass corrections for the relevant particles.

Note that {\tt WallGo} version \WallGoVersion{} does not allow for background temperature-dependent asymptotic masses
in the Liouville operator and the source term.
Figure~\ref{fig:differentMassesIDM} was produced by hardcoding the thermal mass correction into the vacuum mass
using the lowest nucleation temperatures of both benchmarks.
The nucleation temperature ranges for {\tt BM:IDM1} and {\tt BM:IDM3} were $[113.1,124.4]$ and $[118.5,132.1]$ in GeV respectively.

\subsubsection{Apparent superluminal propagation of particles with negative mass squared}
\label{sec:negativeMassSquared}
When analysing how particles affect the wall velocity,
one may encounter negative squared masses,
$m_a^2 < 0$,
for the particles in the Boltzmann equation
due to the non-constant background fields.
Most notably, this always occurs in the Boltzmann equations
for (a linear combination of the) scalar fields forming the wall,
reflecting the instability that drives its propagation through the plasma.
Note that this does not only occur when the particles are described by the vacuum mass,
but also when one uses the thermally corrected asymptotic mass, as discussed in sec.~\ref{sec:asymptoticMasses}.
As far as we are aware,
the occurrence of negative squared masses
has not been addressed in the wall velocity literature
where the focus is typically on the contributions from
fermions and gauge bosons, which do not develop $m_a^2<0$.

Here, we will examine different inconsistencies arising from a negative mass squared.
We will then discuss a Boltzmann equation for wave packets in classical field theory as
an analogue to Boltzmann equations in QFTs.

In the wall frame,
the Boltzmann equation for out-of-equilibrium particles takes
the form of (cf.\ e.g.~\cite{Laurent:2022jrs, Ekstedt:2024fyq} for further details)
\begin{equation}
\label{eq:Boltzmann:outofeq}
  \Bigl(
      p^\mu \partial_\mu
    + \frac{1}{2}\vec{\nabla }m_a^2\cdot \nabla_{\vec{p}}
    \Bigr) \delta f^a(\vec{p}, x^\mu)
    = - \mathcal C_{ab}^{\rm lin}[\delta f^b] + \mathcal S_a
    \,.
\end{equation}
A negative mass-squared term can affect eq.~\eqref{eq:Boltzmann:outofeq}
through the particle energies,
$p^0 = \sqrt{\vec{p}^2 + m^2}$,
which appear both in the time-derivative term and in the equilibrium distribution function
inside the source term, $\mathcal S_a\propto f_\text{eq}'(E)$.
The most evident breakdown of the Boltzmann equation for $m_a^2<0$
arises for particles with low momenta,
$\mathbf{p}^2 + m_a^2 < 0$, whose energies become imaginary.
In this regime, the fluctuations corresponding to the particles are themselves unstable.
For high-momentum particles,
the Boltzmann equation with a negative mass squared exhibits a subtler failure.
Such particles can propagate faster than the speed of light,
{\em viz.}
\begin{align}
\label{eq:superLuminalFirstContact}
  \abs{\vec{v}}=\frac{\abs{\vec{p}}}{p^0}&=\frac{\abs{\vec{p}}}{\sqrt{\vec{p}^2+m_a^2}}>1
  \,,&
  m_a^2&<0
  \,, &
  \vec{p}^2+m_a^2 &\geq0
  \,.
\end{align}
Hence, the Boltzmann equation would imply violations of causality.

A negative mass squared can enter the Boltzmann equation in a number of ways, for instance, if a scalar field has a negative vacuum mass squared, and one chooses to use this vacuum mass for the corresponding scalar particle in the Boltzmann equation.
This indeed arises for some benchmark points of
the IDM introduced in sec.~\ref{sec:models},
where the vacuum mass squared at vanishing field value,
$\mu_2^2$, is negative for the scalar fields in the new doublet.
However, the thermal mass squared (including the vacuum mass and the thermal correction) of the field is positive in any minimum of the thermal effective potential,
so there is no actual underlying instability.
It is important to stress that for scalars
the asymptotic mass discussed in sec.~\ref{sec:asymptoticMasses}
coincides with the static thermal mass.
Consequently, using the scalar field asymptotic mass in the Boltzmann equation
resolves this issue of apparent instabilities for vanishing external field values.
This holds generally in the minima of any effective potential.

However, this is not the end of the story.
Including thermal corrections
does ensure that
the squared masses remain positive in different phases (i.e.\ in minima of the effective potential), but
it does not guarantee positivity throughout the bubble wall.
For the scalar field undergoing the transition, the thermally corrected mass squared
necessarily becomes negative in the bubble wall,
which follows from the instability of the potential describing the bubble wall.
A negative mass squared will also occur
for Goldstone modes appearing in a symmetry breaking transition.

Now that we have identified the negative mass squared appearing for the nucleating fields,
let us step a bit back and discuss how these problems appear, noting the different scales present.
The derivation of the Boltzmann equation requires a scale hierarchy between the temperature
and the wall width, $T\gg L_\text{wall}^{-1}$.
The description of the nucleating quantum field effectively splits into two descriptions:
the Boltzmann equation for the UV thermal particles, $E\sim T$, and the IR classical background field,
$E\sim L_\text{wall}^{-1}$. Thus the possible imaginary energy for the particles is not within
the applicability region of the Boltzmann equation, $E\sim T$.
Rather, it falls into the energy scales of the background field, $E\sim L_\text{wall}^{-1}$.
When handling the field fluctuations consistently,
the imaginary energy does not actually arise~\cite{Dashko:2024anp}.
Thus, the only issue within the intended applicability regime of the Boltzmann equation
for particles with $E \sim T$
is the apparent superluminal propagation of particles in regions with $m_a^2 < 0$.

To interpret the superluminal propagation appearing in the Boltzmann equation,
we will turn to an easier but analogous situation:
a Boltzmann equation in classical field theory for self-similarly propagating wave packets.
The Liouville operator of the Boltzmann equation will be exactly the same as for off-equilibrium QFTs, eq.~\eqref{eq:Boltzmann:outofeq}.
Correspondingly, the Boltzmann equation for the classical wave packets will exhibit
exactly the same superluminal propagation in the regions with negative mass squared
even though the underlying description is causal. Details related to the derivation and to causality can be found in appendix~\ref{app:BoltzmannEquationSuperluminal}.
Note that we will not treat collisions in the current discussion.
Their information must also propagate causally, but we leave a study 
of this to future work.
For simplicity, the computation is done in one dimension .

The gist of the derivation is that one can identify the locations, $\mathbf{x}_n$,
and the wave numbers, $\mathbf{k}_n$, of the wave packets, and show that they evolve according to a local description in terms of velocities,
$\dot{\mathbf{x}}_n=\mathbf{v}_n(\mathbf{x}_n,\mathbf{k}_n)$ and forces,
$\dot{\mathbf{k}}_n=\mathbf{F}_n(\mathbf{x}_n,\mathbf{k}_n)$.
Hence, the description in terms of positions and wave numbers corresponds to a particle-like description.
In addition, the velocities and forces are identical in form to the corresponding quantities of
the particles in off-equilibrium QFTs,
$\dot{\mathbf{x}}_n=\mathbf{v}_n(\mathbf{x}_n,\mathbf{p}_n)$,
$\dot{\mathbf{p}}_n=\mathbf{F}_n(\mathbf{x}_n,\mathbf{p}_n)$.
This correspondence allows us to pinpoint the origin of the superluminal propagation.

The main result is that the wave packet locations, $\mathbf{x}_n$, \textit{do} move superluminally
in the regions with negative squared masses.
Correspondingly, the particles in the Boltzmann equation must also move superluminally.

We want to emphasise that the wave-packet locations moving superluminally does not violate causality.
The reason for the apparent superluminal motion is
that the information of the wave packet is not all at its location, $\mathbf{x}$.
The self-similarly propagating wave packets are global structures, and the information about them is
already spread throughout the space.
Inside the regions of negative squared masses,
the centre of the wave packet just moves superluminally without carrying information itself
due to the group velocity becoming superluminal.
If we identify the wave packet as a particle, the entity can now move superluminally without breaking causality.
The origin of superluminal propagation can thus be understood as
\textit{promoting a global structure into a local, point-like one}.

The Boltzmann equation serves as a proper description for the self-similarly propagating wave packets and their superluminal propagation matches the underlying causal theory.
However, it is not entirely clear if the apparent superluminal propagation of
particles in Boltzmann equations from interacting quantum field theories is always physical.
For example, consider a particle that is created within the region of negative mass squared.
It travels out of the light cone originated at its creation event, and can interact with
other particles in a manner that seems acausal.
Again however, the particle creation is not a localisable event in quantum field theories, since the
particles themselves are fundamentally non-localisable (e.g.\ sec.~6.5 in~\cite{Duncan:2012aja}).
Thus, it is entirely conceivable that a Boltzmann equation with collisions
is still a consistent description.
This scenario warrants further investigation.

\section{Scalar field equation of motion}
\label{sec:scalar_fields}

In this section, we will focus on the dynamical equations for the scalar fields,
having investigated the contributions of Boltzmann particles to the bubble wall speed in sec.~\ref{sec:Boltzmann}.

\subsection{Constructing the equation of motion}

The equation of motion for the scalar fields $\phi_i$, coupled to the Boltzmann particles, is modelled as
\begin{equation}
\label{eq:ScalarEOM}
	\partial^2 \phi_i
	+ \frac{\partial \Veff(\bm\phi, T)}{\partial \phi_i}
	+ \sum_a \frac{\partial m_a^2}{\partial \phi_i}
  \int_\vec{p} \frac{1}{2E} \delta f^a(p^\mu, z)
  =0
	\,,
\end{equation}
where the sum runs over all particle species in the plasma,
$\delta f^a$ denotes the deviation from the corresponding equilibrium distribution function, and
$z$ denotes the coordinate perpendicular to the bubble wall.
The equilibrium contributions of all plasma particles have been
absorbed in $\Veff(\bm\phi, T)$.

Note that the structure of eq.~\eqref{eq:ScalarEOM} is not the most general possible, and may be insufficient to capture all relevant physical effects.
In the literature, a number of modifications to its form have been considered, including
(i)
derivative corrections from near-local-equilibrium Boltzmann particles~\cite{Ares:2021nap}, %
(ii)
nonlocal terms from the contributions of fluctuations of the scalar fields themselves~\cite{Dashko:2024anp},
(iii)
fluctuation and dissipation terms, as in a Langevin equation~\cite{Eriksson:2025owh},
Boltzmann-like corrections related to background-field dependent vertices~\cite{Ai:2025bjw},
and
(iv)
higher-dimensional operator terms induced when constructing the EOM in an EFT picture
by integrating out hard or soft modes~\cite{Chala:2024xll,Chakrabortty:2024wto,Bernardo:2025vkz,Chala:2025aiz}.
However, we will not consider further such modifications to the structure of equation~\eqref{eq:ScalarEOM}.

Our focus in this section will be to test how corrections to the effective potential $\Veff(\bm\phi, T)$ affect the bubble wall speed $\vw$.
The effective potential affects the dynamics of the scalar fields directly through its equation of motion,
but also indirectly by modifying the masses of the coupled Boltzmann particles.
Further, the hydrodynamic equations contain thermodynamic quantities which depend on $\Veff$ evaluated at its minima,
and the nucleation temperature $\Tn$, at which the computation is performed, depends on $\Veff$.

In the wall velocity literature it is common to use for $\Veff$ the finite-temperature one-loop effective potential, potentially with daisy-resummed masses,
cf.~\cite{Moore:1995si, Konstandin:2014zta, Lewicki:2021pgr, Laurent:2022jrs}.
This is also what we have used for the two models studied here.
However, utilising the one-loop effective potential in such a way is ad-hoc,
as the one-loop effective potential is derived for constant, homogeneous background scalar fields, yet a bubble wall is not constant.
One might hope that this could be justified by a derivative expansion, yet this is \textit{never} possible for the fluctuations of the scalar fields themselves;
see sec.~3.5 of~\cite{Croon:2020cgk}.
The breakdown of the derivative expansion is also
related to pathological imaginary parts in $\Veff$,
typically removed by hand either through $m_a^2\to |m_a^2|$ or alternatively
$\Veff \to \re \Veff$.
In the two models studied here, we have chosen the latter option.

Our investigation of $\Veff$ is motivated by the slow convergence of perturbation theory
at high temperature, due to the Bose-enhancement of IR-modes.
Consequently, thermodynamic parameters derived from $\Veff$, such as
the critical temperature $\Tc$,
the nucleation temperature $\Tn$, or
the phase transition strength $\alpha$
can exhibit large
theoretical uncertainties at low perturbative orders~\cite{Kainulainen:2019kyp, Croon:2020cgk, Gould:2021oba}.
This is often reflected in a strong sensitivity to the renormalisation group scale.
In recent years,
a growing body of literature has investigated the impact of higher-order corrections
on phase transition parameters and the resulting GW signals~\cite{%
  Laine:2017hdk,Kainulainen:2019kyp,Croon:2020cgk,Gould:2021oba,Gould:2023jbz,Kierkla:2023von,%
  Gould:2024jjt,Lewicki:2024xan,Ekstedt:2024etx,Kierkla:2025qyz,Kierkla:2025vwp,Bhatnagar:2025jhh,%
  Chala:2025oul}.
These studies consistently find that
the daisy-resummed one-loop effective potential does not provide 
sufficient accuracy for reliable predictions.

For the computation of $\vw$, one might be tempted to simply compute $\Veff$ to higher loop order,
for example, using the two-loop output of {\tt DRalgo}~\cite{Ekstedt:2022bff}, and to plug this potential into the equations of motion.
Unfortunately, there is no guarantee that this brings us closer to the true $\vw$,
as these equations are by no means less ad-hoc than the ones based on the one-loop effective potential.
Effective field theory offers a consistent framework to resolve these issues, whereby only those modes with energies much higher than the scalar fields contribute to the form of $\Veff$~\cite{Gould:2021ccf, Hirvonen:2024rfg, Eriksson:2025owh}.
The effects of lower energy fluctuations remain coupled, and their propagators must be solved for in the presence of the varying scalar background fields~\cite{Dashko:2024anp}.
This avoids the breakdown of the derivative expansion.
The correction can be added perturbatively to the scalar equations of motion,
or the backreaction can be accounted for iteratively;
see e.g.~\cite{Garbrecht:2015yza, Carosi:2024lop}.
Since such a procedure has not been implemented in {\tt WallGo},
we leave a detailed comparison for future work.

In this section, we estimate the uncertainty associated to $\Veff$.
First, we will study the dependence on the nucleation temperature for
a fixed effective potential, and then
we will compute $\vw$ with an effective potential with and without daisy resummation.

\subsubsection{Nucleation temperature}
\label{sec:EOM:Tn}

One of the most important inputs for determining
the bubble wall velocity is the nucleation temperature $\Tn$.
The degree of supercooling,
quantified by $\Tc - \Tn$, is directly related to the pressure difference
$\Delta p(\Tn)$ driving the bubble expansion, and thus to
the strength of the phase transition $\alphan$.
In the limit of vanishing supercooling, $\Tn \approx \Tc$,
the pressure difference vanishes by definition ($\Delta p(\Tc) = 0$), and
the wall cannot propagate, implying $\vw = 0$.
More generally, for small supercooling,
the wall velocity is expected to scale approximately as
$\vw \propto \Tc - \Tn$~\cite{Witten:1984rs, Gouttenoire:2023roe, Cline:2025bwe}.

As discussed in the introduction to this subsection, obtaining accurate results for thermodynamic quantities requires computing $\Veff$ to sufficiently high loop order. 
Once a reliable effective potential is available, the critical temperature $\Tc$ can typically be determined with good accuracy, for example through a strict perturbative expansion~\cite{Laine:1994zq,Gould:2023ovu}.
Indeed, it has been shown that such computations can reproduce lattice results for $\Tc$ with impressive precision~\cite{Kainulainen:2019kyp,Gould:2023ovu,Ekstedt:2024etx}.

In contrast, the nucleation temperature $\Tn$ is more difficult to compute reliably.
Unlike $\Tc$, which depends only on the location of the minima of $\Veff$, the determination of $\Tn$ involves the bubble nucleation rate, which is governed by the full effective action.
At one-loop order, this requires evaluating the fluctuation determinant around
the bounce solution~\cite{Callan:1977pt}.
For simple models, this has been automated in the {\tt BubbleDet} package~\cite{Ekstedt:2023sqc}.
However, comparisons with non-perturbative lattice simulations show that the agreement for
the nucleation rate is significantly worse than for quantities like $\Tc$~\cite{Gould:2025wec},
resulting in an uncertainty in the value of $\Tn$.

Furthermore,
the expected relative error on $\vw$ caused by the uncertainty $\Delta \Tn$
should scale with $\Delta \Tn/(\Tc-\Tn)$,
 since $(\Tc - \Tn)$ characterises the temperature range
over which the thermodynamics of the model changes significantly. 
Therefore, even a relatively small fractional uncertainty $\Delta \Tn / \Tn \lesssim 1$ can lead to a large relative uncertainty in $\vw$ when the supercooling is small, i.e.\ when $\Delta \Tn \sim \Tc - \Tn$.

\begin{figure}[t]
    \centering
    \includegraphics[width=0.51\textwidth]{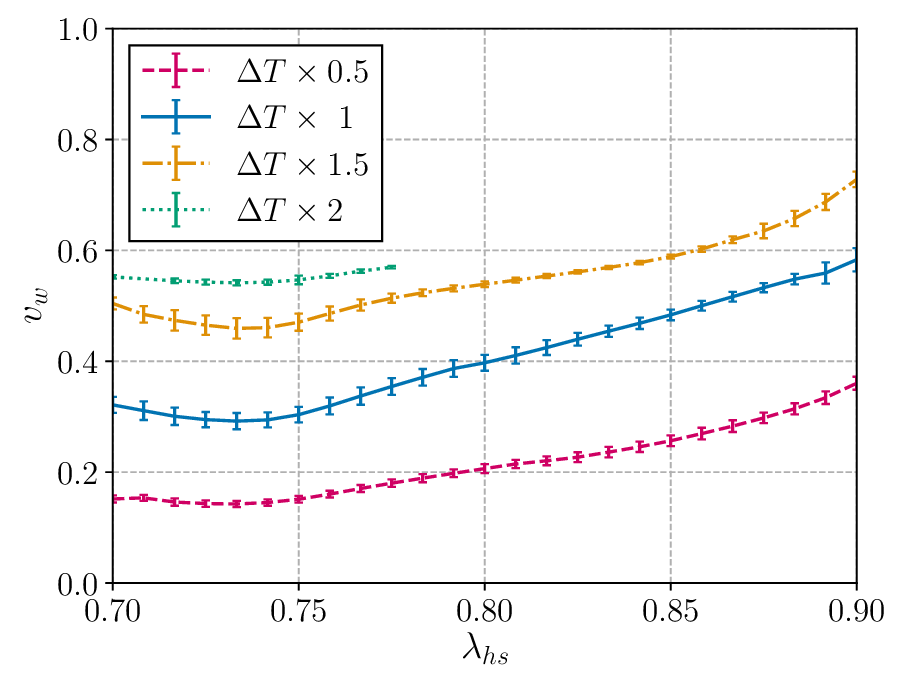}%
    \includegraphics[width=0.49\textwidth]{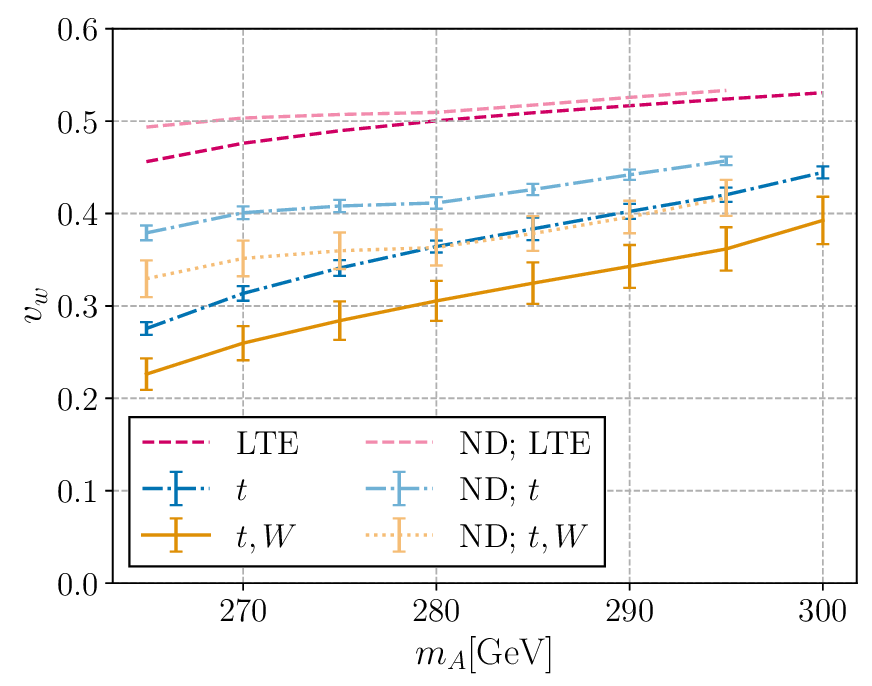}
    \caption{%
      Left:
      Wall velocity in {\tt BM:xSM1} for different amounts of supercooling,
      $\Delta T = \Tc - \Tn$.
      Each curve corresponds to a modified $\Tn$ such that $\Delta T$ is rescaled by
      $\Delta T \to \Delta T \times N$ with a factor $N \in \{0.5, 1, 1.5, 2\}$
      relative to the value obtained in the original scan. 
      In the most extreme case ($N = 2$),
      the false vacuum becomes unstable at large values of $\lambda_{hs}$,
      resulting in an incomplete curve.
      Right:
      Wall velocity of {\tt BM:IDM1} with (dark colors) and without (light colors)
      daisy resummation, for different sets of out-of-equilibrium particles,
      as indicated in the legend.
    }
    \label{fig:supercooling:daisy}
\end{figure}
To obtain a more quantitative estimate of how sensitive $\vw$ is to $\Tn$,
we show in fig.~\ref{fig:supercooling:daisy} (left) the wall velocity in {\tt BM:xSM1}
as a function of the amount of supercooling.
One can see that the error estimate previously obtained is quite accurate for the slowest points, which have the least supercooling;
e.g.\ the solid-blue and dashed-pink lines for $\lambda_{hs}\lesssim 0.8$.
For faster walls, $\vw$ becomes less sensitive on $\Delta \Tn$, in agreement with the general expectation given by eq.~\eqref{eq:vwError}.
Despite this,
$\Tn$ remains a critical parameter, as it can even influence
the qualitative nature of the phase transition.
For instance,
if the supercooling becomes too large,
the false vacuum fraction may even be non-decreasing with time,
preventing a successful phase transition~\cite{Turner:1992tz}.
This is what happens for the dotted-green line when $\lambda_{hs}\gtrsim0.77$.
This underscores the need for a precise evaluation of $\Tn$
to evaluate the wall velocity accurately.

\subsubsection{The effective potential with and without daisy resummation}
\label{sec:EOM:Veff}

Let us also investigate the uncertainty associated with the implementation of the effective potential itself.
In particular, we focus on the role of the daisy resummation,
which captures
partial leading plasma screening effects by resumming the most infrared-sensitive ring diagrams.
This computation is, however, not self-consistent, because the derivative expansion breaks for these loops~\cite{Croon:2020cgk, Gould:2021ccf}.
To quantify the numerical impact of this ambiguity,
we study the benchmark {\tt BM:IDM1} and compare it to a scenario
where the effective potential is computed \emph{without} daisy resummation,
keeping the heavy scalar mass $\mA$ fixed.
In both cases, the nucleation temperature is determined from the condition
$S_b = 127$, where $S_b$ denotes the bounce action.

Figure~\ref{fig:supercooling:daisy} (right) shows the wall velocity, $\vw$,
computed under three different assumptions:
(i) LTE,
(ii) with only the top quark out of equilibrium,
(iii) with both the top quark and the $W$ boson out of equilibrium,
and for $\vcoll = \vn$ in all cases.
The choice of effective potential, $\Veff$, has the largest impact on $\vw$
for smaller values of $\mA$ when either the top quark alone or both the top and $W$ boson 
are out of equilibrium, with deviations in $\vw$ reaching up to $\mathcal{O}(50\%)$.
This enhanced sensitivity can be understood
since the wall velocity $\vw$ becomes more sensitive
to small variations in the pressure when $\vw$ is small.
The pressure difference across the wall changes most rapidly near the Jouguet velocity $\vJ$,
but varies only weakly for $\vw \ll \vJ$.
Consequently, when the phase transition is weaker and out-of-equilibrium particles 
provide significant friction, $\vw$ remains relatively small, and even modest changes 
in $\Veff$ (and hence in the pressure) can lead to disproportionately large variations in $\vw$.

\subsection{Solving the equation of motion}
\label{sec:EOM:tanhAnsatz}

An important approximation made by \WallGo{} is to represent
the profile of the scalar fields with a tanh \textit{ansatz} of the form
\begin{equation}
\label{eq:tanhAnsatz}
  \phi_i(z) =
      v_{\rmii{HT},i}
    + \frac{v_{\rmii{LT},i}-v_{\rmii{HT},i}}{2}
    \Bigl[
      1
    - \tanh\Bigl(\frac{z}{L_i}+\delta_i\Bigr)
  \Bigr]
  \,,
\end{equation}
where
$v_{\rmii{HT},i}$ is the high-$T$ and
$v_{\rmii{LT},i}$ the low-$T$ \vev{} that
are fixed by the minimisation of the effective potential.
This profile is meant to represent the qualitative features of the scalar fields in a simple manner, 
and, as discussed below, it forms the exact solution to the equation of motion
in a cubic-plus-quartic interaction potential at the critical temperature.
The profile interpolates between the scalar field values on both sides of the wall while allowing for arbitrary wall widths $L_i$ and offsets $\delta_i$.%
\footnote{%
  Exploiting translational invariance along the $z$-axis,
  we fix the first wall position by setting $\delta_1 = 0$.
}
These parameters are chosen to minimise the effective action,
yielding the moment equations~\cite{Ekstedt:2024fyq}
\begin{equation}\label{eq:momentsEOM}
    \int {\rm d}z\,\frac{\partial\phi_i}{\partial \delta_i}R_i
  = \int {\rm d}z\,\frac{\partial\phi_i}{\partial L_i}R_i=0
    \,,
\end{equation}
with $R_i$ the residual of the EOM for $\phi_i$
\begin{equation}\label{eq:EOMResidual}
  R_i(z) =
      - \partial_z^2 \phi_i
      + \frac{\partial V_{\rm eq}}{\partial \phi_i}
      + \frac{\partial V_{\rm out}}{\partial \phi_i} \,,
\end{equation}
and $\partial V_{\rm out} / \partial \phi_i$ representing the out-of-equilibrium contributions.
Although the eqs.~(\ref{eq:momentsEOM}) are trivially satisfied when the EOMs are satisfied, they are not sufficient alone to guarantee an exact solution for $\phi_i$. Therefore, one should expect the value of $\vw$ calculated by \WallGo{} to be impacted by the fact that the EOMs are not exactly satisfied when the tanh \textit{ansatz}~\eqref{eq:tanhAnsatz} is used.

\begin{figure}[t]
    \centering
    \includegraphics[width=0.5\textwidth]{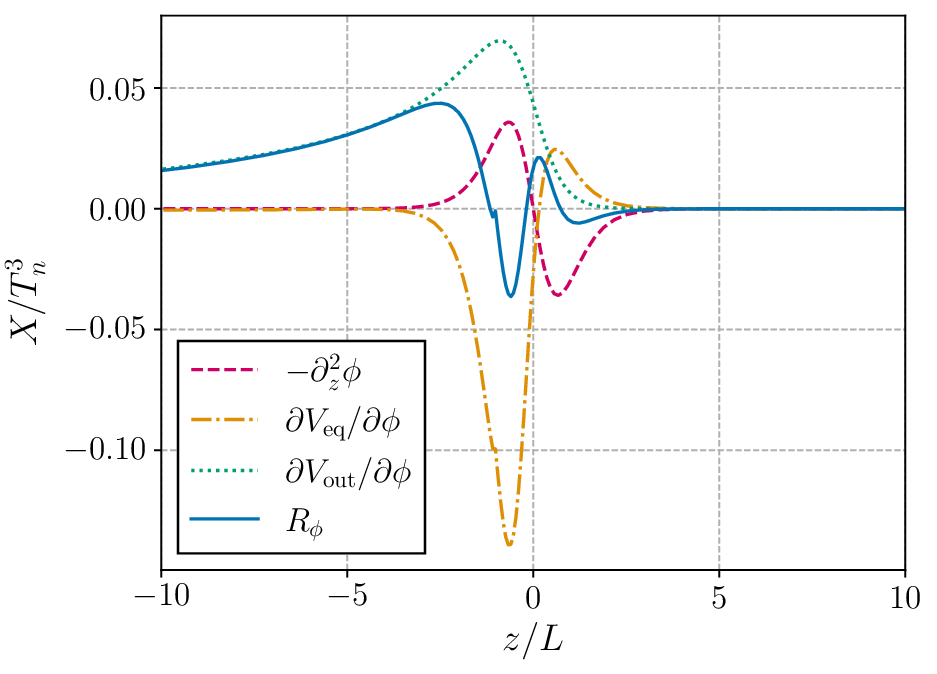}
    \caption{%
      Different terms appearing in the Higgs equation of motion, as well as
      its residual in solid blue, for a solution in the xSM
      with $m_s=120$~GeV,
      $\lambda_{hs}=0.9$ and $\lambda_s=1$ that satisfies
      the moment equations~\eqref{eq:momentsEOM}.
      This solution has a root mean square residual of $R_\rmii{RMS}\approx0.4$.
    }
    \label{fig:tanhError}
\end{figure}

To assess the degree to which the EOMs are violated,
it is useful to examine $R_i(z)$ and each individual term appearing in its definition (\ref{eq:EOMResidual}).
Figure~\ref{fig:tanhError} shows these terms for a representative xSM solution.
Near the wall, the residual $R_i$ is generally smaller than the individual terms,
indicating that 
the tanh \textit{ansatz} offers a qualitatively appropriate solution,
though the residual remains sizeable.
Behind the wall, however, only
$\partial V_{\rm out} / \partial \phi_i$ remains nonzero, and
the EOMs are no longer satisfied.
This is expected, since the moment equations~\eqref{eq:momentsEOM}
involve derivatives such
as $\partial \phi / \partial \delta$ or $\partial \phi / \partial L$,
which vanish rapidly away from the wall.
As a result, the region $|z| \gtrsim L$ does not contribute to the moments and
remains unconstrained by the moment equations.

Ultimately, this issue arises because the EOMs depend on several physical length scales.
For instance,
the kinetic term $\partial_z^2\phi_i$ varies over a scale  $L_i\sim 1/m_i$,
while the out-of-equilibrium term $\partial V_{\rm out} / \partial \phi_i$
typically varies on a different scale $\sim 1/\Gamma_i$, 
with $\Gamma_i$ denoting the typical collision rate appearing in the Boltzmann equation.
In contrast,
the tanh \textit{ansatz}~\eqref{eq:tanhAnsatz} depends only on the single scale $L_i$ and
is therefore unable to resolve additional structure associated with other relevant scales.

For the simple cubic-plus-quartic potential in LTE, the tanh \textit{ansatz} becomes exact as the degree of supercooling goes to zero. At this point, the scalar masses in the two phases are equal, so that there is really only one scale entering the solution. For nonzero supercooling, the leading deviations from this are $\mathcal{O}((\Tc^2-\Tn^2)/(\Tc^2-T_0^2))$, where $T_0$ is the temperature at which the metastable phase becomes absolutely unstable;
cf.\ $\bar{\lambda}$ in~\cite{Enqvist:1991xw}.
Note that, in more complicated models or beyond LTE,
the tanh \textit{ansatz} will not in general become exact as the degree of supercooling goes to zero.

To have a more quantitative measure of how accurate the tanh \textit{ansatz} is,
we define the root mean square residual
\begin{equation}\label{eq:RMSResidual}
  R_{\rmii{RMS},i} = \sqrt{\frac{1}{N_i}\int\! {\rm d}z\, R_i^2}
  \,,
\end{equation}
with $N_i$ a normalisation factor defined to represent the typical scale of each term in the EOM
\begin{equation}
  N_i=\int\! {\rm d}z\, \left[\left(\partial_z^2\phi_i\right)^2+\left(\frac{\partial V_{\rm eq}}{\partial \phi_i}\right)^2+\left(\frac{\partial V_{\rm out}}{\partial \phi_i}\right)^2\right]
  \,.
\end{equation}
An accurate solution of the EOM has a small $R_\rmii{RMS}$, while $R_\rmii{RMS}$ will become of $\mathcal{O}(1)$ for bad solutions. For example, the solution shown in fig.~\ref{fig:tanhError} has $R_\rmii{RMS}\approx 0.4$, reflecting the fact that the EOM is only solved relatively well inside and in front of the wall.

\begin{figure}
    \centering
    \includegraphics[width=0.6\textwidth]{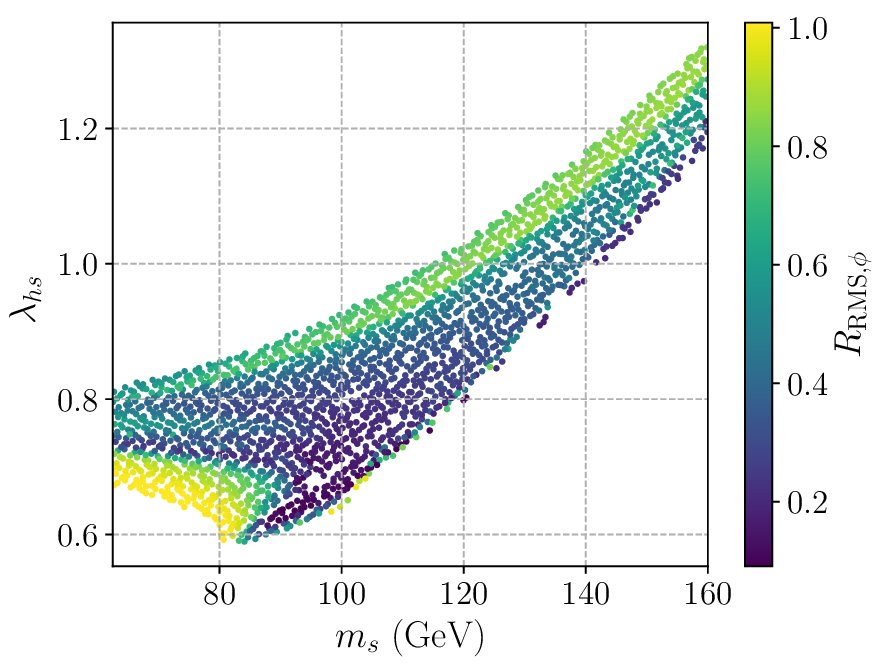}
    \caption{%
      Value of the root mean square residual of the Higgs EOM defined in eq.~\eqref{eq:RMSResidual} for the xSM scan.
      The average $R_\rmii{RMS}$ over the whole scan is approximately 0.5.
      }
    \label{fig:tanhErrorScan}
\end{figure}

The value of $R_\rmii{RMS}$ obtained by \WallGo{} in the xSM is shown in fig.~\ref{fig:tanhErrorScan}.
For many points, $R_\rmii{RMS}$ can be quite large, going up to 1.
The average value over the whole scan is roughly 0.5,
reflecting the fact that the tanh \textit{ansatz} cannot resolve the long tail behind the wall.
This clearly indicates that the tanh \textit{ansatz} is a poor approximation to
represent the profile of the scalar fields undergoing the phase transition.
It is however unclear how this error in $\phi_i(z)$ will propagate to the wall velocity,
which is ultimately what we are interested in.
Determining $\Delta \vw$ coming from this source of uncertainty would necessitate comparing
the EOM's solution with and without the tanh \textit{ansatz}.
This goes beyond the capability of \WallGo{} at the moment.
However, a generalised profile going beyond the tanh \textit{ansatz}
is planned to be implemented in a future version of \WallGo{}.

\section{Discussion}
\label{sec:discussion}

In summary, we have assessed the current state-of-the-art calculations of
the bubble wall velocity and identified key sources of theoretical uncertainty.
To this end, we employed the software package \WallGo{}~\cite{Ekstedt:2024fyq},
which numerically solves the coupled Boltzmann, scalar field, and hydrodynamic equations
for almost arbitrary models, using
a spectral representation of the distribution functions.
While our investigation is not exhaustive,
it represents the most comprehensive analysis to date and offers guidance for future improvements.

In the course of this work,
we also made several theoretical advances.
We demonstrated that linearising the Boltzmann equation is almost always a valid approximation,
as the associated error vanishes both near and far from equilibrium.
We showed that infrared power enhancements can dominate
individual channels of the collision integrals,
and that capturing them correctly at leading order requires including background field dependence and the full momentum dependence of the hard-thermal-loop self-energies.
Additionally, we clarified the role of negative mass-squared regions in
the Boltzmann equations and showed that the resulting superluminal propagation is
a spurious artefact of the quasiparticle approximation.

\begin{table}[t]
\centering
\begin{tabular}{|l|l|c|c|c|}
\hline
\multicolumn{2}{|c|}{\multirow{2}{*}{Source of uncertainty}} 
& \multicolumn{2}{c|}{$\Delta v_w / v_w$} 
& \multirow{2}{*}{$\Delta P / P_{\rm eq}$} \\
\cline{3-4}
\multicolumn{2}{|c|}{} & xSM & IDM & \\
\hline\hline
\multirow{4}{*}{\hyperref[sec:Boltzmann]{Boltzmann}}
& \hyperref[sec:Boltzmann:linearisation]{Linearisation}
& 0.1\% &  0.01\% & $\mathcal{O}\Bigl(\delta f_2/f_\text{eq}\Bigr)$ \\
& \hyperref[sec:powerDiv]{Power enhanced collision integrals}\protect\hyperlink{fn:ddagger}{$^\ddagger$}
& 90\% & 300\%& $\mathcal{O}(\delta f_1/f_\text{eq})$ \\
& \hyperref[sec:NLL:scalar:quartic]{NLL collision integrals}
& --- &  20\% & $\mathcal{O}\Bigl(\delta f_1/\bigl(f_\text{eq} \ln(1/g)\bigr)\Bigr)$\protect\hyperlink{fn:dagger}{$^\dagger$} \\
& \hyperref[sec:Boltzmann:particles:inout]{Particles in and out of equilibrium}
& 30\% &  400\% & $\mathcal{O}\Bigl(\sum_a \delta f_1^a/f_\text{eq}\Bigr)$\\
& \hyperref[sec:Boltzmann:negativeMsq]{Particle masses}
& --- &  20\% & --- \\
\hline \hline
\multirow{3}{*}{\hyperref[sec:scalar_fields]{Scalar field EOM}}
& \hyperref[sec:EOM:Tn]{Nucleation temperature}
& 100\% & --- & $\mathcal{O}(g)$ \\
& \hyperref[sec:EOM:Veff]{Effective potential}
& --- &  20\% & $\mathcal{O}(g)$ \\
& \hyperref[sec:EOM:tanhAnsatz]{Tanh-Ansatz}
& 50\%\protect\hyperlink{fn:star}{$^*$} & --- & --- \\
\hline
\end{tabular}
\caption{%
    Comparison of relative uncertainties from different sources for the xSM and IDM benchmarks.
    For each source of uncertainty,
    we show the median relative error~\eqref{eq:vwError}
    across the parameter points studied.
    Horizontal lines indicate absent data.
    The numerical values should be interpreted with caution as
    they are specific to the benchmarks analysed and are intended only as
    a rough indication of which sources of uncertainty merit further attention.
    \\[-3pt]
    \rule{0.4\textwidth}{0.4pt}\\[-3pt]  %
    {\footnotesize
    \protect\hypertarget{fn:ddagger}{$^\ddagger$}
    These are estimated from the variation in $v_w$ within the grey band in fig.~\ref{fig:collisionMultiplier}}
    \\
    {\footnotesize
    \protect\hypertarget{fn:dagger}{$^\dagger$}
    This order assumes the absence of power-enhanced collision integrals.}
    \\
    {\footnotesize
    \protect\hypertarget{fn:star}{$^*$}
    This value is an estimate of the error in the equation of motion, not a direct estimate of $\Delta \vw/\vw$.}
  }
\label{tab:uncertainties_summarised}
\end{table}
Table~\ref{tab:uncertainties_summarised} summarises our numerical findings,
using two well-studied models as benchmarks: the xSM and the IDM.
For each source of uncertainty, we provide numerical estimates for the magnitude of the error on $\vw$ in these two models.
The quoted value of $\Delta \vw/\vw$ is obtained as
\begin{equation}
	\Delta \vw/\vw = {\rm median}\left(\frac{\text{max}(v_{w})-\text{min}(v_{w})}{\text{min}(v_{w})} \right) \times 100 \%
  \,,
\end{equation}
where the median is taken over all data points displayed in the corresponding graphs, and then rounded to one significant figure.
Further details can be found in the corresponding sections above.
The uncertainties in the IDM are generally larger than in the xSM,
likely due to the stronger couplings we considered.
This, in turn, is related to the one-step nature of the phase transition in
the IDM~\cite{Blinov:2015vma, Laine:2017hdk, Kainulainen:2019kyp}.
These numerical estimates are necessarily specific to the parameter points studied and are intended only as a rough guide.
To complement these numerical values, the right column provides an analytic estimate for the relative uncertainty in the pressure.

Uncertainties arising in the (linearised) Boltzmann equation propagate to
the wall velocity through deviations from equilibrium, $\delta f_1$,
which lead to uncertainties in the pressure.
Since the equilibrium distributions are included separately in the scalar field equation of motion and hydrodynamic equations,
a relative $\mathcal{O}(1)$ error in $\delta f_1$ induces
a relative $\mathcal{O}(\delta f_1/f_\text{eq})$ error in
the pressure, the error in the wall velocity
then follows from eq.~\eqref{eq:vwError}.
We have identified several sources of error that approximately reach this magnitude.
The size of deviations from local equilibrium is determined by
the relative importance of collisions versus the background field in the Boltzmann equation.
In the limit of large collisions,
deviations from equilibrium vanish as
$\delta f/f_\text{eq}\propto 1/\mathcal{C}$, while in
the opposite limit of small collisions,
$\delta f/f_\text{eq}= \mathcal{O}(1)$.

We have also studied uncertainties in the scalar equation of motion itself.
Our conclusions, summarised in the lower half of table~\ref{tab:uncertainties_summarised},
also reveal numerically large uncertainties.
Using a standard one-loop approximation,
the parametric size of these errors is somewhat smaller than those in the Boltzmann equation.
However, uncertainties in the scalar equation of motion more directly affect the wall speed.
Consequently, even though the errors are parametrically suppressed by a single power of the coupling
$g \equiv \sqrt{g^2}$,
this offers little benefit at realistic coupling values.

In summary, we have identified numerous $\mathcal{O}(1)$ uncertainties in
current state-of-the-art calculations of the bubble wall velocity.
Our analysis demonstrates that substantial theoretical progress is required
to achieve predictions more reliable than order-of-magnitude estimates.
The prospect of detecting gravitational waves from cosmological first-order phase transitions
with future experiments such as LISA provides strong motivation to pursue this program.

\section*{Acknowledgements}
We thank
Andreas Ekstedt,
Thanasis Giannakopoulos,
Mark Hindmarsh,
Thomas Konstandin,
Andr\'e Milagre,
Ignacy Na\l{}\k{e}cz,
Tuomas V.I.~Tenkanen,
and
Mateusz Zych
for useful discussions.
We are especially grateful to Jacopo Ghiglieri for clarifications about the leading log approximation.
We thank Nikita Blinov for providing details on the implementation of the Inert Doublet Model.
OG and JH were supported by a Royal Society Dorothy Hodgkin Fellowship.
BL was supported by the Perimeter Institute for Theoretical Physics and the Fonds de recherche du Québec Nature et technologies (FRQNT). Research at Perimeter Institute is supported in part by the Government of Canada through the Department of Innovation, Science and Economic Development Canada and by the Province of Ontario through the Ministry of Colleges and Universities.
PS was supported by
the Swiss National Science Foundation (SNSF) under grant
\href{https://data.snf.ch/grants/grant/215997}{\tt PZ00P2-215997}.
JvdV was supported by the Dutch Research Council (NWO), under project number
\href{https://www.nwo.nl/en/projects/viveni212133}{\tt VI.Veni.212.133}.
\\

\noindent
{\bf Data availability statement.}
The datasets supporting
figs.~\ref{fig:linCriterion}--\ref{fig:spectralConvergence}
and the file used to produce the IDM benchmark points
are publicly available at~\cite{ZenodoData}.
The software versions used were
\WallGo{}~\WallGoVersion{},
\WallGoMatrix{}~\WallGoMatrixVersion{}, and
\WallGoCollision{}~\WallGoCollisionVersion{}.
Diagrams were generated with {\tt Axodraw}~\cite{Collins:2016aya}.

\appendix
\renewcommand{\thesection}{\Alph{section}}
\renewcommand{\thesubsection}{\Alph{section}.\arabic{subsection}}
\renewcommand{\theequation}{\Alph{section}.\arabic{equation}}

\section{New features in \WallGo{} \WallGoVersion{}}
\label{sec:wallGoUpdate}

\subsection{Phase tracing}
When performing parameter scans over different particle physics models in this study,
we found that the phase tracing algorithm of
\WallGo{}~\WallGoOldVersion{} was not always stable.
While no known algorithm can correctly follow the minima of arbitrary non-polynomial potentials,
\WallGoVersion{} provides several updates to address common phase tracing issues and offers a means to circumvent them.

\subsubsection{Previous phase tracing algorithm}
In the original \WallGo{} release \WallGoOldVersion{},
the phase tracing algorithm started from a user-specified approximate minimum of the effective potential,
passed via {\tt WallGo.PhaseInfo},
and used the default Broyden-Fletcher-Goldfarb-Shanno algorithm of \texttt{scipy.optimize.minimize}
to find the precise minimum $\phi_i^{0}$ at the input temperature, $\Tn$.
The temperature dependence of the minimum starting at
$\phi_i^\text{min}(T=\Tn) = \phi_i^{0}$ was then traced by solving
the following first-order initial value problem,
\begin{equation}~\label{eq:phase_tracing_ivp}
      \frac{\partial^2 V_\text{eff}}{\partial \phi_i \partial \phi_j}\bigg|_{\phi=\phi^\text{min}} \frac{\mathrm{d}\phi^\text{min}_j}{\mathrm{d}T}
    + \frac{\partial^2 V_\text{eff}}{\partial \phi_i \partial T}\bigg|_{\phi=\phi^\text{min}}
    = 0 \,.
\end{equation}
The temperature derivative was discretised with the fourth-order accurate Runge-Kutta method
as implemented in \texttt{scipy.integrate.solve\_ivp}.
Phase tracing was stopped if either the temperature step size became too small
or the proposed point was a saddlepoint or maximum of the potential.
Once completed,
the resulting temperature-dependent minimum was interpolated with
a cubic spline for use in other parts of the wall speed computation.

Similar phase tracing algorithms have been adopted in
\texttt{CosmoTransitions}~\cite{Wainwright:2011kj} and
\texttt{PhaseTracer}~\cite{Athron:2020sbe,Athron:2024xrh}.
In contrast,
\texttt{BSMPT}~\cite{Basler:2024aaf} and
\texttt{Vevacious}~\cite{Camargo-Molina:2013qva}
iteratively call a local minimisation algorithm at $T+\delta T$,
starting from the minimum at $T$, to build up the temperature dependence.
This approach is more sensitive to the choice of $\delta T$:
large values can cause jumps between different minima, while
small values result in slow phase tracing.

\subsubsection{Updated phase tracing algorithm}

In \WallGo{}~\WallGoVersion{}, the phase tracing algorithm has been updated as follows.
Rather than interpolating $V_{\rm eff}(T)$, $\phi_i^{\min}(T)$, and their derivatives with cubic splines,
the interpolation degree is now specified by users through the configuration parameter
{\tt configThermodynamics.interpolationDegree}.
The default value is 1, which, despite lower interpolation accuracy,
improves stability around temperatures where a phase disappears
and $\phi_i^{\rm min}(T)$ exhibits nonanalytical behaviour.
Cubic splines were sometimes unreliable around these points, leading to unphysical results.
If no such points occur in the thermal history, users can increase the interpolation degree for higher accuracy.

The computation of derivatives of $V_{\rm eff}(T)$ and $\phi_i^{\min}(T)$ has also been updated.
In \WallGoOldVersion{}, derivatives were computed from the interpolating splines,
introducing a nontrivial dependence on the minimisation of $\phi_i$,
which can be subject to larger errors $\Delta\phi_i$ than expected.
These errors are amplified when taking derivatives,
leading to $\mathcal{O}(\Delta\phi_i^2/\Delta T^2)$ errors in the second derivative,
where $\Delta T$ is the spacing between interpolation points.
Such errors can easily reach $\mathcal{O}(1)$ if the minimisation procedure performs worse than expected,
producing unreliable results.

In \WallGoVersion{},
the derivatives are computed directly from the potential with finite differences.
Since the points used to evaluate the finite differences are all evaluated with the same $\phi_i^{\min}$,
the error $\Delta\phi_i$ does not get enhanced by $1/\Delta T^2$ and the accuracy stays under control.
The total derivatives are computed using the formulas
\begin{align}
  \frac{{\rm d}V_{\rm eff}}{{\rm d}T}&= \frac{\partial V_{\rm eff}}{\partial T}
  \,,\\
  \label{eq:d2VdT2}
  \frac{{\rm d}^2V_{\rm eff}}{{\rm d}T^2}&=
        \frac{\partial^2 V_{\rm eff}}{\partial T^2}
      + \frac{\partial\phi_i^{\min}}{\partial T}
        \frac{\partial^2 V_{\rm eff}}{\partial\phi_i\partial T}
  = 
      \frac{\partial^2 V_{\rm eff}}{\partial T^2}
    - \frac{\partial^2 V_{\rm eff}}{\partial\phi_i\partial T}
      \biggl(\frac{\partial^2V_{\rm eff}}{\phi_i\phi_j}\biggr)^{-1}
      \frac{\partial^2 V_{\rm eff}}{\partial\phi_j\partial T}
    \,,
\end{align}
where we have used eq.~\eqref{eq:phase_tracing_ivp} to derive the last line.

Finally, we have improved the algorithm for finding the end of a phase to make it more robust
by including a new criterion that detects discrepancies between eq.~\eqref{eq:d2VdT2} and
the finite difference evaluation of ${\rm d}^2V_{\rm eff}/{\rm d}T^2$.
The latter uses values of $\phi_i^{\rm min}$ evaluated at several temperatures and
jumps discontinuously from one phase to the other when a phase disappears.
Hence, the finite difference approach breaks down at phase disappearances,
while eq.~(\ref{eq:d2VdT2}) remains accurate,
allowing large deviations between the two results to identify phase disappearances.
We have found that this substantially increases the overall stability of \WallGo{}. 

\subsubsection{User-provided traced phases}
In certain situations, users may prefer to bypass phase tracing and directly specify
the positions of the minima and the corresponding potential values.
This can accelerate the program's initialisation or address potential issues with the phase tracing process.
Since \WallGo{} does not compute the nucleation temperature,
users likely already rely on an external implementation
that could also provide the positions of the minima as a function of temperature.

The temperatures, minima positions, potential values, and allowed discrepancy
between the provided and internally computed effective potential values
can be passed as a {\tt WallGo.FreeEnergyArrays} object to {\tt Manager.setupThermodynamicsHydrodynamics()}.
If no arrays are passed, the program continues by tracing the phases. 
If arrays are passed but do not have the required shape, or if the difference between
the provided and internal effective potential values exceeds the allowed discrepancy, 
the program throws an error. 

A warning is printed if the temperature steps exceed the estimated maximum step size for accurate interpolation,
\begin{equation}
  \Delta T_{\rm max} =
    \texttt{temperatureVariationScale} \times
    \bigl(\texttt{configThermodynamics.phaseTracerTol}\bigr)^{1/4}
  \,,
\end{equation}
where {\tt temperatureVariationScale} is user-provided and
{\tt configThermodynamics.phaseTracerTol} a configuration parameter.

\subsection{Aliasing in the spectral expansion}

In ref.~\cite{Ekstedt:2024fyq},
we showed the exponential convergence of the spectral expansion for one parameter point in the xSM.
While this is expected for a converging spectral expansion,
the process of iterating the coupled Boltzmann, scalar field, and hydrodynamic equations
can introduce nonlinearities that lead to ``aliasing instabilities''~\cite{boyd2001chebyshev}.

Such instabilities may occur for various reasons,
including unresolved small length scales and insufficient numerical viscosity in hydrodynamic simulations.
In the context of \WallGo{}, we have found aliasing instabilities in a fraction of our example models and parameter points.

To make \WallGo{} less sensitive to aliasing instabilities, we have implemented two optional spectral truncation schemes.
The first follows Orszag's two-thirds rule~\cite{orszag1971elimination}, where the number of nonzero modes is truncated to two-thirds of the maximum after solving the Boltzmann equation, with higher modes padded with zeros.
The second scheme automatically identifies aliasing instabilities
by fitting the exponential slope of the last third of the spectral coefficients.
If this slope is positive, Orszag's two-thirds rule is applied; otherwise, the spectral expansion is not truncated.
In \WallGo{}, the truncation scheme is chosen through \texttt{BoltzmannSolver.truncationOption},
which can be set to \texttt{NONE}, \texttt{THIRD}, or \texttt{AUTO}.
The default is \texttt{AUTO}, and all results presented in this work were obtained with this setting.

\begin{figure}[t]
  \centering
  \includegraphics[width=1.0\textwidth]{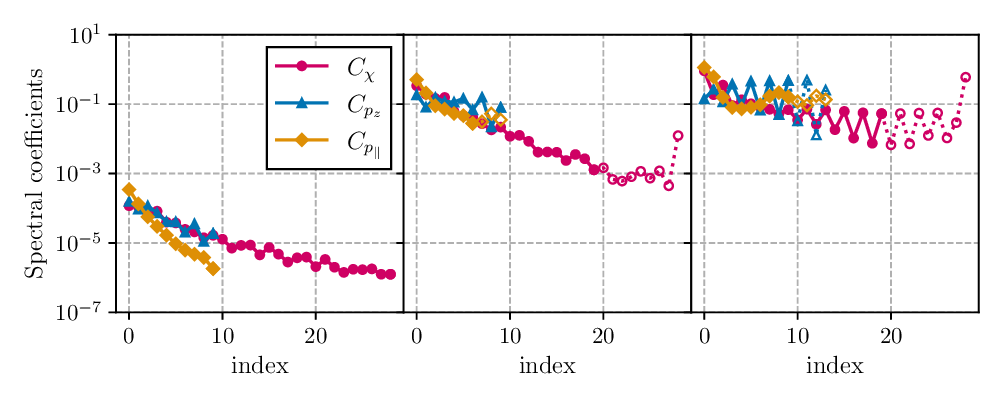}%
    \caption{%
      Spectral convergence on the first iteration at three different model parameter points.
      The quantities plotted are
      $(C_\chi)_i \equiv \sum_{ajk}|\delta f^a_{ijk}|$,
      $(C_{p_z})_j \equiv \sum_{aik}|\delta f^a_{ijk}|$ and
      $(C_{p_\parallel}) \equiv \sum_{aij}|\delta f^a_{ijk}|$ (see~\cite{Ekstedt:2024fyq} for the notation),
      each of which is expected to decrease (exponentially) with the index.
      The plot on the left shows a parameter point in the xSM where the convergence is good.
      The plots in the middle and on the right show a parameter point in the SM with light Higgs,
      treated with {\tt N=11} and {\tt N=15}, respectively.
      The middle plot shows convergence failing at large spectral indices.
      Here the open circles and dotted lines show the points removed with
      \texttt{truncationOption=AUTO} to de-alias the expansion.
      On the right, the spectral expansion appears broken and the truncation is not expected to help significantly.
    }
    \label{fig:spectralConvergence}
\end{figure}
Some examples of the application of the automatic method are shown in fig.~\ref{fig:spectralConvergence}.
On the left-most plot, the spectral expansion is converging well;
the slope of the last 1/3 of the coefficients is negative so the automatic method does not truncate the expansion.
In the middle plot, the last 1/3 of the expansion coefficients are growing in the $\chi$ and $p_\parallel$ directions,
so these are truncated by the automatic method.
The remaining untruncated coefficients appear to converge,
indicating that the truncation has resolved the aliasing instabilities in this case.
In the right-most plot, the spectral expansion suffers from more significant aliasing instabilities,
which are truncated but not resolved by the spectral truncation algorithm.
These are likely due to unresolved scales at this parameter point.

Note that while these spectral truncation schemes may lead to apparent convergence onto smooth solutions,
there is no guarantee that the results are correct.
In cases where aliasing instabilities are present,
further investigation of \texttt{WallGoResults.deltaF} may be required to ensure that the results are physically meaningful.
For further discussion in the context of hydrodynamics,
see Chapter~11 of~\cite{boyd2001chebyshev}.

\subsection{Error estimates}
We have added new estimates for sources of uncertainty and improved existing error estimates.

\subsubsection{Error in the wall velocity}
In \WallGo{} \WallGoOldVersion{}, when the tolerance for the root-finding algorithm was not limiting,
the error estimate for the wall speed was chosen in a relatively ad hoc way:
$|v_\rmii{LTE}-\vw|\delta$, where $\delta$ is the relative error in the particle distribution functions due to spectral truncation.
In \WallGoVersion{}, the error estimate uses eq.~\eqref{eq:vwError} directly,
thereby taking into account the derivative of the pressure with respect to the wall velocity.
Concretely, we now use
\begin{equation}
\texttt{WallGoResults.wallVelocityError} =
\mbox{max}\left(\texttt{errTol} ,\ \left|\Delta P \frac{{\rm d}\vw}{{\rm d}P}\right| \delta \right)
  \,,
\end{equation}
with $\delta$ defined as before and \texttt{errTol} given in {\tt ConfigEOM}.

\subsubsection{Linearisation criterion}
The second-order linearisation criterion has been implemented and is saved as
\begin{equation}
\texttt{WallGoResults.linearizationCriterion2} = \frac{P[\delta f_2]}{P[f_\texttt{WallGo}]} \,.
\end{equation}
As argued in sec.~\ref{sec:Boltzmann:linearisation}, if this is much smaller than 1,
the linearisation approximation should be valid, even if \texttt{WallGoResults.linearizationCriterion1},
the first-order linearisation criterion, is not small.

\subsubsection{Tanh \textit{ansatz} error estimate}
\texttt{WallGo} continues to rely on the tanh \textit{ansatz} to solve the scalar equations of motion.
The residual due to the inexactness of this \textit{ansatz} is computed as discussed in sec.~\ref{sec:EOM:tanhAnsatz}.
In particular,
we have added the following output:
\begin{equation}
  \texttt{WallGoResults.eomResidual} = R_{\rmii{RMS},i} \,,
\end{equation}
where the right-hand side is defined in eq.~\eqref{eq:RMSResidual}.

\subsubsection{Violation of energy-momentum conservation}
When solving for the wall velocity with \WallGo{}, the temperature and fluid profiles
are determined from energy-momentum conservation, see eq.~(18) of~\cite{Laurent:2022jrs}.
As a next step, the Boltzmann equation is solved. Since the temperature and fluid profiles were determined
using the \emph{previous} solution to the Boltzmann equation, the new solution will introduce a small violation of energy-momentum conservation.
In the bubble wall frame, energy-momentum conservation implies that the components
$T^{30}(z)$ and $T^{33}(z)$ are equal to constants
$c_1$ and $c_2$~\cite{Laurent:2022jrs}.
In \WallGo{},
We quantify deviations from this by
\begin{align}
  {\tt WallGoResults.violationOfEMConservation} &=
  \nn[2mm] &
  \hspace{-5cm}
  \left\{
    \frac{1}{L_w}\sqrt{\int \mathrm{d}z\, (T^{30}(z) - c_1)^2}
    \;,
    \frac{1}{L_w}\sqrt{\int \mathrm{d}z\, (T^{33}(z) - c_2)^2} \right\}
    \,,
\end{align}
where $L_w=\text{max}(L_i)$, the maximum wall width amongst the scalar fields.

\subsection{New tests}
We have added the Standard Model with a light Higgs to our benchmark models for testing.
We added a test for the determination of the critical temperature and the end of the broken phase in the phase tracing.
We also added a test for the accuracy of the tanh approximation in the limit of small supercooling, and added module tests for new features.

\subsection{Example file for output to {\tt PTTools}}

We have added an example file, {\tt exampleOutputThermodynamics.py}, to the documentation
demonstrating how \WallGo{} can be used to construct and save the thermodynamic information needed for input to other calculations.
This can be integrated with the software package {\tt PTTools}~\cite{pttools,Maki:2025}, for example,
to compute the gravitational wave spectrum in the Sound Shell Model~\cite{Hindmarsh:2016lnk,Hindmarsh:2019phv}.

\section{New features in \WallGoMatrix{} \WallGoMatrixVersion{}}
\label{sec:wallGoMatrixUpdate}

\subsection{Leading-logarithmic truncation}
\label{sec:leadingLogAlgorithm}

\begin{algorithm}[t]
  \caption{Classification of enhanced structures
  in $ab \to cd$ scattering channels
  with distributions $f^a = f^a_\rmi{eq} + \delta f^a$
  used when {\tt TruncateAtLeadingLog -> True}
  in {\tt ExportMatrixElements}.
  \WallGoMatrix{} adopts this algorithm
  by identifying particles directly instead of distribution functions.}
\label{alg:divergence-structure}
\begin{algorithmic}[0]

\If{particle distributions
  $\delta f^a \equiv \delta f^c$ and
  $\delta f^b \equiv \delta f^d$}
\Comment{Forward channel}
  \If{particles distributions
    $\delta f^a \equiv \delta f^d$}
  \Comment{Fully identical particles}
    \State No power enhancement.
    \State Logarithmic enhancement from $1/t^2$ and $1/u^2$.
    \State Terms proportional to $1/u$ or $1/t$ are finite.
  \Else
    \State Power enhancement from $1/u^2$.
    \State Logarithmic enhancement from $1/t^2$ or $1/u$.
  \EndIf

\ElsIf{particle distributions
  $\delta f^a \equiv \delta f^d$ and
  $\delta f^b \equiv \delta f^c$}
\Comment{Crossed channel}
  \State Power enhancement from $1/t^2$.
  \State Logarithmic enhancement from $1/u^2$ or $1/t$.

\Else
\Comment{Non-identical particles}
  \State Power enhancement from $1/t^2$ and $1/u^2$.
  \State Logarithmic enhancement from $1/t$ or $1/u$.
\EndIf

\end{algorithmic}
\end{algorithm}

In \WallGoMatrix{}~\WallGoMatrixVersion{},
we have implemented algorithm~\ref{alg:divergence-structure},
which identifies identical particles to determine whether a matrix element
is infrared enhanced through a power or logarithmic divergence,
as discussed in sec.~\ref{sec:Boltzmann:enhancements}.
In principle, this algorithm should be applied at the level of distribution functions,
but \WallGoMatrix{} can only identify external particles in a given matrix element.

The option {\tt lightParticleList} allows grouping
particles assumed to be in equilibrium with similar properties,
which is useful in QCD when electroweak processes are neglected.
However, all particles in {\tt lightParticleList} must possess identical
quantum numbers, particularly spin and gauge charges.
Therefore, keeping representations explicit using {\tt ParticleList} is advised.
Similar to \WallGoMatrix~\WallGoMatrixOldVersion~\cite{Ekstedt:2024fyq,Fonseca:2020vke},
the generation of matrix elements through {\tt ExportMatrixElements}
can be truncated at leading-logarithmic order,
but is now based on algorithm~\ref{alg:divergence-structure}.
Optionally, one may also tag logarithmically and power-enhanced contributions
via
\begin{lstlisting}[language=Mathematica]
{TruncateAtLeadingLog>True, TagLeadingLog->False}
\end{lstlisting}
which are the default settings.
With this configuration, both power-enhanced and leading-logarithmic terms are kept,
while tagging of enhanced contributions is disabled by default.

\subsection{Changelog}

The most relevant changes affect
NLL channels that, while possible to generate in
\WallGoMatrix{}~\WallGoMatrixOldVersion,
were not included in our analyses before this article.
We list all the changes and improvements
in the accompanying
\WallGoMatrix{}~\WallGoMatrixVersion{}. Here, particles are labelled as
$V$ for vector bosons,
$F$ for fermions, and
$S$ for scalars.
\begin{itemize}
  \item[(i)]
    Corrected factors $2$ in
    the propagating vector-scalar interference terms of
    $F_1 F_2 \to F_3 F_4$
    that were previously suppressed in eq.~(B.15) of~\cite{Ekstedt:2024fyq}.
  \item[(ii)]
    Corrected the implementation of
    $F_1 F_2 \to V_1 V_2$,
    where, in the presence of multiple propagating vector groups,
    some propagator masses were incorrectly assigned.
  \item[(iii)]
    Corrected a factor in the squared $s$-channel of
    $S_1 S_2 \to V_1 V_2$.
  \item[(iv)]
    Included missing terms in the interference terms between
    $s$-, $t$-, and $u$-channels of $S_1 S_2 \to F_1 F_2$
    that were previously suppressed in eq.~(B.20) of~\cite{Ekstedt:2024fyq}.
\end{itemize}

Additionally, the updated code includes an extensive testing suite 
that validates a variety of models against 
{\tt FeynCalc}~\cite{Shtabovenko:2023idz} implementations in the massless limit. 
The tested models comprise:
\begin{itemize}
  \item 
    the most general renormalisable model of 5 interacting scalar fields,
  \item
    a U(1) gauge sector including a fundamental scalar and fermion with Yukawa coupling,
  \item
    ${\rm SU}(N)$ gauge sectors with $N = 2, 3, 5$, each containing 
    a left-handed fermion in the $\{{\bm N}\}$ representation, 
    a right-handed fermion in $\{{\bm 1}\}$, 
    and a scalar in $\{{\bm N}\}$, coupled through a Yukawa interaction,
  \item
    ${\rm SU}(2)$ and ${\rm SU}(3)_\rmii{$L$}$ gauge sectors with 
    a left-handed fermion in $\{{\bf 3}, {\bf 2}\}$, 
    a right-handed fermion in $\{{\bf 3}, {\bf 1}\}$, 
    and a scalar in $\{{\bf 1}, {\bf 2}\}$ with Yukawa interaction,
  \item
    ${\rm SU}(3)$ and ${\rm SU}(5)_\rmii{$L$}$ gauge sectors with 
    a left-handed fermion in $\{{\bf 5}, {\bf 3}\}$, 
    a right-handed fermion in $\{{\bf 5}, {\bf 1}\}$, 
    and a scalar in $\{{\bf 1}, {\bf 3}\}$ with Yukawa interaction,
  \item
    QCD as implemented in~\cite{Arnold:2003zc}, 
    using the {\tt FeynArts}~\cite{Hahn:2000kx} model file {\tt SMQCD},
  \item
    the Standard Model content using the {\tt FeynArts}~\cite{Hahn:2000kx} model file {\tt SMQCD},
  \item
    the inert doublet model with LL matrix elements from~\cite{Jiang:2022btc},
  \item
    the singlet-extended Standard Model with LL matrix elements from~\cite{Kozaczuk:2015owa}.
\end{itemize}
Here, the corresponding group representations are shown in curly brackets.
All test configurations are available in the
\href{https://github.com/Wall-Go/WallGoMatrix/tree/main/tests}{\tt tests} folder
of the \WallGoMatrix{} repository,
and the associated {\tt FeynCalc} implementations are collected in
\href{https://github.com/Wall-Go/WallGoMatrix/tree/main/tests/FeynCalc}{\tt tests/FeynCalc}.

The model creation in \WallGoMatrix{}~\WallGoMatrixVersion{}
is based on {\tt GroupMath}~{\tt v1.1.3}~\cite{Fonseca:2020vke},
which represents an update from the initial release of
\WallGoMatrix{}~\WallGoMatrixOldVersion{} that supported
{\tt GroupMath}~{\tt v1.1.2}.

\section{New features in \WallGoCollision{} \WallGoCollisionVersion{}}
\label{sec:wallGoCollisionUpdate}
\subsection{Computing collision terms for specific particle pair only}
In some cases, one might not want to compute the full set of collision integrals, but only compute the collision integrals for one specific particle pair.
This possibility was already included in {\tt v1.0.0} of \WallGoCollision{}.
In \WallGoCollisionVersion{}, we have added {\tt Python} bindings,
such that this functionality can be used within \WallGo{}.
For illustration, we have added a new example file (available at {\tt \WallGoUrl{}})
for the Yukawa model.

\section{Boltzmann equation for an ideal gas with superluminal propagation}\label{app:BoltzmannEquationSuperluminal}

To gain intuition about the superluminal propagation in the Boltzmann equation,
we derive a Boltzmann equation for an ideal gas of wave packets
in a classical field theory with a phase separation.
Even though the underlying description is causal,
the Boltzmann equation will nonetheless permit superluminal propagation.
This allows us to
pinpoint the origin of the seemingly non-causal particle propagation,
understand that such propagation might be allowed for
the particles described by their Boltzmann equation.
Moreover, the classical picture highlights the separation
of scales between the instability of the background wall and the Boltzmann particles, $T\gg L_\text{wall}^{-1}$ 
discussed in sec.~\ref{sec:negativeMassSquared}.

To derive the Boltzmann equation, we will present a field theory with phase separation and study fluctuations
on the background of a kink solution interpolating between the phases.
We will utilise the Wentzel–Kramers–Brillouin (WKB) approximation to find an approximate solution
for the fluctuations.
This approximation relies on the aforementioned separation of scales. We are also able to
confirm that the WKB approximation is not the source of superluminal propagation.
Then, we will consider the special case of a wave packet with a definite momentum
and location. This corresponds to the Boltzmann
particles in quantum field theories with definite locations and momenta.
We show that the momentum and location of the wave packet form
a closed system of equations. Consequently, we can describe the wave packets effectively
as particles with locations and momenta.
Finally, we can obtain the Boltzmann equation for the ideal gas of classical wave
packets within the effective particle description. The Boltzmann equation has
exactly the same form as in eq.~\eqref{eq:Boltzmann:outofeq} from QFTs.

We consider a toy model that contains a real scalar field in
$D=1+1$ dimensions described by
\begin{align}
  \partial_t^2\phi-\partial_x^2\phi+V'(\phi)&=0
    \,, &
    \text{with} \quad
    V(\phi)&=\frac{\kappa}{4}(\phi^2-v^2)^2
    \,.
\end{align}
The potential has two minima at $\phi=\pm v$ and a maximum at $\phi=0$.
Between the minima, the background field is taken to be the kink solution
\begin{equation}
    \phi_\text{b}(x)=v\tanh\sqrt{\frac{\kappa}{2}}v x
    \,,
\end{equation}
which is analogous to a bubble wall,
but unlike the dynamic bubble wall, the kink is a stable, static solution.
Inside the kink solution,
the second derivative of the potential, $V''(\phi)$, becomes negative,
giving rise to a region with negative squared mass.

Let us look at the fluctuations that we will use to construct wave packets, $\delta\phi=\phi-\phi_\text{b}$.%
\footnote{
  We will also use these fluctuations to eventually derive a Boltzmann equation. The procedure
  looks quite different from the one in out-of-equlibrium QFT, where one usually uses two-point
  functions. The reason for being able to use just the field variable
  is that we can handle fluctuations directly in classical field theory, whereas the 
  usual QFT derivation relies on averaged-out quantities. After this averaging process, the information of the 
  particle fluctuations is in the two-point function, rather than the one-point function.
}
As long as these fluctuations are small,
they are described by the linearised equation of motion,
\begin{align}
\label{eq:diffEquationForTheFluctuations}
    \big(\partial_t^2-\partial_x^2+U(x)\big)\delta\phi
    & = 0
    \,, \\[1mm]
    \big(-\partial_x^2+U(x)\big)\widetilde{\delta\phi}
    & = \omega^2 \widetilde{\delta\phi}
    \,,
\end{align}
where we used
a temporal Fourier transform in the second line and
$U(x) \equiv V''(\phi_\text{b}(x))$.

We emphasise the causal structure of
the linear differential equation~\eqref{eq:diffEquationForTheFluctuations}
for the perturbations $\delta \phi$ that dictates that information can only propagate within light cones.
To that end, we use a deconstruction into first-order differential equations.
The differential equation~\eqref{eq:diffEquationForTheFluctuations}
has a unique solution if we set the initial conditions on a spatial hypersuface for $\delta\phi$ and $\partial_t \delta\phi$. In this case, we can deconstruct the original, hyperbolic equation into a set of three first-order differential equations for three modes,
\begin{align}
\label{eq:ODE:fluct:Boltzmann}
    \partial_t \Phi_0&=\frac{1}{2}(\Phi_++\Phi_-)
    \,,&
    (\partial_t+\partial_x) \Phi_+&=-U(x)\Phi_0
    \,,&
    (\partial_t-\partial_x) \Phi_-&=-U(x)\Phi_0
    \,,
\end{align}
with initial conditions
\begin{align}
\label{eq:ODE:fluct:Boltzmann:initial}
  \Phi_0&=\delta\phi\,,&
  \Phi_+&=(\partial_t-\partial_x)\delta\phi\,,&
  \Phi_-&=(\partial_t+\partial_x)\delta\phi\,,
\end{align}
on the hypersurface.
The relations given on the hypersurface, eq.~\eqref{eq:ODE:fluct:Boltzmann:initial},
will hold for the entire spacetime due to the first-order differential equations,
eq.~\eqref{eq:ODE:fluct:Boltzmann}.
This can be seen from the first equation and by subtracting and adding
the last two equations of eq.~\eqref{eq:ODE:fluct:Boltzmann}
\begin{align}
\label{eq:ODE:fluct:Boltzmann:combine}
  \partial_t \Phi_0&=\frac{\Phi_{+}+\Phi_{-}}{2}
    \,,&&
    \Big(\partial_t\delta\phi=\partial_t\delta\phi\Big)
    \,,\\
    \partial_t \frac{\Phi_{-}-\Phi_{+}}{2}&=\partial_x \frac{\Phi_{+}+\Phi_{-}}{2}
    \,, &&
    \Big(\partial_t\partial_x\delta\phi=\partial_x\partial_t\delta\phi\Big)
    \,,
    \\
    \partial_t \frac{\Phi_{+}+\Phi_{-}}{2}&=\partial_x \frac{\Phi_{-}-\Phi_{+}}{2} -U(x)\Phi_0
    \,, &&\Big(\partial_t^2\delta\phi=\partial_x^2\delta\phi-U(x)\delta\phi\Big)
    \,.
    \label{eq:ODE:fluct:Boltzmann:lastForReference}
\end{align}
All of the first-order equations above, eqs.~\eqref{eq:ODE:fluct:Boltzmann:combine}--\eqref{eq:ODE:fluct:Boltzmann:lastForReference}, correspond to true statements regarding the time evolution of
the field fluctuations, $\delta\phi$: The equations govern the time evolution of $\delta\phi$,
$\partial_x\phi$, and $\partial_t\delta\phi$ respectively, as indicated in the brackets.
Consequently, the solution to the original second-order differential equation
can be recovered from $\phi(t,x) = \Phi_0(t,x)$.
This deconstruction reveals the causal structure of the theory.
Two propagating modes, $\Phi_+$ and $\Phi_-$, travel at the speed of light, while
a third mode, $\Phi_0$, is non-propagating.
Consequently, information cannot travel faster than light.
Equivalently, modes at a space-time point are only connected through the derivatives
to neighbouring points along the light cones, $\partial_t\pm\partial_x$, and
the temporal direction, $\partial_t$.

We now turn to the dynamics of wave packets using
the WKB approximation.
This approach yields explicit approximate formulae for the wave packet velocities and
the forces acting on them.
A nice feature of this model is that the spectrum,
the set of eigenmodes, $f_i$, and their eigenvalues, $\omega_i^2$, in
\begin{equation}
    \bigl(-\partial_x^2+U(x)\bigr)f_i=\omega_i^2f_i
    \,,
\end{equation}
are also known analytically~\cite{Vachaspati:2006zz}.
Hence, we can confirm that the WKB approximation is not the reason
for the non-causal nature of the Boltzmann equation by numerically constructing the wave packets and their dynamics.

The spectrum can be summarised as follows~\cite{Vachaspati:2006zz}: there are no modes with $\omega^2 < 0$,
indicating the absence of instabilities.
The lowest eigenmode corresponds to the translational mode of the kink and thus
has $\omega^2 = 0$.
Another discrete mode represents a deformation of the kink,
while the remaining modes form a continuum of scattering states with $\omega^2 > 2\kappa v^2$.
The WKB approximation applies in the regime
$\omega^2 \gg 2\kappa v^2 \geq \abs{U(x)}$,
precisely where the modes represent well-defined, stable excitations rather than
wall fluctuations, which are potentially unstable in supercooled phase transitions.
The modes captured by the WKB approximation are the scattering states mentioned above.
Notice that $\omega^2 \gg 2\kappa v^2 \geq \abs{U(x)}$ ensures that the scattering states live 
on a shorter length scale and on a higher energy scale
than the background wall. %

The solution to
the time-dependent equation in the WKB approximation with modes propagating to the positive $x'$ direction
takes the form
\begin{equation}
\label{eq:WKBresultToTheRight}
    \delta\phi(t,x')\approx
    \int_{-\infty}^\infty
    {\rm d}\omega'\, A(\omega')
    \exp\biggl\{i\biggl(-\omega' t+\omega' x'-\frac{\int_{-\infty}^{x'} {\rm d} x''\, U(x'')}{2\omega'}\biggr)\biggr\}
    \,.
\end{equation}
For the approximation to be valid,
the spectrum $A(\omega')$ should not contain low frequencies,
$(\omega')^2 \lesssim \kappa v^2$.
Moreover, reality of the field requires $A^*(-\omega') = A(\omega')$.
Note that from now on we will notationally reserve $x$ for the eventual location of
the wave packet. Hence, we are using $x'$ for coordinate.

A particle in the Boltzmann equation has a definite location
and a definite momentum.
We can translate these into requirements for the wave packet,
and then find the corresponding Boltzmann equation for these wave packets.
The definite (time-dependent) location of a wave packet, $x(t)$,
requires that the uncertainty in the location, $\sigma_x$, is much smaller than the length scale, $L_\rmi{w}$,
of the background variations,
\begin{equation}
\label{eq:wavePacket:location:approx}
  \sigma_x \sim \sigma_\omega^{-1}\ll L_\rmi{w}^{ }\sim (\sqrt{\kappa}v)^{-1}
  \,,
\end{equation}
where $\sigma_\omega$ is the width of the frequency spectrum.%
\footnote{%
  The $\sigma_x \sim \sigma_\omega^{-1}$ relation corresponds to the minimal uncertainty in location given
  a frequency spectrum, analogous to the uncertainty principle of a wave function in one-particle
  quantum mechanics, $\sigma_x \sigma_p\geq\frac{1}{2}$.
  The former relation will eventually break down, becoming $\sigma_x \gg \sigma_\omega^{-1}$ during the time evolution of
  the wave packets due to dispersion.
}
With the assumption~\eqref{eq:wavePacket:location:approx},
we can expand the exponent in $\Delta x\equiv x'-x(t)$,
and obtain
\begin{equation}\label{eq:wavepacketExpandx}
    \delta\phi(t,x') \approx
    \int_{-\infty}^\infty
    \dd\omega'\, A(\omega')
    \exp\biggl\{i\biggl[-\omega' t+\omega'x(t)-\frac{\int_{-\infty}^{x(t)} \dd x''\, U(x'')}{2\omega'}
    + \biggl(\omega' -\frac{U(x)}{2\omega'}\biggr)\Delta x\biggr]\biggr\}
    \,.
\end{equation}
The definite momentum of the wave packet requires that the peak frequency of the wave packet, $\omega$, is much larger than
the width of the spectrum, $\omega\gg\sigma_\omega$.
This introduces an additional hierarchy beyond the WKB condition
$\omega^2 \gg \kappa v^2$, namely
$\omega^2 \gg \sigma_\omega^2 \gg \kappa v^2$, due to the width of the wave packet in the $x'$ direction.
We can already find the definite wave number of the wave packet
\begin{equation}\label{eq:approximatek}
  k=\omega -\frac{U(x)}{2\omega}\quad \bigl(=p\bigr)
  \,,
\end{equation}
from the spatial dependence of the wave packet through $\Delta x$.
In the language of particles, this corresponds to the one-dimensional momentum $p$, shown in brackets.

The velocity of the wave packet is not yet manifest from eq.~(\ref{eq:wavepacketExpandx}),
but we will be able to extract the velocity from the collective motion of the envelope of the wave packet.
To achieve this,
we expand the exponent in $\Delta\omega \equiv \omega' - \omega$,
taking advantage of the sharply peaked spectrum:
\begin{align}\label{eq:fullyMassagedWavePacket}
    \delta\phi(t,x')&\approx
    \exp\Bigl[i\bigl(-\omega t+\omega x(t)-\frac{\int_{-\infty}^{x(t)} \dd x''\, U(x'')}{2\omega}+k\Delta x\bigr)\Bigr]\\
    &\quad\times
    \int_{-\infty}^\infty
    \dd\omega'\,A(\omega')
    \exp\biggl\{i\Delta\omega\biggl(-t+x(t)+\frac{\int_{-\infty}^{x(t)} \dd x''\, U(x'')}{2\omega^2}+\pdv{k}{\omega}\Delta x\biggr)\biggr\}
    \,.
\end{align}
The magnitude of the self-similarly propagating wave packet only depends
on $\Delta x$, i.e.\ the distance from the centre at any given time:
\begin{align}
\label{eq:selfSimilarWaveEnvelope}
    \abs{\delta\phi}(\Delta x)&=\abs{
      \int_{-\infty}^\infty
      \dd\omega'\,A(\omega')
      \exp\biggl\{i\Delta\omega\biggl(-t+x(t)+\frac{\int_{-\infty}^{x(t)} \dd x''\, U(x'')}{2\omega^2}+\pdv{k}{\omega}\Delta x\biggr)\biggr\}}\,.
\end{align}
Since the magnitude depends only on $\Delta x$, the time derivative  in the direction
at which $\Delta x$ stays constant,
$\dv{\Delta x}{t}=0$, is zero:
$\dv{ }{t}\abs{\delta\phi}(\Delta x)=0$.
We can enforce this with the argument in the exponent:
\begin{align}
  0&=\dv{}{t}\biggl(-t+x(t)+\frac{\int_{-\infty}^{x(t)} \dd x''\, U(x'')}{2\omega^2}+\pdv{k}{\omega}\Delta x\biggr)\\
  &=-1+\biggl(1+\frac{U(x(t))}{2\omega^2}+\frac{U'(x(t))}{2\omega^2}\Delta x\biggr)\dv{x}{t}\,.
\end{align}
The last term on the last line inside the parenthesis is smaller due to the hierarchy in length scales.
Thus, we can identify the velocity of the wave packet:
\begin{align}\label{eq:velocityOfWavePacket}
    v&=\dv{x}{t}\approx1-\frac{U(x)}{2\omega^2} 
    \quad\biggl(= 
    1-\frac{m^2}{2E^2}\approx\frac{p}{\sqrt{p^2+m^2}}\biggr)
    \,.
\end{align}
In the parenthesis,
we have identified the wave quantities with the particle ones:
$m^2=U(x)=V''(\phi_\text{b}(x))$ and
$E=\omega$.

The wave packet velocity indeed becomes superluminal when $U(x) < 0$.
We have numerically confirmed that the formula accurately estimates
the propagation of the wave packet if $\omega\gg\sigma_\omega\gg \kappa v^2$ using both
the WKB formula in eq.~\eqref{eq:WKBresultToTheRight} and
the exact solution to the time-dependent equation using the exact spectrum.
Importantly, this superluminal propagation does not violate causality,
as the underlying time-dependent equation enforces information propagation at
or below the speed of light.
The superluminal propagation is actually just an illusion.
Since the wave packet is a global structure stretching out to spatial infinities,
the information of an incoming wave packet is actually already in front of it to begin with ---
outside the light cones centered at $x(t)$.
The centre point moves
superluminally following the group velocity, but
the full information of the wave packet is not carried by the centre point.
This creates the false impression of superluminal motion, without violating causality.

Let us finally complete the journey to the Boltzmann equation. We will realise that 
the location and momentum of the wave packet form a closed system of equations, and
can thus be described separately.
We can also show that the wave packets can be
superposed due to the linearity of the time evolution of the fluctuations
to form an ideal gas of wave packets. Finally, we will construct a one-particle
distribution function and show that it is governed by the same Boltzmann equation
as the particles in the corresponding QFT.

The force term acting on the wave packet takes the same form as with a particle,
namely
\begin{align}\label{eq:forceOnWavePacket}
    F&=\dv{k}{t}=
      \dv{x}{t}\pdv{k}{x}
    + \overbrace{\dv{\omega}{t}}^{=0}\pdv{k}{\omega}
    \approx
    -\frac{U'(x)}{2\omega}
    \quad\biggl(=-\frac{\partial_x m^2}{E}=-\partial_x E\biggr)
    \,.
\end{align}
The corresponding particle quantities are again presented in the parenthesis.

Now, note that the time-evolution of the location, $x$,
and momentum, $k$, only depend on the location and momentum themselves:
\begin{align}
  \dv{x}{t}&\approx1-\frac{U(x)}{2k^2}
  \,, &
  \dv{k}{t}&\approx-\frac{U'(x)}{2k}\,.
\end{align}
Here, we have used the approximate relations in eqs.~\eqref{eq:approximatek},
\eqref{eq:velocityOfWavePacket}, \eqref{eq:forceOnWavePacket}.
Due to being a closed system of equations, we can describe the location and momentum
independently of the internal degrees of freedom in the wave packet. The location and momentum are just the
degrees of freedom of a point particle. Hence, we have derived an effective description
for the wave packet that corresponds to a particle.

So far, we have only looked at a single wave packet, $\delta\phi_\text{wp}(t,x,k)$,
where we have now suppressed the internal degrees of freedom
(i.e.\ the complete spectrum $\Bar{A}(\omega')$ in eq.~\eqref{eq:fullyMassagedWavePacket}).
Forming the gas is very straightforward due to the linearity of the equation of motion
for the fluctuations in eq.~\eqref{eq:diffEquationForTheFluctuations}:
\begin{equation}
  \delta\phi_\text{gas}=\sum_{n=1}^N \delta\phi_{\text{wp},n}(t,x_n,k_n)\,.
\end{equation}
Notice that all of the locations and momenta evolve independently of the other
wave-packet particles due to the linearity. Hence, it is an ideal gas of effective
particles.

As the locations and momenta of the wave-packet particles can be described separately,
we can describe the ideal gas with a Boltzmann equation~\cite{LIFSHITZ19811} that is collisionless due to the non-interacting nature of the wave packets in the
linear approximation.
Here, we give a brief sketch of the derivation for a one-particle density function, 
$f(x,k)$. If the wave packets have certain initial locations, $x_n(0), k_n(0)$,
the density function takes the form
\begin{equation} \label{eq:wave_packet_ensemble}
  f(t, x, k) = \sum_{n=1}^N \delta(x - x_n(t)) \delta(k- k_n(t))
  \,.
\end{equation}
The time evolution of the probability distribution comes from the time evolution
of the locations and momenta of the wave packets, $x_n(t), k_n(t)$,
which can be solved from eqs.~\eqref{eq:velocityOfWavePacket}, \eqref{eq:forceOnWavePacket}.
We can obtain our Boltzmann equation by taking the time derivative of the probability distribution
\begin{align}
  \partial_t f(t, x, k) &= -\sum_n \left[\dot{x}_n(t)\partial_x \delta(x - x_n(t)) \delta(k- k_n(t)) + \delta(x - x_n(t)) \dot{k}_n(t)\cdot \partial_k \delta(k- k_n(t))\right] \,, \\
  &= -\left(v\, \partial_x + F\,\partial_k \right) f(t, x, k)
  \,.
\end{align}
Thus, the positions and momenta of the wave packets obey a Boltzmann equation with the same Liouville operator as
the particles:
\begin{equation}
  \biggl[
        \omega\partial_t
      + \biggl(k-\frac{U(x)}{2k}\biggr)\partial_{x}
      - \frac{1}{2}U'(x)\,\partial_{k}
    \biggr] f(t, x, k)=0
    \,.
\end{equation}

{\small
\bibliographystyle{utphys}
\bibliography{ref.bib}

@article{Ekstedt:2024fyq,
    author = "Ekstedt, Andreas and Gould, Oliver and Hirvonen, Joonas and Laurent, Benoit and Niemi, Lauri and Schicho, Philipp and van de Vis, Jorinde",
    title = "{How fast does the WallGo? A package for computing wall velocities in first-order phase transitions}",
    eprint = "2411.04970",
    archivePrefix = "arXiv",
    primaryClass = "hep-ph",
    reportNumber = "CERN-TH-2024-174, DESY-24-162, HIP-2024-21/TH",
    doi = "10.1007/JHEP04(2025)101",
    journal = "JHEP",
    volume = "04",
    pages = "101",
    year = "2025"
}

@misc{pttools,
  author       = {Hindmarsh, Mark and Mäki, Mika and Al-Ajmi, Mudhahir and Bail, Danny and Cutting, Daniel and Gowling, Chloe and Hijazi, Mulham and Lindsay, Jacky and Soughton, Mike},
  title        = {{\tt PTtools}},
  year         = {2015--2024},
  url          = {https://github.com/CFT-HY/pttools},
  note         = {Available at \url{https://github.com/CFT-HY/pttools}},
  language     = {English}
}

@phdthesis{Maki:2025,
  author       = {Mika Mäki},
  title        = {The effect of sound speed on the gravitational wave spectrum of first order phase transitions in the early universe},
  school       = {University of Helsinki},
  year         = {2025},
  month        = {January},
  url          = {http://hdl.handle.net/10138/591514},
  language     = {English}
}

@article{Blinov:2015vma,
    author = "Blinov, Nikita and Profumo, Stefano and Stefaniak, Tim",
    title = "{The Electroweak Phase Transition in the Inert Doublet Model}",
    eprint = "1504.05949",
    archivePrefix = "arXiv",
    primaryClass = "hep-ph",
    doi = "10.1088/1475-7516/2015/07/028",
    journal = "JCAP",
    volume = "07",
    pages = "028",
    year = "2015"
}

@article{Arnold:2002ja,
    author = "Arnold, Peter Brockway and Moore, Guy D. and Yaffe, Laurence G.",
    title = "{Photon and gluon emission in relativistic plasmas}",
    eprint = "hep-ph/0204343",
    archivePrefix = "arXiv",
    reportNumber = "UW-PT-02-06",
    doi = "10.1088/1126-6708/2002/06/030",
    journal = "JHEP",
    volume = "06",
    pages = "030",
    year = "2002"
}

@article{Bodeker:2009qy,
    author = "Bodeker, Dietrich and Moore, Guy D.",
    title = "{Can electroweak bubble walls run away?}",
    eprint = "0903.4099",
    archivePrefix = "arXiv",
    primaryClass = "hep-ph",
    doi = "10.1088/1475-7516/2009/05/009",
    journal = "JCAP",
    volume = "05",
    pages = "009",
    year = "2009"
}

@article{Enqvist:1991xw,
    author = "Enqvist, K. and Ignatius, J. and Kajantie, K. and Rummukainen, K.",
    title = "{Nucleation and bubble growth in a first order cosmological electroweak phase transition}",
    reportNumber = "HU-TFT-91-35",
    doi = "10.1103/PhysRevD.45.3415",
    journal = "Phys. Rev. D",
    volume = "45",
    pages = "3415--3428",
    year = "1992"
}

@article{DeCurtis:2024hvh,
    author = "De Curtis, Stefania and Delle Rose, Luigi and Guiggiani, Andrea and Gil Muyor, \'Angel and Panico, Giuliano",
    title = "{Non-linearities in cosmological bubble wall dynamics}",
    eprint = "2401.13522",
    archivePrefix = "arXiv",
    primaryClass = "hep-ph",
    doi = "10.1007/JHEP05(2024)009",
    journal = "JHEP",
    volume = "05",
    pages = "009",
    year = "2024"
}

@article{Wainwright:2011kj,
    author = "Wainwright, Carroll L.",
    title = "{CosmoTransitions: Computing Cosmological Phase Transition Temperatures and Bubble Profiles with Multiple Fields}",
    eprint = "1109.4189",
    archivePrefix = "arXiv",
    primaryClass = "hep-ph",
    doi = "10.1016/j.cpc.2012.04.004",
    journal = "Comput. Phys. Commun.",
    volume = "183",
    pages = "2006--2013",
    year = "2012"
}

@article{Athron:2020sbe,
    author = "Athron, Peter and Bal\'azs, Csaba and Fowlie, Andrew and Zhang, Yang",
    title = "{PhaseTracer: tracing cosmological phases and calculating transition properties}",
    eprint = "2003.02859",
    archivePrefix = "arXiv",
    primaryClass = "hep-ph",
    reportNumber = "CoEPP-MN-20-3",
    doi = "10.1140/epjc/s10052-020-8035-2",
    journal = "Eur. Phys. J. C",
    volume = "80",
    number = "6",
    pages = "567",
    year = "2020"
}

@article{Athron:2024xrh,
    author = "Athron, Peter and Bal\'azs, Csaba and Fowlie, Andrew and Morris, Lachlan and Searle, William and Xiao, Yang and Zhang, Yang",
    title = "{PhaseTracer2: from the effective potential to gravitational waves}",
    eprint = "2412.04881",
    archivePrefix = "arXiv",
    primaryClass = "astro-ph.CO",
    doi = "10.1140/epjc/s10052-025-14258-y",
    journal = "Eur. Phys. J. C",
    volume = "85",
    number = "5",
    pages = "559",
    year = "2025"
}

@article{Basler:2024aaf,
    author = {Basler, Philipp and Biermann, Lisa and M{\"u}hlleitner, Margarete and M{\"u}ller, Jonas and Santos, Rui and Viana, Jo{\~a}o},
    title = "{BSMPT v3 a tool for phase transitions and primordial gravitational waves in extended Higgs sectors}",
    eprint = "2404.19037",
    archivePrefix = "arXiv",
    primaryClass = "hep-ph",
    reportNumber = "KA-TP-08-2024",
    doi = "10.1016/j.cpc.2025.109766",
    journal = "Comput. Phys. Commun.",
    volume = "316",
    pages = "109766",
    year = "2025"
}

@article{Eriksson:2025owh,
    author = "Eriksson, M. and Laine, M.",
    title = "{Entropy production at electroweak bubble walls from scalar field fluctuations}",
    eprint = "2507.07755",
    archivePrefix = "arXiv",
    primaryClass = "hep-ph",
    doi = "10.1088/1475-7516/2025/09/027",
    journal = "JCAP",
    volume = "09",
    pages = "027",
    year = "2025"
}

@article{Camargo-Molina:2013qva,
    author = "Camargo-Molina, J. E. and O'Leary, B. and Porod, W. and Staub, F.",
    title = "{$\mathbf{Vevacious}$: A Tool For Finding The Global Minima Of One-Loop Effective Potentials With Many Scalars}",
    eprint = "1307.1477",
    archivePrefix = "arXiv",
    primaryClass = "hep-ph",
    doi = "10.1140/epjc/s10052-013-2588-2",
    journal = "Eur. Phys. J. C",
    volume = "73",
    number = "10",
    pages = "2588",
    year = "2013"
}

@article{orszag1971elimination,
  title="{On the elimination of aliasing in finite-difference schemes by filtering high-wavenumber components}",
  author="Orszag, Steven A",
  journal={Journal of Atmospheric Sciences},
  volume={28},
  number={6},
  pages={1074--1074},
  year={1971}
}

@article{Arnold:2000dr,
    author = "Arnold, Peter Brockway and Moore, Guy D. and Yaffe, Laurence G.",
    title = "{Transport coefficients in high temperature gauge theories. 1. Leading log results}",
    eprint = "hep-ph/0010177",
    archivePrefix = "arXiv",
    reportNumber = "UW-PT-00-15",
    doi = "10.1088/1126-6708/2000/11/001",
    journal = "JHEP",
    volume = "11",
    pages = "001",
    year = "2000"
}

@article{Arnold:2003zc,
    author = "Arnold, Peter Brockway and Moore, Guy D and Yaffe, Laurence G.",
    title = "{Transport coefficients in high temperature gauge theories. 2. Beyond leading log}",
    eprint = "hep-ph/0302165",
    archivePrefix = "arXiv",
    doi = "10.1088/1126-6708/2003/05/051",
    journal = "JHEP",
    volume = "05",
    pages = "051",
    year = "2003"
}

@article{Arnold:2001ms,
    author = "Arnold, Peter Brockway and Moore, Guy D. and Yaffe, Laurence G.",
    title = "{Photon emission from quark gluon plasma: Complete leading order results}",
    eprint = "hep-ph/0111107",
    archivePrefix = "arXiv",
    reportNumber = "UW-PT-01-22",
    doi = "10.1088/1126-6708/2001/12/009",
    journal = "JHEP",
    volume = "12",
    pages = "009",
    year = "2001"
}

@article{Arnold:2002zm,
    author = "Arnold, Peter Brockway and Moore, Guy D. and Yaffe, Laurence G.",
    title = "{Effective kinetic theory for high temperature gauge theories}",
    eprint = "hep-ph/0209353",
    archivePrefix = "arXiv",
    doi = "10.1088/1126-6708/2003/01/030",
    journal = "JHEP",
    volume = "01",
    pages = "030",
    year = "2003"
}

@article{Jiang:2022btc,
    author = "Jiang, Siyu and Huang, Fa Peng and Wang, Xiao",
    title = "{Bubble wall velocity during electroweak phase transition in the inert doublet model}",
    eprint = "2211.13142",
    archivePrefix = "arXiv",
    primaryClass = "hep-ph",
    doi = "10.1103/PhysRevD.107.095005",
    journal = "Phys. Rev. D",
    volume = "107",
    number = "9",
    pages = "095005",
    year = "2023"
}

@article{Kainulainen:2019kyp,
    author = "Kainulainen, Kimmo and Keus, Venus and Niemi, Lauri and Rummukainen, Kari and Tenkanen, Tuomas V. I. and Vaskonen, Ville",
    title = "{On the validity of perturbative studies of the electroweak phase transition in the Two Higgs Doublet model}",
    eprint = "1904.01329",
    archivePrefix = "arXiv",
    primaryClass = "hep-ph",
    doi = "10.1007/JHEP06(2019)075",
    journal = "JHEP",
    volume = "06",
    pages = "075",
    year = "2019"
}

@article{Goudelis:2013uca,
    author = "Goudelis, A. and Herrmann, B. and St\r{a}l, O.",
    title = "{Dark matter in the Inert Doublet Model after the discovery of a Higgs-like boson at the LHC}",
    eprint = "1303.3010",
    archivePrefix = "arXiv",
    primaryClass = "hep-ph",
    reportNumber = "LAPTH-006-13",
    doi = "10.1007/JHEP09(2013)106",
    journal = "JHEP",
    volume = "09",
    pages = "106",
    year = "2013"
}

@article{Guada:2020xnz,
    author = "Guada, Victor and Nemev\v{s}ek, Miha and Pintar, Matev\v{z}",
    title = "{FindBounce: Package for multi-field bounce actions}",
    eprint = "2002.00881",
    archivePrefix = "arXiv",
    primaryClass = "hep-ph",
    doi = "10.1016/j.cpc.2020.107480",
    journal = "Comput. Phys. Commun.",
    volume = "256",
    pages = "107480",
    year = "2020"
}

@article{Caprini:2019egz,
    author = "Caprini, Chiara and others",
    title = "{Detecting gravitational waves from cosmological phase transitions with LISA: an update}",
    eprint = "1910.13125",
    archivePrefix = "arXiv",
    primaryClass = "astro-ph.CO",
    reportNumber = "DESY-19-159, IPPP/19/27, HIP-2019-14/TH, MITP/19-066, IFT-UAM/CSIC-19-139",
    doi = "10.1088/1475-7516/2020/03/024",
    journal = "JCAP",
    volume = "03",
    pages = "024",
    year = "2020"
}

@article{Wang:2024wcs,
    author = "Wang, Dian-Wei and Yan, Qi-Shu and Huang, Mei",
    title = "{Bubble wall velocity and gravitational wave in the minimal left-right symmetric model}",
    eprint = "2405.01949",
    archivePrefix = "arXiv",
    primaryClass = "gr-qc",
    doi = "10.1103/PhysRevD.110.076011",
    journal = "Phys. Rev. D",
    volume = "110",
    number = "7",
    pages = "076011",
    year = "2024"
}

@article{BarrosoMancha:2020fay,
    author = "Barroso Mancha, Marc and Prokopec, Tomislav and Swiezewska, Bogumila",
    title = "{Field-theoretic derivation of bubble-wall force}",
    eprint = "2005.10875",
    archivePrefix = "arXiv",
    primaryClass = "hep-th",
    doi = "10.1007/JHEP01(2021)070",
    journal = "JHEP",
    volume = "01",
    pages = "070",
    year = "2021"
}

@article{Liu:1992tn,
    author = "Liu, Bao-Hua and McLerran, Larry D. and Turok, Neil",
    title = "{Bubble nucleation and growth at a baryon number producing electroweak phase transition}",
    reportNumber = "TPI-MINN-92-18-T",
    doi = "10.1103/PhysRevD.46.2668",
    journal = "Phys. Rev. D",
    volume = "46",
    pages = "2668--2688",
    year = "1992"
}

@article{Ai:2024btx,
    author = "Ai, Wen-Yuan and Laurent, Benoit and van de Vis, Jorinde",
    title = "{Bounds on the bubble wall velocity}",
    eprint = "2411.13641",
    archivePrefix = "arXiv",
    primaryClass = "hep-ph",
    reportNumber = "CERN-TH-2024-198, KCL-PH-TH/2024-57",
    doi = "10.1007/JHEP02(2025)119",
    journal = "JHEP",
    volume = "02",
    pages = "119",
    year = "2025"
}

@article{Laurent:2022jrs,
    author = "Laurent, Benoit and Cline, James M.",
    title = "{First principles determination of bubble wall velocity}",
    eprint = "2204.13120",
    archivePrefix = "arXiv",
    primaryClass = "hep-ph",
    doi = "10.1103/PhysRevD.106.023501",
    journal = "Phys. Rev. D",
    volume = "106",
    number = "2",
    pages = "023501",
    year = "2022"
}

@book{boyd2001chebyshev,
  title={Chebyshev and Fourier spectral methods},
  author={Boyd, John P},
  year={2001},
  publisher={Courier Corporation}
}

@article{Ekstedt:2022bff,
    author = "Ekstedt, Andreas and Schicho, Philipp and Tenkanen, Tuomas V. I.",
    title = "{DRalgo: A package for effective field theory approach for thermal phase transitions}",
    eprint = "2205.08815",
    archivePrefix = "arXiv",
    primaryClass = "hep-ph",
    reportNumber = "HIP-2022-11/TH, NORDITA 2022-030",
    doi = "10.1016/j.cpc.2023.108725",
    journal = "Comput. Phys. Commun.",
    volume = "288",
    pages = "108725",
    year = "2023"
}

@article{Dashko:2024anp,
    author = "Dashko, Andrii and Ekstedt, Andreas",
    title = "{Bubble-wall speed with loop corrections}",
    eprint = "2411.05075",
    archivePrefix = "arXiv",
    primaryClass = "hep-ph",
    reportNumber = "DESY-24-169",
    doi = "10.1007/JHEP03(2025)024",
    journal = "JHEP",
    volume = "03",
    pages = "024",
    year = "2025"
}

@article{Gouttenoire:2023roe,
    author = "Gouttenoire, Yann and Kuflik, Eric and Liu, Di",
    title = "{Heavy baryon dark matter from SU(N) confinement: Bubble wall velocity and boundary effects}",
    eprint = "2311.00029",
    archivePrefix = "arXiv",
    primaryClass = "hep-ph",
    reportNumber = "LAPTH-055/23",
    doi = "10.1103/PhysRevD.109.035002",
    journal = "Phys. Rev. D",
    volume = "109",
    number = "3",
    pages = "035002",
    year = "2024"
}

@article{Moore:1995si,
    author = "Moore, Guy D. and Prokopec, Tomislav",
    title = "{How fast can the wall move? A Study of the electroweak phase transition dynamics}",
    eprint = "hep-ph/9506475",
    archivePrefix = "arXiv",
    reportNumber = "PUPT-1544, PUP-TH-1544, LANCS-TH-9517",
    doi = "10.1103/PhysRevD.52.7182",
    journal = "Phys. Rev. D",
    volume = "52",
    pages = "7182--7204",
    year = "1995"
}

@article{Lewicki:2021pgr,
    author = "Lewicki, Marek and Merchand, Marco and Zych, Mateusz",
    title = "{Electroweak bubble wall expansion: gravitational waves and baryogenesis in Standard Model-like thermal plasma}",
    eprint = "2111.02393",
    archivePrefix = "arXiv",
    primaryClass = "astro-ph.CO",
    doi = "10.1007/JHEP02(2022)017",
    journal = "JHEP",
    volume = "02",
    pages = "017",
    year = "2022"
}

@article{Dorsch:2023tss,
    author = "Dorsch, Glauber C. and Pinto, Daniel A.",
    title = "{Bubble wall velocities with an extended fluid Ansatz}",
    eprint = "2312.02354",
    archivePrefix = "arXiv",
    primaryClass = "hep-ph",
    doi = "10.1088/1475-7516/2024/04/027",
    journal = "JCAP",
    volume = "04",
    pages = "027",
    year = "2024"
}

@article{Branchina:2025adj,
    author = "Branchina, Carlo and Conaci, Angela and De Curtis, Stefania and Delle Rose, Luigi",
    title = "{Bubble wall velocity with out-of-equilibrium corrections}",
    eprint = "2510.21942",
    archivePrefix = "arXiv",
    primaryClass = "hep-ph",
    month = "10",
    year = "2025"
}

@article{Cline:2021dkf,
    author = "Cline, James M. and Laurent, Benoit",
    title = "{Electroweak baryogenesis from light fermion sources: A critical study}",
    eprint = "2108.04249",
    archivePrefix = "arXiv",
    primaryClass = "hep-ph",
    doi = "10.1103/PhysRevD.104.083507",
    journal = "Phys. Rev. D",
    volume = "104",
    number = "8",
    pages = "083507",
    year = "2021"
}

@article{Cline:2020jre,
    author = "Cline, James M. and Kainulainen, Kimmo",
    title = "{Electroweak baryogenesis at high bubble wall velocities}",
    eprint = "2001.00568",
    archivePrefix = "arXiv",
    primaryClass = "hep-ph",
    reportNumber = "CERN-TH-2019-227",
    doi = "10.1103/PhysRevD.101.063525",
    journal = "Phys. Rev. D",
    volume = "101",
    number = "6",
    pages = "063525",
    year = "2020"
}

@article{vandeVis:2025efm,
    author = "van de Vis, Jorinde and de Vries, Jordy and Postma, Marieke",
    title = "{Bubble Trouble: a Review on Electroweak Baryogenesis}",
    eprint = "2508.09989",
    archivePrefix = "arXiv",
    primaryClass = "hep-ph",
    reportNumber = "CERN-TH-2025-161, Nikhef 2025-012",
    month = "8",
    year = "2025"
}

@article{Dorsch:2021ubz,
    author = "Dorsch, Glauber C. and Huber, Stephan J. and Konstandin, Thomas",
    title = "{On the wall velocity dependence of electroweak baryogenesis}",
    eprint = "2106.06547",
    archivePrefix = "arXiv",
    primaryClass = "hep-ph",
    reportNumber = "DESY-21-089, DESY 21-089",
    doi = "10.1088/1475-7516/2021/08/020",
    journal = "JCAP",
    volume = "08",
    pages = "020",
    year = "2021"
}

@article{Azatov:2021ifm,
    author = "Azatov, Aleksandr and Vanvlasselaer, Miguel and Yin, Wen",
    title = "{Dark Matter production from relativistic bubble walls}",
    eprint = "2101.05721",
    archivePrefix = "arXiv",
    primaryClass = "hep-ph",
    reportNumber = "SISSA 03/2021/FISI",
    doi = "10.1007/JHEP03(2021)288",
    journal = "JHEP",
    volume = "03",
    pages = "288",
    year = "2021"
}

@article{Baker:2019ndr,
    author = "Baker, Michael J. and Kopp, Joachim and Long, Andrew J.",
    title = "{Filtered Dark Matter at a First Order Phase Transition}",
    eprint = "1912.02830",
    archivePrefix = "arXiv",
    primaryClass = "hep-ph",
    doi = "10.1103/PhysRevLett.125.151102",
    journal = "Phys. Rev. Lett.",
    volume = "125",
    number = "15",
    pages = "151102",
    year = "2020"
}

@article{Giudice:2024tcp,
    author = "Giudice, Gian F. and Lee, Hyun Min and Pomarol, Alex and Shakya, Bibhushan",
    title = "{Nonthermal heavy dark matter from a first-order phase transition}",
    eprint = "2403.03252",
    archivePrefix = "arXiv",
    primaryClass = "hep-ph",
    reportNumber = "CERN-TH-2024-031, DESY-24-031",
    doi = "10.1007/JHEP12(2024)190",
    journal = "JHEP",
    volume = "12",
    pages = "190",
    year = "2024"
}

@article{Konstandin:2014zta,
    author = "Konstandin, Thomas and Nardini, Germano and Rues, Ingo",
    title = "{From Boltzmann equations to steady wall velocities}",
    eprint = "1407.3132",
    archivePrefix = "arXiv",
    primaryClass = "hep-ph",
    reportNumber = "DESY-14-127, NSF-KITP-14-089",
    doi = "10.1088/1475-7516/2014/09/028",
    journal = "JCAP",
    volume = "09",
    pages = "028",
    year = "2014"
}

@article{Dorsch:2024jjl,
    author = "Dorsch, Gl{\'a}uber C. and Konstandin, Thomas and Perboni, Enrico and Pinto, Daniel A.",
    title = "{Non-singular solutions to the Boltzmann equation with a fluid Ansatz}",
    eprint = "2412.09266",
    archivePrefix = "arXiv",
    primaryClass = "hep-ph",
    reportNumber = "DESY-24-193",
    doi = "10.1088/1475-7516/2025/04/033",
    journal = "JCAP",
    volume = "04",
    pages = "033",
    year = "2025"
}

@article{Croon:2020cgk,
    author = "Croon, Djuna and Gould, Oliver and Schicho, Philipp and Tenkanen, Tuomas V. I. and White, Graham",
    title = "{Theoretical uncertainties for cosmological first-order phase transitions}",
    eprint = "2009.10080",
    archivePrefix = "arXiv",
    primaryClass = "hep-ph",
    reportNumber = "HIP-2020-26/TH",
    doi = "10.1007/JHEP04(2021)055",
    journal = "JHEP",
    volume = "04",
    pages = "055",
    year = "2021"
}

@article{Gould:2021ccf,
    author = "Gould, Oliver and Hirvonen, Joonas",
    title = "{Effective field theory approach to thermal bubble nucleation}",
    eprint = "2108.04377",
    archivePrefix = "arXiv",
    primaryClass = "hep-ph",
    reportNumber = "HIP-2020-19/TH",
    doi = "10.1103/PhysRevD.104.096015",
    journal = "Phys. Rev. D",
    volume = "104",
    number = "9",
    pages = "096015",
    year = "2021"
}

@article{Garbrecht:2015yza,
    author = "Garbrecht, Bjorn and Millington, Peter",
    title = "{Self-consistent solitons for vacuum decay in radiatively generated potentials}",
    eprint = "1509.08480",
    archivePrefix = "arXiv",
    primaryClass = "hep-ph",
    reportNumber = "TUM-HEP-1017-15",
    doi = "10.1103/PhysRevD.92.125022",
    journal = "Phys. Rev. D",
    volume = "92",
    pages = "125022",
    year = "2015"
}

@article{Carosi:2024lop,
    author = {Carosi, Matthias and Garbrecht, Bj{\"o}rn},
    title = "{False vacuum decay beyond the quadratic approximation: Summation of nonlocal self-energies}",
    eprint = "2411.18421",
    archivePrefix = "arXiv",
    primaryClass = "hep-th",
    reportNumber = "TUM-HEP-1540/24",
    doi = "10.1103/PhysRevD.111.085002",
    journal = "Phys. Rev. D",
    volume = "111",
    number = "8",
    pages = "085002",
    year = "2025"
}

@article{Gould:2021oba,
    author = "Gould, Oliver and Tenkanen, Tuomas V. I.",
    title = "{On the perturbative expansion at high temperature and implications for cosmological phase transitions}",
    eprint = "2104.04399",
    archivePrefix = "arXiv",
    primaryClass = "hep-ph",
    reportNumber = "NORDITA 2021-010",
    doi = "10.1007/JHEP06(2021)069",
    journal = "JHEP",
    volume = "06",
    pages = "069",
    year = "2021"
}

@article{Lewicki:2024xan,
    author = "Lewicki, Marek and Merchand, Marco and Sagunski, Laura and Schicho, Philipp and Schmitt, Daniel",
    title = "{Impact of theoretical uncertainties on model parameter reconstruction from GW signals sourced by cosmological phase transitions}",
    eprint = "2403.03769",
    archivePrefix = "arXiv",
    primaryClass = "hep-ph",
    doi = "10.1103/PhysRevD.110.023538",
    journal = "Phys. Rev. D",
    volume = "110",
    number = "2",
    pages = "023538",
    year = "2024"
}

@article{Kierkla:2023von,
    author = "Kierkla, Maciej and Swiezewska, Bogumila and Tenkanen, Tuomas V. I. and van de Vis, Jorinde",
    title = "{Gravitational waves from supercooled phase transitions: dimensional transmutation meets dimensional reduction}",
    eprint = "2312.12413",
    archivePrefix = "arXiv",
    primaryClass = "hep-ph",
    doi = "10.1007/JHEP02(2024)234",
    journal = "JHEP",
    volume = "02",
    pages = "234",
    year = "2024"
}

@article{Kierkla:2025qyz,
    author = "Kierkla, Maciej and Schicho, Philipp and Swiezewska, Bogumila and Tenkanen, Tuomas V. I. and van de Vis, Jorinde",
    title = "{Finite-temperature bubble nucleation with shifting scale hierarchies}",
    eprint = "2503.13597",
    archivePrefix = "arXiv",
    primaryClass = "hep-ph",
    reportNumber = "CERN-TH-2025-046, HIP-2024-27/TH",
    doi = "10.1007/JHEP07(2025)153",
    journal = "JHEP",
    volume = "07",
    pages = "153",
    year = "2025"
}

@article{Ekstedt:2024etx,
    author = "Ekstedt, Andreas and Schicho, Philipp and Tenkanen, Tuomas V. I.",
    title = "{Cosmological phase transitions at three loops: The final verdict on perturbation theory}",
    eprint = "2405.18349",
    archivePrefix = "arXiv",
    primaryClass = "hep-ph",
    reportNumber = "HIP-2024-15/TH",
    doi = "10.1103/PhysRevD.110.096006",
    journal = "Phys. Rev. D",
    volume = "110",
    number = "9",
    pages = "096006",
    year = "2024"
}

@article{Gould:2024jjt,
    author = "Gould, Oliver and Saffin, Paul M.",
    title = "{Perturbative gravitational wave predictions for the real-scalar extended Standard Model}",
    eprint = "2411.08951",
    archivePrefix = "arXiv",
    primaryClass = "hep-ph",
    doi = "10.1007/JHEP03(2025)105",
    journal = "JHEP",
    volume = "03",
    pages = "105",
    year = "2025"
}

@article{Gould:2023jbz,
    author = "Gould, Oliver and Xie, Cheng",
    title = "{Higher orders for cosmological phase transitions: a global study in a Yukawa model}",
    eprint = "2310.02308",
    archivePrefix = "arXiv",
    primaryClass = "hep-ph",
    doi = "10.1007/JHEP12(2023)049",
    journal = "JHEP",
    volume = "12",
    pages = "049",
    year = "2023"
}

@article{Laine:2017hdk,
    author = "Laine, M. and Meyer, M. and Nardini, G.",
    title = "{Thermal phase transition with full 2-loop effective potential}",
    eprint = "1702.07479",
    archivePrefix = "arXiv",
    primaryClass = "hep-ph",
    doi = "10.1016/j.nuclphysb.2017.04.023",
    journal = "Nucl. Phys. B",
    volume = "920",
    pages = "565--600",
    year = "2017"
}

@article{Gould:2023ovu,
    author = "Gould, Oliver and Tenkanen, Tuomas V. I.",
    title = "{Perturbative effective field theory expansions for cosmological phase transitions}",
    eprint = "2309.01672",
    archivePrefix = "arXiv",
    primaryClass = "hep-ph",
    reportNumber = "NORDITA 2023-037",
    doi = "10.1007/JHEP01(2024)048",
    journal = "JHEP",
    volume = "01",
    pages = "048",
    year = "2024"
}

@article{Laine:1994zq,
    author = "Laine, M.",
    title = "{Gauge dependence of the high temperature two loop effective potential for the Higgs field}",
    eprint = "hep-ph/9411252",
    archivePrefix = "arXiv",
    reportNumber = "HU-TFT-94-46",
    doi = "10.1103/PhysRevD.51.4525",
    journal = "Phys. Rev. D",
    volume = "51",
    pages = "4525--4532",
    year = "1995"
}

@article{Callan:1977pt,
    author = "Callan, Jr., Curtis G. and Coleman, Sidney R.",
    title = "{The Fate of the False Vacuum. 2. First Quantum Corrections}",
    reportNumber = "HUTP-77-A032",
    doi = "10.1103/PhysRevD.16.1762",
    journal = "Phys. Rev. D",
    volume = "16",
    pages = "1762--1768",
    year = "1977"
}

@article{Ekstedt:2023sqc,
    author = "Ekstedt, Andreas and Gould, Oliver and Hirvonen, Joonas",
    title = "{BubbleDet: a Python package to compute functional determinants for bubble nucleation}",
    eprint = "2308.15652",
    archivePrefix = "arXiv",
    primaryClass = "hep-ph",
    doi = "10.1007/JHEP12(2023)056",
    journal = "JHEP",
    volume = "12",
    pages = "056",
    year = "2023"
}

@article{Gould:2025wec,
    author = "Gould, Oliver and Kormu, Anna and Weir, David J.",
    title = "{Testing nucleation calculations for strong phase transitions}",
    eprint = "2502.04185",
    archivePrefix = "arXiv",
    primaryClass = "hep-lat",
    reportNumber = "HIP-2025-7/TH",
    doi = "10.22323/1.466.0365",
    journal = "PoS",
    volume = "LATTICE2024",
    pages = "365",
    year = "2025"
}

@article{Blaizot:2001nr,
    author = "Blaizot, Jean-Paul and Iancu, Edmond",
    title = "{The Quark gluon plasma: Collective dynamics and hard thermal loops}",
    eprint = "hep-ph/0101103",
    archivePrefix = "arXiv",
    reportNumber = "SACLAY-T01-005, CERN-TH-2000-272",
    doi = "10.1016/S0370-1573(01)00061-8",
    journal = "Phys. Rept.",
    volume = "359",
    pages = "355--528",
    year = "2002"
}

@article{Jeon:1994if,
    author = "Jeon, Sangyong",
    title = "{Hydrodynamic transport coefficients in relativistic scalar field theory}",
    eprint = "hep-ph/9409250",
    archivePrefix = "arXiv",
    reportNumber = "UW-PT-94-09",
    doi = "10.1103/PhysRevD.52.3591",
    journal = "Phys. Rev. D",
    volume = "52",
    pages = "3591--3642",
    year = "1995"
}

@book{Vachaspati:2006zz,
    author = "Vachaspati, Tanmay",
    title = "{Kinks and Domain Walls : An Introduction to Classical and Quantum Solitons}",
    doi = "10.1017/9781009290456",
    isbn = "978-1-009-29045-6, 978-1-009-29041-8, 978-1-009-29042-5, 978-0-521-14191-8, 978-0-521-83605-0, 978-0-511-24290-8",
    publisher = "Oxford University Press",
    year = "2007",
    pages="38"
}

@article{Witten:1984rs,
    author = "Witten, Edward",
    title = "{Cosmic Separation of Phases}",
    reportNumber = "PRINT-84-0400 (IAS,PRINCETON)",
    doi = "10.1103/PhysRevD.30.272",
    journal = "Phys. Rev. D",
    volume = "30",
    pages = "272--285",
    year = "1984"
}

@article{Navarrete:2025yxy,
    author = {Navarrete, Pablo and Paatelainen, Risto and Sepp{\"a}nen, Kaapo and Tenkanen, Tuomas V. I.},
    title = "{Cosmological phase transitions without high-temperature expansions}",
    eprint = "2507.07014",
    archivePrefix = "arXiv",
    primaryClass = "hep-ph",
    reportNumber = "HIP-2025-20/TH",
    month = "7",
    year = "2025"
}

@article{Biondini:2025ihi,
    author = "Biondini, S. and Eriksson, M. and Laine, M.",
    title = "{Computing singlet scalar freeze-out with plasmon and plasmino states}",
    eprint = "2505.05206",
    archivePrefix = "arXiv",
    primaryClass = "hep-ph",
    doi = "10.1007/JHEP08(2025)197",
    journal = "JHEP",
    volume = "08",
    pages = "197",
    year = "2025"
}

@article{Ghiglieri:2023ies,
    author = "Ghiglieri, Jacopo and Schicho, Philipp and Schlusser, Niels and Weitz, Eamonn",
    title = "{The force-force correlator at the hard thermal scale of hot QCD}",
    eprint = "2312.11731",
    archivePrefix = "arXiv",
    primaryClass = "hep-ph",
    doi = "10.1007/JHEP03(2024)111",
    journal = "JHEP",
    volume = "03",
    pages = "111",
    year = "2024"
}

@article{Ghiglieri:2021bom,
    author = "Ghiglieri, Jacopo and Moore, Guy D. and Schicho, Philipp and Schlusser, Niels",
    title = "{The force-force-correlator in hot QCD perturbatively and from the lattice}",
    eprint = "2112.01407",
    archivePrefix = "arXiv",
    primaryClass = "hep-ph",
    reportNumber = "HIP-2021-43/TH",
    doi = "10.1007/JHEP02(2022)058",
    journal = "JHEP",
    volume = "02",
    pages = "058",
    year = "2022"
}

@article{Bhatnagar:2025jhh,
    author = "Bhatnagar, Ansh and Croon, Djuna and Schicho, Philipp",
    title = "{Interpreting the 95 GeV resonance in the Two Higgs Doublet Model: Implications for the Electroweak Phase Transition}",
    eprint = "2506.20716",
    archivePrefix = "arXiv",
    primaryClass = "hep-ph",
    reportNumber = "IPPP/25/39",
    month = "6",
    year = "2025"
}

@article{Ares:2021nap,
    author = {Ares, F{\"e}anor Reuben and Henriksson, Oscar and Hindmarsh, Mark and Hoyos, Carlos and Jokela, Niko},
    title = "{Gravitational Waves at Strong Coupling from an Effective Action}",
    eprint = "2110.14442",
    archivePrefix = "arXiv",
    primaryClass = "hep-th",
    reportNumber = "HIP-2021-24/TH",
    doi = "10.1103/PhysRevLett.128.131101",
    journal = "Phys. Rev. Lett.",
    volume = "128",
    number = "13",
    pages = "131101",
    year = "2022"
}

@article{Ai:2025bjw,
    author = "Ai, Wen-Yuan and Carosi, Matthias and Garbrecht, Bjorn and Tamarit, Carlos and Vanvlasselaer, Miguel",
    title = "{Bubble wall dynamics from nonequilibrium quantum field theory}",
    eprint = "2504.13725",
    archivePrefix = "arXiv",
    primaryClass = "hep-ph",
    reportNumber = "MITP-25-029",
    doi = "10.1007/JHEP08(2025)077",
    journal = "JHEP",
    volume = "08",
    pages = "077",
    year = "2025"
}

@article{Turner:1992tz,
    author = "Turner, Michael S. and Weinberg, Erick J. and Widrow, Lawrence M.",
    title = "{Bubble nucleation in first order inflation and other cosmological phase transitions}",
    reportNumber = "FERMILAB-PUB-91-334-A, CU-TP-558, IASSNS-HEP-92-21",
    doi = "10.1103/PhysRevD.46.2384",
    journal = "Phys. Rev. D",
    volume = "46",
    pages = "2384--2403",
    year = "1992"
}

@misc{ZenodoData,
    author = "van de Vis, Jorinde and Schicho, Philipp and Niemi, Lauri and Hirvonen, Joonas and Gould, Oliver",
    title = "{Estimating theoretical uncertainties in the bubble wall velocity [Data set]}",
  year         = "2025",
  publisher    = "Zenodo",
  doi          = "10.5281/zenodo.16761243",
  url          = "https://doi.org/10.5281/zenodo.16761243"
}

@article{Collins:2016aya,
    author = "Collins, John C. and Vermaseren, J. A. M.",
    title = "{Axodraw Version 2}",
    eprint = "1606.01177",
    archivePrefix = "arXiv",
    primaryClass = "cs.OH",
    month = "5",
    year = "2016"
}

@article{Arnold:1992rz,
    author = "Arnold, Peter Brockway and Espinosa, Olivier",
    title = "{The Effective potential and first order phase transitions: Beyond leading-order}",
    eprint = "hep-ph/9212235",
    archivePrefix = "arXiv",
    reportNumber = "UW-PT-92-18, USM-TH-60",
    doi = "10.1103/PhysRevD.47.3546",
    journal = "Phys. Rev. D",
    volume = "47",
    pages = "3546",
    year = "1993",
    note = "[Erratum: Phys.Rev.D 50, 6662 (1994)]"
}

@article{Espinosa:2010hh,
    author = "Espinosa, Jose R. and Konstandin, Thomas and No, Jose M. and Servant, Geraldine",
    title = "{Energy Budget of Cosmological First-order Phase Transitions}",
    eprint = "1004.4187",
    archivePrefix = "arXiv",
    primaryClass = "hep-ph",
    reportNumber = "CERN-PH-TH-2010-027",
    doi = "10.1088/1475-7516/2010/06/028",
    journal = "JCAP",
    volume = "06",
    pages = "028",
    year = "2010"
}

@article{Hindmarsh:2017gnf,
    author = "Hindmarsh, Mark and Huber, Stephan J. and Rummukainen, Kari and Weir, David J.",
    title = "{Shape of the acoustic gravitational wave power spectrum from a first order phase transition}",
    eprint = "1704.05871",
    archivePrefix = "arXiv",
    primaryClass = "astro-ph.CO",
    reportNumber = "HIP-2017-02-TH, HIP-2017-02/TH",
    doi = "10.1103/PhysRevD.96.103520",
    journal = "Phys. Rev. D",
    volume = "96",
    number = "10",
    pages = "103520",
    year = "2017",
    note = "[Erratum: Phys.Rev.D 101, 089902 (2020)]"
}

@article{Hindmarsh:2019phv,
    author = "Hindmarsh, Mark and Hijazi, Mulham",
    title = "{Gravitational waves from first order cosmological phase transitions in the Sound Shell Model}",
    eprint = "1909.10040",
    archivePrefix = "arXiv",
    primaryClass = "astro-ph.CO",
    reportNumber = "NORDITA-2019-083, HIP-2019-29/TH",
    doi = "10.1088/1475-7516/2019/12/062",
    journal = "JCAP",
    volume = "12",
    pages = "062",
    year = "2019"
}

@article{Hindmarsh:2016lnk,
    author = "Hindmarsh, Mark",
    title = "{Sound shell model for acoustic gravitational wave production at a first-order phase transition in the early Universe}",
    eprint = "1608.04735",
    archivePrefix = "arXiv",
    primaryClass = "astro-ph.CO",
    doi = "10.1103/PhysRevLett.120.071301",
    journal = "Phys. Rev. Lett.",
    volume = "120",
    number = "7",
    pages = "071301",
    year = "2018"
}

@article{Kajantie:1995dw,
    author = "Kajantie, K. and Laine, M. and Rummukainen, K. and Shaposhnikov, Mikhail E.",
    title = "{Generic rules for high temperature dimensional reduction and their application to the standard model}",
    eprint = "hep-ph/9508379",
    archivePrefix = "arXiv",
    reportNumber = "CERN-TH-95-226, HU-TFT-95-50, IUHET-312",
    doi = "10.1016/0550-3213(95)00549-8",
    journal = "Nucl. Phys. B",
    volume = "458",
    pages = "90--136",
    year = "1996"
}

@article{Hirvonen:2024rfg,
    author = "Hirvonen, Joonas",
    title = "{Real-time nucleation and off-equilibrium effects in high-temperature quantum field theories}",
    eprint = "2403.07987",
    archivePrefix = "arXiv",
    primaryClass = "hep-ph",
    reportNumber = "HIP-2024-6/TH",
    doi = "10.1103/4p7n-nyln",
    journal = "Phys. Rev. D",
    volume = "111",
    number = "11",
    pages = "116020",
    year = "2025"
}

@incollection{LIFSHITZ19811,
title = "{Chapter I - Kinetic theory of gases}",
editor = {E.M. Lifshitz and L.P. Pitaevski},
booktitle = {Physical Kinetics},
publisher = {Pergamon},
address = {Amsterdam},
pages = {1-88},
year = {1981},
volume = {10},
series = {Course of Theoretical Physics},
isbn = {978-0-08-026480-6},
doi = {https://doi.org/10.1016/B978-0-08-026480-6.50006-0},
url = {https://www.sciencedirect.com/science/article/pii/B9780080264806500060},
author = {E.M. Lifshitz and L.P. Pitaevski}
}

@article{Anisimov:2010gy,
    author = "Anisimov, Alexey and Besak, Denis and Bodeker, Dietrich",
    title = "{Thermal production of relativistic Majorana neutrinos: Strong enhancement by multiple soft scattering}",
    eprint = "1012.3784",
    archivePrefix = "arXiv",
    primaryClass = "hep-ph",
    reportNumber = "BI-TP-2010-48",
    doi = "10.1088/1475-7516/2011/03/042",
    journal = "JCAP",
    volume = "03",
    pages = "042",
    year = "2011"
}

@article{Ghiglieri:2016xye,
    author = "Ghiglieri, J. and Laine, M.",
    title = "{Neutrino dynamics below the electroweak crossover}",
    eprint = "1605.07720",
    archivePrefix = "arXiv",
    primaryClass = "hep-ph",
    doi = "10.1088/1475-7516/2016/07/015",
    journal = "JCAP",
    volume = "07",
    pages = "015",
    year = "2016"
}

@article{Ghiglieri:2014kma,
    author = "Ghiglieri, Jacopo and Moore, Guy D.",
    title = "{Low Mass Thermal Dilepton Production at NLO in a Weakly Coupled Quark-Gluon Plasma}",
    eprint = "1410.4203",
    archivePrefix = "arXiv",
    primaryClass = "hep-ph",
    doi = "10.1007/JHEP12(2014)029",
    journal = "JHEP",
    volume = "12",
    pages = "029",
    year = "2014"
}

@article{Cline:2025bwe,
    author = "Cline, James M. and Laurent, Benoit",
    title = "{Bubble wall velocity for first-order QCD phase transition}",
    eprint = "2502.12321",
    archivePrefix = "arXiv",
    primaryClass = "hep-ph",
    doi = "10.1103/PhysRevD.111.083522",
    journal = "Phys. Rev. D",
    volume = "111",
    number = "8",
    pages = "083522",
    year = "2025"
}

@article{Ghiglieri:2015ala,
    author = "Ghiglieri, Jacopo and Moore, Guy D. and Teaney, Derek",
    title = "{Jet-Medium Interactions at NLO in a Weakly-Coupled Quark-Gluon Plasma}",
    eprint = "1509.07773",
    archivePrefix = "arXiv",
    primaryClass = "hep-ph",
    doi = "10.1007/JHEP03(2016)095",
    journal = "JHEP",
    volume = "03",
    pages = "095",
    year = "2016"
}

@article{Kozaczuk:2015owa,
    author = "Kozaczuk, Jonathan",
    title = "{Bubble Expansion and the Viability of Singlet-Driven Electroweak Baryogenesis}",
    eprint = "1506.04741",
    archivePrefix = "arXiv",
    primaryClass = "hep-ph",
    doi = "10.1007/JHEP10(2015)135",
    journal = "JHEP",
    volume = "10",
    pages = "135",
    year = "2015"
}

@article{Shtabovenko:2023idz,
    author = "Shtabovenko, Vladyslav and Mertig, Rolf and Orellana, Frederik",
    title = "{FeynCalc 10: Do multiloop integrals dream of computer codes?}",
    eprint = "2312.14089",
    archivePrefix = "arXiv",
    primaryClass = "hep-ph",
    reportNumber = "P3H-23-089, TTP23-056, SI-HEP-2023-27",
    doi = "10.1016/j.cpc.2024.109357",
    journal = "Comput. Phys. Commun.",
    volume = "306",
    pages = "109357",
    year = "2025"
}

@article{Hahn:2000kx,
    author = "Hahn, Thomas",
    title = "{Generating Feynman diagrams and amplitudes with FeynArts 3}",
    eprint = "hep-ph/0012260",
    archivePrefix = "arXiv",
    reportNumber = "KA-TP-23-2000",
    doi = "10.1016/S0010-4655(01)00290-9",
    journal = "Comput. Phys. Commun.",
    volume = "140",
    pages = "418--431",
    year = "2001"
}

@article{tHooft:1978jhc,
    author = "'t Hooft, Gerard and Veltman, M. J. G.",
    title = "{Scalar One Loop Integrals}",
    reportNumber = "PRINT-79-0134 (UTRECHT)",
    doi = "10.1016/0550-3213(79)90605-9",
    journal = "Nucl. Phys. B",
    volume = "153",
    pages = "365--401",
    year = "1979"
}

@article{Denner:1991kt,
    author = "Denner, Ansgar",
    title = "{Techniques for calculation of electroweak radiative corrections at the one loop level and results for W physics at LEP-200}",
    eprint = "0709.1075",
    archivePrefix = "arXiv",
    primaryClass = "hep-ph",
    reportNumber = "PRINT-91-0349 (WURZBURG)",
    doi = "10.1002/prop.2190410402",
    journal = "Fortsch. Phys.",
    volume = "41",
    pages = "307--420",
    year = "1993"
}

@book{Duncan:2012aja,
    author = "Duncan, Anthony",
    title = "{The Conceptual Framework of Quantum Field Theory}",
    doi = "10.1093/acprof:oso/9780199573264.001.0001",
    isbn = "978-0-19-880765-0, 978-0-19-880765-0, 978-0-19-957326-4",
    publisher = "Oxford University Press",
    month = "8",
    year = "2012"
}

@article{Fonseca:2020vke,
    author = "Fonseca, Renato M.",
    title = "{GroupMath: A Mathematica package for group theory calculations}",
    eprint = "2011.01764",
    archivePrefix = "arXiv",
    primaryClass = "hep-th",
    doi = "10.1016/j.cpc.2021.108085",
    journal = "Comput. Phys. Commun.",
    volume = "267",
    pages = "108085",
    year = "2021"
}

@article{Kierkla:2025vwp,
    author = "Kierkla, Maciej and Ramberg, Nicklas and Schicho, Philipp and Schmitt, Daniel",
    title = "{Theoretical uncertainties for primordial black holes from cosmological phase transitions}",
    eprint = "2506.15496",
    archivePrefix = "arXiv",
    primaryClass = "hep-ph",
    reportNumber = "SISSA 07/2025/FISI",
    month = "6",
    year = "2025"
}

@article{Ekstedt:2023oqb,
    author = "Ekstedt, Andreas",
    title = "{Propagation of gauge fields in hot and dense plasmas at higher orders}",
    eprint = "2304.09255",
    archivePrefix = "arXiv",
    primaryClass = "hep-ph",
    month = "4",
    year = "2023"
}

@article{Bernardo:2025vkz,
    author = "Bernardo, Fabio and Klose, Philipp and Schicho, Philipp and Tenkanen, Tuomas V. I.",
    title = "{Higher-dimensional operators at finite temperature affect gravitational-wave predictions}",
    eprint = "2503.18904",
    archivePrefix = "arXiv",
    primaryClass = "hep-ph",
    reportNumber = "HIP-2025-6/TH",
    doi = "10.1007/JHEP08(2025)109",
    journal = "JHEP",
    volume = "08",
    pages = "109",
    year = "2025"
}

@article{Chala:2025aiz,
    author = "Chala, Mikael and Guedes, Guilherme",
    title = "{The high-temperature limit of the SM(EFT)}",
    eprint = "2503.20016",
    archivePrefix = "arXiv",
    primaryClass = "hep-ph",
    doi = "10.1007/JHEP07(2025)085",
    journal = "JHEP",
    volume = "07",
    pages = "085",
    year = "2025"
}

@article{Chala:2024xll,
    author = "Chala, Mikael and Criado, Juan Carlos and Gil, Luis and Miras, Javier L{\'o}pez",
    title = "{Higher-order-operator corrections to phase-transition parameters in dimensional reduction}",
    eprint = "2406.02667",
    archivePrefix = "arXiv",
    primaryClass = "hep-ph",
    doi = "10.1007/JHEP10(2024)025",
    journal = "JHEP",
    volume = "10",
    pages = "025",
    year = "2024"
}

@article{Chala:2025oul,
    author = "Chala, Mikael and Gil, Luis and Ren, Zhe",
    title = "{Phase transitions in dimensional reduction up to three loops}",
    eprint = "2505.14335",
    archivePrefix = "arXiv",
    primaryClass = "hep-ph",
    doi = "10.1088/1674-1137/adf322",
    journal = "Chin. Phys.",
    volume = "49",
    number = "12",
    pages = "123105",
    year = "2025"
}

@article{Chakrabortty:2024wto,
    author = "Chakrabortty, Joydeep and Mohanty, Subhendra",
    title = "{One Loop Thermal Effective Action}",
    eprint = "2411.14146",
    archivePrefix = "arXiv",
    primaryClass = "hep-th",
    doi = "10.1016/j.nuclphysb.2025.117165",
    journal = "Nucl. Phys. B",
    volume = "1020",
    pages = "117165",
    year = "2025"
}
}
\end{document}